\documentclass[aps,prd,twocolumn,showpacs,floats,nofootinbib,longbibliography,10pt]{revtex4-2}

\usepackage{multirow}
\usepackage{graphicx}
\usepackage{amsmath,amsfonts}
\usepackage{dcolumn}
\usepackage{hyperref}
\usepackage{orcidlink}

\mathchardef\minus = "002D

\DeclareSymbolFont{toneletters}{T1}{\familydefault}{m}{it}
\DeclareMathSymbol\edthm{\mathord}{toneletters}{"F0}

\DeclareSymbolFont{EulerFraktur}{U}{euf}{m}{n}
\DeclareMathSymbol\WigD{\mathord}{EulerFraktur}{"44}
\DeclareMathSymbol\Wigd{\mathord}{EulerFraktur}{"64}
\let\Re\relax
\DeclareMathOperator{\Re}{{Re}}
\let\Im\relax
\DeclareMathOperator{\Im}{{Im}}

\newcommand{\swY}[4][]{{}_{{}_{#2}}\!Y^{#1}_{#3}(#4)}   
\newcommand{\swSH}[5][]{{}_{{}_{#2}}S^{#1}_{#3}(#4;#5)} 
\newcommand{\swsS}[4][]{{}_{{}_{#2}}S^{#1}_{#3}(#4)}  
\newcommand{\swS}[5][]{{}_{{}_{#2}}S^{#1}_{#3}(#4;#5)}  
\newcommand{\scA}[4][]{{}_{{}_{#2}}A^{#1}_{#3}(#4)} 
\newcommand{\YSH}[4][]{{}_{{}_{#2}}\mathcal{A}^{#1}_{#3}(#4)}

\newcommand{\Phase}[1]{\mathcal{P}_{\text{#1}}}
\newcommand{\Pfactor}[1]{P_{\text{#1}}}

\newcolumntype{.}{D{.}{.}{-1}}
\newcolumntype{d}[1]{D{.}{.}{#1}}

\DeclareMathOperator{\sign}{sign}

\begin{document}

\title{Choosing the phase for the spin-weighted spheroidal functions}

\author{Gregory B. Cook\,\orcidlink{0000-0002-4395-7617}}\email{cookgb@wfu.edu}
\affiliation{Department of Physics, Wake Forest University,
		 Winston-Salem, North Carolina 27109, USA}
\author{Xiyue Wang\,\orcidlink{0009-0006-3331-0832}}\email{xiyuew@caltech.edu}
\affiliation{Department of Physics, Wake Forest University,
		 Winston-Salem, North Carolina 27109, USA}

\date{\today}

\begin{abstract}
The spin-weighted spheroidal functions are the eigenfunctions of the angular Teukolsky equation.  They are a generalization of the widely used spin-weighted spherical functions, and are extremely important in the area of black-hole perturbation theory.  Like other special functions, they have an inherent phase ambiguity and need to be phase fixed to be uniquely defined.  Clearly, such a phase choice does not have a direct physical impact.  But, without specifying phase choices, meaningful comparison between results reported in various works becomes more difficult.  To date, possible phase choices for the spin-weighted spheroidal functions have received little attention.  Here, we clearly define and extensively explore two useful phase fixing schemes, and we propose that the spherical-limit phase-fixing scheme be adopted as the default phase-fixing scheme for the spin-weighted spheroidal functions.
\end{abstract}

\maketitle

\section{Introduction}
First introduced by Teukolsky\cite{teukolsky-1973}, the spin-weighted spheroidal harmonics(SWSHs) $\swSH{s}{\ell m}{\theta,\phi}{c}$ have their primary application in the arena of black-hole perturbation theory, although they are useful in other contexts.  They are generalizations of both the scalar spheroidal harmonics $\swSH{0}{\ell m}{\theta,\phi}{c}$ and the spin-weighted spherical harmonics $\swY{s}{\ell m}{\theta,\phi}=\swSH{s}{\ell m}{\theta,\phi}{0}$, both of which have much broader application.  The spin-weighted spherical harmonics\cite{NewmanPenrose-1966,Goldberg-etal-1967,thorne80,Dray-1985} $\swY{s}{\ell m}{\theta,\phi}$ provide a complete set of orthonormal basis functions for representing tensor- or spinor-valued functions\cite{Campbell-1971} on the surface of a sphere, and their broad use in physics is facilitated by the fact that their functional representation and properties are known analytically.  The scalar spheroidal harmonics $\swSH{0}{\ell m}{\theta,\phi}{c}$ are generalizations of the standard scalar spherical harmonics $\swY{0}{\ell m}{\theta,\phi}$ and are useful in a range of situations\cite{Flammer}.  For example, they arise naturally when separation of variables is performed in spheroidal rather than spherical coordinates.  In this case, the additional argument $c$ is related to the oblateness of the spheroidal coordinates. They are somewhat more difficult to use than spherical harmonics because the closed-form functional representations of their eigenvalues and eigenvectors are known only for special cases\cite{Barrowes-etal-2004}.  Beyond the scalar case, the general SWSHs are even more complicated to use with even less known about the general behavior of their eigensolutions\cite{berticardosocasals-2006,VickersCook2022}.

An especially important application of the SWSHs is in determining the various modes of the Kerr geometry\cite{teukolsky-1973,leaver-1985,cook-zalutskiy-2014}.  While considering the linear perturbation of various fields near a rotating black hole, Teukolsky found that a single equation, known as the Teukolsky master equation, could describe the linear perturbation of a field of any spin weight.  Moreover, the equation proved separable with the angular behavior being described by the SWSHs.  With appropriate boundary conditions, the radial and angular Teukolsky equations can be solved to compute both the quasinormal modes (QNMs) and total-transmission modes (TTMs) of the Kerr geometry.  Of these modes, so far, the QNMs have proven to be most useful.  The behavior of an isolated but perturbed rotating black hole is well described by a linear combination of QNMs\footnote{The QNMs do not form a complete set of functions.  The set of modes associated with a linearly-perturbed black hole are known to include power-law tails and a prompt response\cite{Leaver-1986b}.  Furthermore, quadratic modes from second-order perturbation theory can be important.}.  The gravitational-wave signal produced by a remnant black hole resulting from a compact-binary coalescence is referred to as a ringdown signal and it can be well fit by such a linear combination of QNMs.  This is the foundation of black-hole spectroscopy(BHS)\cite{Dreyer-etal-2004,BHspectroscopy-2026} which is a general program aimed at extracting as much physical information as possible from a black hole's ringdown signal.  In its simplest incarnation, calibrating BHS begins by fitting the ringdown waveforms from a large set of numerical simulations covering some portion of the parameter space of possible progenitor systems, and extracting a set of QNM expansion coefficients as a function of the known progenitor parameters.  Extracting and exploring these QNM expansion coefficients from various catalogs of numerical simulations has been undertaken in several works\cite{kamaretsos-et-al-2012b,londonShoemakerhealy-2014,baibhav-etal-2018,London-2020,Borhanian-etal-2020,Cheung-etal-2024,Pacilio-etal-2024,Zertuche-etal-2025,Mitman-etal-2025a,Zhu-etal-2025,DyerMoore-2026a}.  The earliest such explorations focused on the magnitude of these expansion coefficients, but more recently the full complex coefficients have been explored.  However, the fact that the SWSHs are inherently complex angular functions with phase freedom has not been carefully considered.

The spin-weighted spherical harmonics $\swY{s}{\ell m}{\theta,\phi}$ are complex functions which consist of a real function multiplied by a factor of $e^{im\phi}$.  In a similar way, we can isolate the spin-weighted spheroidal functions(SWSFs) $\swS{s}{\ell m}{x}{c}$ using the relation\footnote{The definition in Eq.~(\ref{eqn:harmonic def}) differs slightly from similar definitions used in Refs.~\cite{cook-zalutskiy-2014,VickersCook2022} where a different normalization of the SWSFs was used.}
\begin{align}\label{eqn:harmonic def}
	\swSH{s}{\ell m}{\theta,\phi}{c}\equiv\frac1{\sqrt{2\pi}}\swS{s}{\ell m}{\cos\theta}{c}e^{im\phi}.
\end{align}
The SWSFs $\swS{s}{\ell m}{x}{c}$ are inherently complex functions of the complex oblateness parameter $c$.  So, any phase information extracted from the QNM expansion coefficients will be affected by the phase choice made for the SWSF associated with each mode.  At the most basic level, the considerable phase freedom present in the SWSHs means that authors should be explicit about their phase convention when using SWSHs and SWSFs.

The main purpose of this paper is to propose possible choices for fixing the phase of the SWSFs.  To do this, we will begin in Sec.~\ref{sec:properties} by examining the properties of various angular functions and harmonics.  In Sec.~\ref{sec:Phase choices}, we discuss the simple phase choice which is currently in use and propose a better motivated phase choice for the SWSFs.  In Sec.~\ref{sec:examples}, we explore the behavior of different phase choices by applying them to the SWSFs which are part of each mode in a large set of publicly available QNM and TTM modes sequences\cite{KerrModeData-cook-2025}.  Following a brief summary and discussion in Sec.~\ref{sec:discussion}, we include two Appendices.  Appendix~\ref{sec:additional_figures} includes figures associated with the examples in Sec.~\ref{sec:examples}. In Appendix~\ref{sec:SWSpheroidal}, we provide an overview of the publicly available {\em Mathematica} Paclet {\tt SWSpheroidal} which can be used to compute high-precision solutions of the spin-weighted spheroidal eigenvalue problem.

\section{Properties of the angular functions and harmonics}
\label{sec:properties}

The SWSFs satisfy the following differential equation:
\begin{align}\label{eqn:Angular Teukolsky Equation}
\partial_x \Big[ (1-x^2)\partial_x [\swS{s}{\ell{m}}{x}{c}]\Big] 
& \nonumber \\ 
    + \bigg[(cx)^2 - 2 csx + s& + \scA{s}{\ell m}{c} 
 \\ 
      & - \frac{(m+sx)^2}{1-x^2}\bigg]\swS{s}{\ell{m}}{x}{c} = 0,
\nonumber
\end{align}
often referred to as the angular Teukolsky equation\cite{teukolsky-1973,cook-zalutskiy-2014}.  The physically meaningful solutions are those for which the solution is regular at $x=\pm1$, and are defined over the range $-1\le x\le1$.  

The differential equation is parameterized by the quantities $c$, $s$, $m$, and $\scA{s}{\ell m}{c}$.  The complex constant $c$ is often referred to as the oblateness parameter.  When $s=0$ and $c^2$ is real and positive(negative), then the solutions $\swS{0}{\ell{m}}{x}{c}$ are known as the oblate(prolate) angular spheroidal functions.  The parameter $s$ determines the spin weight of the solutions, and functions of various spin weight can be used to represent the components of tensor- or spinor-valued functions on the surface of a sphere.  $s$ takes on either integer or half-integer values.  That is $s=\pm \frac{N}2$ where $N$ is any non-negative integer.  Scalar valued functions on the sphere have spin-weight $s=0$, while vectors on the sphere can be represented in terms of $s=\pm1$ functions.  The $m$ parameter is the azimuthal index which serves as the separation constant for the azimuthal($\phi$) coordinate and takes on integer(half-odd integer) values when $s$ is an integer(half-odd integer).  Finally, $\scA{s}{\ell m}{c}$ is the angular separation constant.  It is the eigenvalue of Eq.~(\ref{eqn:Angular Teukolsky Equation}) with the associated eigenfunction $\swS{s}{\ell{m}}{x}{c}$, and there exist a countably infinite number of eigensolutions determined by the regularity of the solution $\swS{s}{\ell{m}}{x}{c}$ at $x=\pm1$.  

The set of solutions can be most easily enumerated by an integer index $L\ge0$ with solutions ordered in terms of increasing values of either the real part of the eigenvalue $\Re[\scA{s}{\ell m}{c}]$, or the magnitude $|\scA{s}{\ell m}{c}+s|$.\footnote{The necessity of ordering based on the magnitude of this offset value will be discussed below in Sec.~\ref{sec:general sw spheroidal}.}  However, a more standard convention is to use the polar index $\ell\equiv L+\max(|m|,|s|)$, in which case $\ell$ takes on integer(half-odd integer) values when $s$ is an integer(half-odd integer).  We caution that care must always be taken when dealing with either $L$ or $\ell$ in enumerating eigensolutions\cite{VickersCook2022,Barrowes-etal-2004}.  For the commonly used scalar or spin-weighted {\em spherical} harmonics, the eigenvalues can be directly computed from $\ell$ and $s$, and for fixed $m$ these eigenvalues are real and nondegenerate.  In general, however, the $\scA{s}{\ell m}{c}$ are complex.  If we consider $\Re[\scA{s}{\ell m}{c}]$ and $\Im[\scA{s}{\ell m}{c}]$ as functions of $c$, then continuous surfaces over the complex plane of a given eigenvalue can cross each other.  Furthermore, full degeneracies leading to branch points and branch cuts can exist\cite{HartleWilkins1974,Barrowes-etal-2004}.  Because of such complications, a globally consistent labeling of sequences of eigensolutions may not always be possible.

We will assume that the eigenfunctions $\swS{s}{\ell{m}}{x}{c}$ are always normalized such that
\begin{align}\label{eqn:normalized SWSF}
	\int_{-1}^1{|\swS{s}{\ell{m}}{x}{c}|^2dx}=1.
\end{align}
By the definition of Eq.~(\ref{eqn:harmonic def}) the associated SWSHs are also normalized over the unit sphere
\begin{align}\label{eqn:normalized SWSH}
	\oint{|\swSH{s}{\ell{m}}{\theta,\phi}{c}|^2d\Omega}=1.
\end{align}
Of course, there remains an overall phase freedom in the definitions of both the SWSFs and SWSHs.  The primary goal of this work is to explore useful choices for fixing this phase freedom.

For fixed values of $c$, $s$, and $m$, the SWSFs, while they can be normalized, are not orthogonal in general\cite{berticardosocasals-2006,London-2023}\footnote{Stewart\cite{Stewart-1975} states incorrectly, but without proof, that the SWSFs are orthonormal.}. That is, for normalized SWSFs,
\begin{align}\label{eqn:non-othogonal}
	\int_{-1}^1{\swS[*]{s}{\acute\ell{m}}{x}{c}\swS{s}{\ell{m}}{x}{c}dx}= {}_{{}_s}\alpha_{m\acute\ell\ell}(c)\not\propto\delta_{\acute\ell\ell}.
\end{align}
Similarly, for the normalized SWSHs, we have
\begin{align}\label{eqn:non-othogonal-harm}
	\oint{\swSH[*]{s}{\acute\ell\acute{m}}{\theta,\phi}{c}\swSH{s}{\ell{m}}{\theta,\phi}{c}d\Omega}= {}_{{}_s}\alpha_{m\acute\ell\ell}(c)\delta_{\acute{m}m}.
\end{align}
However, for the special cases of $c=0$, or $s=0$ with $c^2$ real, the functions and harmonics are orthogonal.

Interestingly, Ref.~\cite{London-2023} recently pointed out that, for fixed values of $c$, $s$, and $m$, the SWSFs are biorthogonal and complete, which for the SWSFs means that 
\begin{align}\label{eqn:biothogonal}
	\int_{-1}^1{\swS{s}{\acute\ell{m}}{x}{c}\swS{s}{\ell{m}}{x}{c}dx}\propto\delta_{\acute\ell\ell},
	\intertext{and}\label{eqn:c completess}
	{}_{{}_s}f(x)=\sum_\ell{\beta_{\ell}\swS{s}{\ell{m}}{x}{c}}.
\end{align}
Apparently, these results have been known to functional analysis for some time\cite{Brauer-1964}.  From Eq.~(\ref{eqn:c completess}), it immediately follows that a general spin-weight $s$ function ${}_sf(\theta,\phi)$ on the unit sphere can be represented as a linear combination of SWSHs which all have the same value of $c$
\begin{align}
	{}_sf(\theta,\phi)=\sum_{\ell{m}}{\beta_{\ell{m}}\swSH{s}{\ell{m}}{\theta,\phi}{c}}.
\end{align}

To avoid confusion, it is important to emphasize that the completeness of the SWSFs discussed so far requires that the same value of $c$ be used for all SWSFs.  This is in contrast, for example, to the interesting case of expansions in terms of sets of QNMs (cf.~\cite{cook-ringdown-2020}) where each SWSF in the expansion is evaluated at a different value of $c$.  For this case, it has been shown\cite{London-2023} that particular subsets of SWSFs with different values of $c$ are complete.

\subsection{Basic symmetries and phase freedoms}
\label{sec:basicsymmetries}

Three basic symmetries of Eq.~(\ref{eqn:Angular Teukolsky Equation}) constrain the eigensolutions.  Simultaneous exchange of $\{s\to-s,x\to-x\}$ in Eq.~(\ref{eqn:Angular Teukolsky Equation}), leads to
\begin{subequations}\label{eqn:swSF sx idents}
\begin{align}
\label{eqn:swSF sx S ident}
	\swS{-s}{\ell{m}}{x}{c} &= e^{i\varphi_1}\swS{s}{\ell{m}}{-x}{c}, \\
\label{eqn:swSF sx A ident}
	\scA{-s}{\ell{m}}{c} &= \scA{s}{\ell{m}}{c} + 2s,
\end{align}
\end{subequations}
where $\varphi_1$ is an as yet unspecified real number representing a phase freedom.  Simultaneous exchange of $\{m\to-m,x\to-x,c\to-c\}$ in Eq.~(\ref{eqn:Angular Teukolsky Equation}), leads to
\begin{subequations}\label{eqn:swSF mxc idents}
\begin{align}
\label{eqn:swSF mxc S ident}
	\swS{s}{\ell{(-m)}}{x}{c} &= e^{i\varphi_2}\swS{s}{\ell{m}}{-x}{-c}, \\
\label{eqn:swSF mxc A ident}
	\scA{s}{\ell{(-m)}}{c} &= \scA{s}{\ell{m}}{-c},
\end{align}
\end{subequations}
where $\varphi_2$ is another as yet unspecified real number representing a phase freedom.  Finally, complex conjugation of Eq.~(\ref{eqn:Angular Teukolsky Equation}), leads to
\begin{subequations}\label{eqn:swSF cc idents}
\begin{align}
\label{eqn:swSF cc S ident}
	\swS[*]{s}{\ell{m}}{x}{c} &= e^{i\varphi_3}\swS{s}{\ell{m}}{x}{c^*}, \\
\label{eqn:swSF cc A ident}
	\scA[*]{s}{\ell{m}}{c} &= \scA{s}{\ell{m}}{c^*},
\end{align}
\end{subequations}
where $\varphi_3$ is a third as yet unspecified real number representing a phase freedom.

\subsection{Spin-weighted spherical functions and harmonics}
\label{sec:sw-spherical}

When $c=0$, the SWSHs reduce to the spin-weighted spherical harmonics $\swY{s}{\ell{m}}{\theta,\phi}$ which are orthonormal and complete.  These functions will serve two purposes.  First we can use them, and their related angular functions $\swsS{s}{\ell{m}}{x}$, as a basis for an expansion of the SWSHs and SWSFs.  We will define
\begin{subequations}\label{eqn:sw-expansion}
\begin{align}\label{eqn:swSH-expansion}
	\swSH{s}{\ell{m}}{\theta,\phi}{c} = \!\!\!\!\!\sum_{\hat\ell=\max(|m|,|s|)}^\infty{\!\!\!\!\!\YSH{s}{\hat\ell\ell{m}}{c}\swY{s}{\hat\ell{m}}{\theta,\phi}}, \\
	\intertext{and equivalently}\label{eqn:swSF-expansion}
	\swS{s}{\ell{m}}{x}{c} = \!\!\!\!\!\sum_{\hat\ell=\max(|m|,|s|)}^\infty{\!\!\!\!\!\YSH{s}{\hat\ell\ell{m}}{c}\swsS{s}{\hat\ell{m}}{x}}.
\end{align}
\end{subequations}
Second, they will serve as a foundation for defining our phase conventions for the SWSFs and SWSHs.  We also note that we are careful to use special notation to distinguish between the $c=0$ limit of the SWSFs $\swS{s}{\ell{m}}{x}{0}$ and the spin-weighted spherical functions $\swsS{s}{\ell{m}}{x}$, as they are not required to share the same phase choice.

The spin-weighted spherical harmonics $\swY{s}{\ell{m}}{\theta,\phi}$ are proportional to Wigner's $\WigD^{(\ell)}_{m,-s}(\phi,\theta,0)$ matrices\cite{Goldberg-etal-1967,boyle-2016a}, and the spin-weighted spherical functions $\swsS{s}{\ell{m}}{x}$ are similarly proportional to Wigner's $\Wigd^{(\ell)}_{m,-s}(\cos^{-1}x)$ matrices where
\begin{align}\label{eqn:Wigner-D-d-def}
	\WigD^{(\ell)}_{m,n}(\phi,\theta,\gamma) = e^{-im\phi}\Wigd^{(\ell)}_{m,n}(\theta)e^{-in\gamma}.
\end{align}
The $\Wigd^{(\ell)}_{m,n}(\theta)$ matrix elements can be expressed as
\begin{widetext}
	\begin{align}\label{eqn:Wigner-d-form}
		\Wigd^{(\ell)}_{m,n}(\theta)=\sqrt{\frac{(\ell+m)!(\ell-m)!}{(\ell+n)!(\ell-n)!}}\left[\sin\left(\frac{\theta}2\right)\right]^{2\ell}\sum_{r=0}^{\ell-n}(-1)^{\ell-r-n}\left(\begin{array}{c}\ell-n\\r\end{array}\right)\left(\begin{array}{c}\ell+n\\r+n+m\end{array}\right)\left[\cot\left(\frac{\theta}2\right)\right]^{2r+n+m},
	\end{align}
\end{widetext}
where $\left(\begin{array}{c}a\\b\end{array}\right)$ is a binomial coefficient.\footnote{Given Eq.~(\ref{eqn:Wigner-d-form}) the Wigner-$\WigD$ matrix elements are related to the {\em Mathematica} function {\tt WignerD} by $\WigD^{(\ell)}_{m,n}(\phi,\theta,\gamma)=\text{\tt WignerD[$\{\ell,-m,-n\},\phi,\theta,\gamma$]}$.}  Given the definitions in Eqs.~(\ref{eqn:Wigner-d-form}) and (\ref{eqn:Wigner-D-d-def}) we choose to define the spin-weighted spherical harmonics by
\begin{align}\label{eqn:swY-def}
	\swY{s}{\ell{m}}{\theta,\phi} = (-1)^{s-\eta}\sqrt{\frac{2\ell+1}{4\pi}}\WigD^{(\ell)*}_{m,-s}(\phi,\theta,0),
\end{align}
where $\eta=0$ when $s$ takes on integer values, and $\eta=\frac12$ when $s$ takes on half-odd integer values.  The spin-weighted spherical functions, related by Eq.~(\ref{eqn:harmonic def}) with $c=0$, can then be expressed as
\begin{subequations}\label{eqn:spherical-function def}
\begin{align}\label{eqn:spherical-func-s}
	\swsS{s}{\ell{m}}{x} &= (-1)^{s-\eta}\sqrt{\ell+\frac12}\Wigd^{(\ell)}_{m,-s}(\cos^{-1}x), \\
	\label{eqn:spherical-func-m}
	&= (-1)^{-m-\eta}\sqrt{\ell+\frac12}\Wigd^{(\ell)}_{-m,s}(\cos^{-1}x).
\end{align}
\end{subequations}
The factor of $(-1)^{s-\eta}$ in Eq.~(\ref{eqn:swY-def}) is a phase choice and enforces the modern definition of the Condon-Shortley phase convention for integer values of $s$, but the factor of $\eta$ also guarantees that the spin-weighted spherical functions remain real for half-odd integer values of $s$.  In this work, we are careful to write all phase factors so that they are correct for both integer and half-odd integer index values.

Choices for the phase freedoms given in Eqs.~(\ref{eqn:swSF sx S ident}), (\ref{eqn:swSF mxc S ident}), and (\ref{eqn:swSF cc S ident}) are informed by the basic symmetries of the Wigner-$\Wigd$ matrices defined via
\begin{align}
	\Wigd^{(\ell)}_{\acute{m},m}(\theta)=\langle\ell\acute{m}|e^{-i\theta{J_y}}|\ell{m}\rangle.
\end{align}
Here, $J_x$, $J_y$, and $J_z$ are the usual quantum mechanical angular momentum operators defined by $[J_i,J_j]=i\epsilon_{ijk}J_k$ with $\hbar=1$; $J^2|\ell{m}\rangle=\ell(\ell+1)|\ell{m}\rangle$ and $J_z|\ell{m}\rangle=m|\ell{m}\rangle$; and $|\ell{m}\rangle$ are normalized eigenstates.  It is straightforward to confirm the following properties:
\begin{subequations}\label{eqn:Wigner-d props all}
\begin{align}\label{eqn:Wigner-d prop 1}
	\Im[\Wigd^{(\ell)}_{m,n}(\theta)] &= 0 ,\\
	\label{eqn:Wigner-d prop 2}
	\Wigd^{(\ell)}_{m,n}(\theta) &= \Wigd^{(\ell)}_{n,m}(-\theta) ,\\
	\label{eqn:Wigner-d prop 3}
	\Wigd^{(\ell)}_{m,n}(\theta) &= (-1)^{n-m}\Wigd^{(\ell)}_{n,m}(\theta) ,\\
	\label{eqn:Wigner-d prop 4}
	\Wigd^{(\ell)}_{m,n}(\theta) &= \Wigd^{(\ell)}_{-n,-m}(\theta) ,\\
	\label{eqn:Wigner-d prop 5}
	\Wigd^{(\ell)}_{m,n}(0) &= \delta_{m,n} \ ,\\
	\label{eqn:Wigner-d prop 6}
	\Wigd^{(\ell)}_{m,n}(\pi) &= (-1)^{\ell-n}\delta_{m,-n} \ ,\\
	\label{eqn:Wigner-d prop 7}
	\Wigd^{(\ell)}_{m,n}(2\pi) &= (-1)^{2\ell}\delta_{m,n} \ .
\end{align}
\end{subequations}
It then follows from the definition of $\Wigd^{(\ell)}_{m,n}(\theta)$ as elements of a rotation matrix that
\begin{subequations}\label{eqn:Wigner-d rot all}
	\begin{align}\label{eqn:Wigner-d rot 1}
	\Wigd^{(\ell)}_{m,n}(\pi+\theta) &= (-1)^{\ell-n}\Wigd^{(\ell)}_{m,-n}(\theta) ,\\
	\label{eqn:Wigner-d rot 2}
	\Wigd^{(\ell)}_{m,n}(2\pi+\theta) &= (-1)^{2\ell}\Wigd^{(\ell)}_{m,n}(\theta) .
	\end{align}
\end{subequations}
Using only the properties in Eqs.~(\ref{eqn:Wigner-d prop 1})--(\ref{eqn:Wigner-d prop 4}) and (\ref{eqn:Wigner-d rot 1}), and the phase choice specified in Eq.~(\ref{eqn:swY-def})\footnote{By Eq.~(\ref{eqn:harmonic def}), the same phase choice is represented by Eqs.~(\ref{eqn:spherical-func-s}) and (\ref{eqn:spherical-func-m}).}, the phase freedoms represented by $\varphi_1$, $\varphi_2$, and $\varphi_3$ in Eqs.~(\ref{eqn:swSF sx S ident}), (\ref{eqn:swSF mxc S ident}), and (\ref{eqn:swSF cc S ident}) are completely fixed, giving
\begin{subequations}\label{eqn:swY S idents}
\begin{align}
\label{eqn:swY sx S ident}
	\swsS{-s}{\ell{m}}{x} &= (-1)^{\ell-m}\swsS{s}{\ell{m}}{-x}, \\
\label{eqn:swY mx S ident}
	\swsS{s}{\ell{(-m)}}{x} &= (-1)^{\ell+s}\swsS{s}{\ell{m}}{-x}, \\
\label{eqn:swY cc S ident}
	\swsS[*]{s}{\ell{m}}{x} &= \swsS{s}{\ell{m}}{x},
\end{align}
\end{subequations}
when $c=0$.  In fact, Eqs.~(\ref{eqn:swY S idents}) together fix the phase of $\swsS{s}{\ell{m}}{x}$ up to an overall sign choice, while Eqs.~(\ref{eqn:swY sx S ident}) and (\ref{eqn:swY mx S ident}) fixes their phase up to an overall constant complex phase.

Equations~(\ref{eqn:swY sx S ident})--(\ref{eqn:swY cc S ident}) can be expressed in the more familiar forms
\begin{subequations}\label{eqn:swY idents}
	\begin{align}\label{eqn:swY cc Y ident}
		\swY[*]{s}{\ell{m}}{\theta,\phi} = (-1)^{s-m}\swY{-s}{\ell{(-m)}}{\theta,\phi}, \\
		\label{eqn:swY par Y ident}
		\swY{s}{\ell{m}}{\pi-\theta,\pi+\phi} = (-1)^\ell\swY{-s}{\ell{m}}{\theta,\phi},
	\end{align}
\end{subequations}
which fixes the phase of the spin-weighted spherical harmonics up to an overall sign choice.

\subsubsection{Spin-weighted derivatives}
\label{sec:sw derivatives}
Given a scalar ${}_{{}_s}\!Q$ of spin-weight $s$, we follow Refs.~\cite{NewmanPenrose-1966,MoxonScheelTeukolsky-2020} and define spin-weighted derivatives, which raise or lower the spin weight by one, as
\begin{subequations}\label{eqn:sw derivs}
\begin{align}
\label{eqn:edth}
	\edthm{}_{{}_s}\!Q &= -(\sin\theta)^s\left(\frac{\partial}{\partial\theta} 
			+\frac{i}{\sin\theta}\frac{\partial}{\partial\phi}\right)\left[(\sin\theta)^{-s}{}_{{}_s}\!Q\right], \\
	\bar{\edthm}{}_{{}_s}\!Q &= -(\sin\theta)^{-s}\left(\frac{\partial}{\partial\theta} 
\label{eqn:edthbar}
			-\frac{i}{\sin\theta}\frac{\partial}{\partial\phi}\right)\left[(\sin\theta)^s{}_{{}_s}\!Q \right].
\end{align}
\end{subequations}
From our definition of $\swY{s}{\ell{m}}{\theta,\phi}$, we find the particularly useful relations
\begin{subequations}\label{eqn:edth on Y}
\begin{align}\label{eqn:edth Y}
	\edthm\swY{s}{\ell{m}}{\theta,\phi} &= \sqrt{(\ell-s)(\ell+s+1)}\swY{s+1}{\ell{m}}{\theta,\phi}, \\
	\label{eqn:edthbar Y}
	\bar\edthm\swY{s}{\ell{m}}{\theta,\phi} &= -\sqrt{(\ell+s)(\ell-s+1)}\swY{s-1}{\ell{m}}{\theta,\phi}.
\end{align}
\end{subequations}
And, from Eqs.~(\ref{eqn:sw derivs}) and (\ref{eqn:edth on Y}) it directly follows that
\begin{align}
	\frac{\partial}{\partial{x}}\swsS{s}{\ell{m}}{x} &= \frac{1}{2\sqrt{1-x^2}}\Bigl[ \nonumber \\
	&\sqrt{(\ell-s)(\ell+s+1)}\swsS{s+1}{\ell{m}}{x} \\ &- \sqrt{(\ell+s)(\ell-s+1)}\swsS{s-1}{\ell{m}}{x}\Bigr],\nonumber
\end{align}
which we will make use of below.

\subsection{Scalar spheroidal functions}
\label{sec:scalar spheroidal}

The scalar spheroidal functions, also known as spheroidal wave functions of the first kind\cite{NIST:DLFM:SWF}, occur when $s=0$ and are most commonly used with $c^2$ real, in which case they are orthogonal and complete.  The generalization to fully complex values of $c$ has received less attention, but methods to compute them in this regime are known\cite{Barrowes-etal-2004}.  In the general case, the functions are complete for fixed values of $c$, but are not orthogonal.  Instead, they are biorthogonal\cite{London-2023}.  Properly categorizing the solutions for general complex $c$ is significantly more difficult than for the spin-weighted spherical functions.  So much so, that it appears that the routines in {\em Mathematica} cannot be fully trusted\cite{Barrowes-etal-2004} unless $\arg(c)\in\{0,\pi/4,\pi/2\}$.

When $c^2\in\Re$, a standard phase choice \cite{NIST:DLFM:SWF} is to take 
\begin{subequations}\label{eqn:s0 phase}
\begin{align}\label{eqn:s0 phase val}
	\sign[\swS{0}{\ell{m}}{0}{c}] &= (-1)^{(\ell+m)/2} \quad\ :\  \ell-m\text{\ even}, \\
\label{eqn:s0 phase deriv}
	\sign\left[\frac{d\swS{0}{\ell{m}}{x}{c}}{dx}\right]_{x=0} \!\!\!\!\!\!\!&= (-1)^{(\ell+m-1)/2} \ :\  \ell-m\text{\ odd}.
\end{align}
\end{subequations}  
This is reasonable since $\swS{0}{\ell{m}}{x}{c}$ can be chosen to be purely real when $c^2\in\Re$, and Eqs.~(\ref{eqn:swSF sx idents}) guarantee $\swS{0}{\ell{m}}{x}{c}$ to be either an even or an odd function of $x$.  The phase choices in Eqs.~(\ref{eqn:s0 phase}) are consistent with Eqs.~(\ref{eqn:swY S idents}) when $c=0$.  

For general values of $c$, Eqs.~(\ref{eqn:s0 phase}) are no longer sufficient to fix the phase of $\swS{0}{\ell{m}}{x}{c}$ because it is a complex function of $x$.  In this case, Eqs.~(\ref{eqn:swSF sx idents}) still guarantee $\swS{0}{\ell{m}}{x}{c}$ may be chosen to be either an even or an odd complex function of $x$.\footnote{$\swS{0}{\ell{m}}{x}{c}$ and $\swS{0}{\ell{m}}{-x}{c}$ are both solutions of the same differential equation with the same eigenvalue $\scA{0}{\ell{m}}{c}$.}  In Ref.~\cite{Barrowes-etal-2004}, asymptotic expansions of the scalar spheroidal functions are normalized by setting them equal to either the value of the related associated Legendre function or its derivative at $x=0$.  These requirements fix both the normalization and the phase choice.  An equivalent choice which fixes only the phase is to demand that the even scalar spheroidal functions be real at $x=0$ and satisfy Eq.~(\ref{eqn:s0 phase val}), or for the odd functions that their derivatives be real at $x=0$ and satisfy Eq.~(\ref{eqn:s0 phase deriv}).

If a single choice of the phase factors $\varphi_{1,2,3}$ from Eqs.~(\ref{eqn:swSF sx S ident}), (\ref{eqn:swSF mxc S ident}), and (\ref{eqn:swSF cc S ident}) holds, then from the spherical limit behavior in Eqs.~(\ref{eqn:swY S idents}) we should expect the scalar spheroidal functions to satisfy
\begin{subequations}\label{eqn:ssF S idents}
\begin{align}
\label{eqn:ssF sx S ident}
	\swS{0}{\ell{m}}{x}{c} &= (-1)^{\ell-m}\swS{0}{\ell{m}}{-x}{c}, \\
\label{eqn:ssF mx S ident}
	\swS{0}{\ell{(-m)}}{x}{c} &= (-1)^{\ell}\swS{0}{\ell{m}}{-x}{-c}, \\
\label{eqn:ssF cc S ident}
	\swS[*]{0}{\ell{m}}{x}{c} &= \swS{0}{\ell{m}}{x}{c^*}.
\end{align}
\end{subequations}

\subsection{General spin-weighted spheroidal functions}
\label{sec:general sw spheroidal}
The angular Teukolsky equation (\ref{eqn:Angular Teukolsky Equation}) typically must be solved numerically.  There are many possible approaches, but the most commonly used methods are Leaver's method\cite{leaver-1985} and the spectral method developed by Cook and Zalutskiy\cite{cook-zalutskiy-2014} based on the expansion in Eq.~(\ref{eqn:swSF-expansion}).  More recent advances in methods to evaluate the confluent Heun function\cite{Motygin-2018} have made direct solutions possible\cite{Fiziev-2010,Chen-etal-2025}.  Both the spectral and confluent Heun numerical approaches are discussed in more depth in Appendix~\ref{sec:SWSpheroidal}.  The spherical limit is well understood through perturbative analysis\cite{berticardosocasals-2006,Flammer,SpheroidalFunctions-1954,PressTeukolsky-1973,fackerell-1976,Seidel-1989,Breuer-etal-1977,ShahWhiting-2016} as the limit is the analytically known spin-weighted spherical harmonics discussed in Sec.~\ref{sec:sw-spherical}.  There is some understanding of the eigensolutions in the asymptotic limit as $|c|$ becomes large\cite{VickersCook2022,CookLu2023,berticardosocasals-2006,Casalas-oblate-2005,Breuer-etal-1977}.  The eigensolutions appear to fall into two distinct families in the asymptotic limit.  In the first family, the eigenvalue, or separation constant $\scA{s}{\ell{m}}{c}$, has leading order behavior $\pm ic(2\bar{L}+1)$, where the meaning of $\bar{L}$ is defined below via Eq.~(\ref{eqn:sph L def})\cite{CookLu2023,VickersCook2022,berticardosocasals-2006}.  The plus sign holds when $\Im[c]<0$, the minus sign when $\Im[c]>0$, and this asymptotic behavior is not seen for $\Im[c]=0$. In the second family, the leading order behavior is $-c^2$\cite{VickersCook2022,berticardosocasals-2006,Casalas-oblate-2005,Breuer-etal-1977}.  Furthermore, some pairs of the asymptotically $-c^2$ eigenvalues become 2-fold degenerate.

In general, the SWSFs do not exhibit any even/odd symmetry in $x$.  The exceptions are, of course the case where $s=0$, and also the case where $m=0$ in the spherical limit $c=0$.  Therefore, a phase choice similar to Eqs.~(\ref{eqn:s0 phase}) is not possible.  Furthermore, unless $c$ is real, $\swS{s}{\ell{m}}{x}{c}$ is a complex function of $x$.  Together, these facts complicate any approach to fix the phase of the SWSFs.  A further complication relates to the definition of the polar index $\ell$.

It is important to keep in mind that, while the spin weight $s$ and azimuthal index $m$ are parameters of Eq.~(\ref{eqn:Angular Teukolsky Equation}), the polar index $\ell$ is not.  It is a label which distinguishes unique eigensolutions within the infinite set of such solutions for fixed values of $s$, $m$, and $c$.  For the spin-weighted spherical functions $\swsS{s}{\ell{m}}{x}$, the separation constant takes on the non-negative values $\scA{s}{\ell{m}}{0}=\ell(\ell+1)-s(s+1)$ which provide a natural association between the polar index $\ell$ and the eigensolutions.  A useful alternative label used to distinguish elements in the set of eigensolutions is the zero-based index $L$ which orders the solutions in terms of either $\Re[\scA{s}{\ell{m}}{c}]$, or the magnitude $|\scA{s}{\ell{m}}{c}+s|$, from smallest to largest.  For the spin-weighted spherical functions $\swsS{s}{\ell{m}}{x}$, both orderings are identical and we have
\begin{align}\label{eqn:sph L def}
	\ell = L + \max(|s|,|m|).
\end{align}
An additional meaning for $L$ is that it gives the number of zero-crossings of the spin-weighted spherical functions $\swsS{s}{\ell{m}}{x}$.  

In the asymptotic regime, the two families of solutions mentioned above decouple from each other in the following sense.  The family of solutions with asymptotic leading behavior $\pm ic(2\bar{L}+1)$ has eigensolutions which can be labeled by a zero-based index $\bar{L}$ which is obtained from this asymptotic behavior.  The family of solutions with asymptotic leading behavior $-c^2$ is less well understood\cite{VickersCook2022}, but these solutions can be labeled by a zero based index $\hat{L}$ based on the behavior of the next to leading order term, again proportional to $c$, which can be expressed as a function of $\hat{L}$.\footnote{See Eqs.~(6) and (27) in Ref.~\cite{VickersCook2022}, but note that the specialized form given in Eq.~(27) only applies to anomalous asymptotic prolate sequences.  In general, Eqs.~(6) should be used.}  As $|c|\to\infty$, any 2 finite sets of eigenvalues from these two families separate, with $\Re[\scA{s}{\ell{m}}{c}]$ from all members of one family above or below the others depending on $\arg(c)$.  In this sense, the two families of solutions decouple.  But at finite values of $|c|$ in the asymptotic regime, the eigenvalues from members of the two families cross.  In some cases the crossings are true degeneracies, creating branch points and requiring branch cuts\cite{Barrowes-etal-2004}.

\begin{figure}
	\vspace{0.15in}
	\centering
	\includegraphics[width=\linewidth,clip]{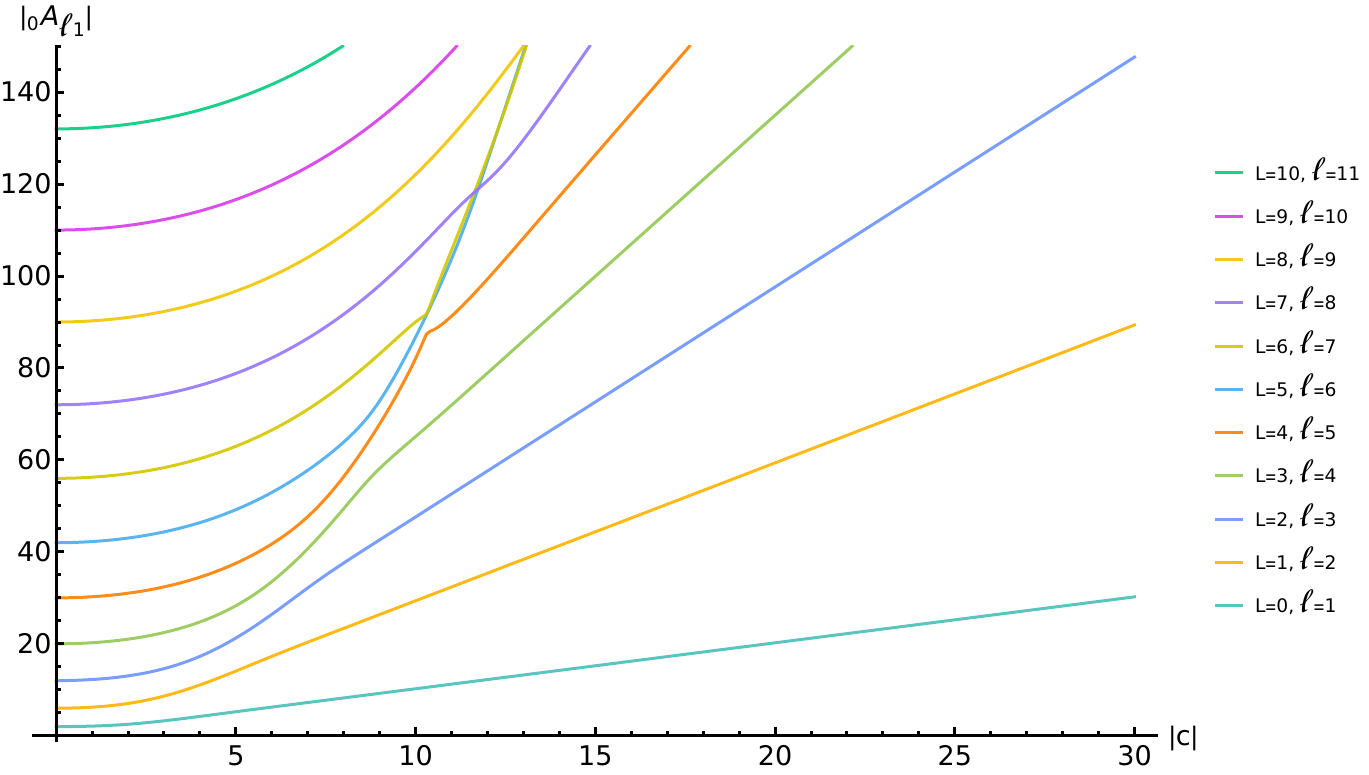}
	\caption{The magnitude of the first 11 eigenvalues for $s=0$ and $m=1$ along a sequence of values of $c=|c|e^{-i\frac{\pi}3}$.  The bottom sequence corresponds to $L=0$($\ell=1$), while the top sequence corresponds to $L=10$($\ell=11$).  Notice that the sequences that are labeled $L=5$ and $L=6$ in the spherical limit($c=0$) belong to the family of solutions which have leading order asymptotic behavior $-c^2$ and could be labeled as $\hat{L}=0$ and $\hat{L}=1$.  The sequences that would be labeled $L=0-4$ and $L=7$, in the spherical limit have entered the asymptotic regime and belong to the family of solutions which have leading order asymptotic behavior $ic(2\bar{L}+1)$ and could be labeled $\bar{L}=0-5$.  The upper 3 sequences have not entered the asymptotic regime in this figure and cannot yet be classified.}
	\label{fig:s0m1AbsA}
\end{figure}

For general complex values of $c$, there seems to be no single indexing scheme which is universally convenient or informative.  For sufficiently small values of $|c|$, the eigenvalues retain the same ordering as in the spherical limit $c=0$ based on either the real part or the magnitude of $\scA{s}{\ell{m}}{c}$, and either $\ell$ or $L$ can be used.  But, as illustrated in Fig.~\ref{fig:s0m1AbsA}, as $|c|$ increases, the ordering of the eigensolutions can change.  In this example, where $s=0$ and $m=1$ and we display $|\scA{0}{\ell1}{c}|$ and label the eigenvalue sequences by $L$ in the spherical limit ($c\to0$).  Beyond $|c|=13$, all of the sequences seen in the figure are clearly displaying their asymptotic behavior.  The lowest 6 sequences at $|c|=13$ correspond to $L=0-4$ and $L=7$.  All of these sequences belong to the family of solutions with asymptotic leading behavior $\pm ic(2\bar{L}+1)$ and could be labeled by $\bar{L}=0-5$ based on their asymptotic behavior.  The 2 sequences labeled $L=5$ and $L=6$, however, are seen to be an asymptotically degenerate pair of solution in the family with leading order asymptotic behavior $-c^2$.  Although not visible in this plot, these 2 sequences are seen to switch order with each other as they transition to their degenerate asymptotic behavior, but more importantly, they are seen to cross the sequence which would be labeled by $L=7$ in the spherical limit.  These two solutions could be labeled as $\hat{L}=0$ and $\hat{L}=1$ based on their next to leading order behavior.

Ultimately, a choice must be made as to how the set of eigensolutions will be ordered.  For sufficiently large $|c|$, $\Re[\scA{s}{\ell{m}}{c}]$ for solutions with asymptotic leading behavior $\pm ic(2\bar{L}+1)$ is always non-negative.  However this is not the case for solutions with asymptotic leading behavior $-c^2$.  This has a significant impact on the ordering of eigensolutions when the ordering is based on the real part of the eigensolutions.  Primarily for this reason, we have chosen to always order the eigensolutions based on the magnitude of the eigensolutions $|\scA{s}{\ell{m}}{c}+s|$.

We choose the ordering based on $|\scA{s}{\ell{m}}{c}+s|$ rather than simply $|\scA{s}{\ell{m}}{c}|$ in order to allow for a consistent choice for $\varphi_1$ in Eq.~(\ref{eqn:swSF sx S ident}), and Eq.~(\ref{eqn:swSF sx S phase}) below, for the basic symmetry $\{s\to-s,x\to-x\}$.  Equation~(\ref{eqn:swSF sx A ident}) means that $\scA{-s}{\ell{m}}{c}-s=\scA{s}{\ell{m}}{c}+s$.  In order to obtain a consistent ordering of solutions which should be related by the basic symmetry $\{s\to-s,x\to-x\}$, we choose the ordering of solutions based on $|\scA{s}{\ell{m}}{c}+s|$.\footnote{This is not the only approach to yield a consistent ordering.  We could also choose $|\scA{\pm|s|}{\ell{m}}{c}|$, but this requires giving preference to positive or negative values of $s$.}

\section{Phase choices for the spin-weighted spheroidal functions}
\label{sec:Phase choices}

The detailed consideration of the ordering and labeling of solutions given in Sec.~\ref{sec:general sw spheroidal} was necessary because we wish to consider methods for fixing the phase freedom of the SWSFs.  The basic symmetries of the SWSFs given in Eqs.~(\ref{eqn:swSF sx S ident}), (\ref{eqn:swSF mxc S ident}), and (\ref{eqn:swSF cc S ident}) take on the restricted forms given by Eqs.~(\ref{eqn:swY S idents}) for the spin-weighted spherical functions with the phase choice given in Eqs.~(\ref{eqn:spherical-func-s}).  We have found that the scalar spheroidal functions produced by the spectral solution scheme described below satisfy the restricted forms in Eqs.~(\ref{eqn:ssF S idents}) with no additional phase correction.  The full generalization of these restricted forms would be
\begin{subequations}\label{eqn:swSF S phase}
\begin{align}
\label{eqn:swSF sx S phase}
	\swS{-s}{\ell{m}}{x}{c} &= (-1)^{\ell-m}\swS{s}{\ell{m}}{-x}{c}, \\
\label{eqn:swSF mx S phase}
	\swS{s}{\ell{(-m)}}{x}{c} &= (-1)^{\ell+s}\swS{s}{\ell{m}}{-x}{-c}, \\
\label{eqn:swSF cc S phase}
	\swS[*]{s}{\ell{m}}{x}{c} &= \swS{s}{\ell{m}}{x}{c^*}.
\end{align}
\end{subequations}
These symmetry conditions have been given in Ref.~\cite{cook-zalutskiy-2014},\footnote{In Ref.~\cite{cook-zalutskiy-2014} the conditions were written for integer indices only.} and were used in Ref.~\cite{cook-ringdown-2020} to relate two QNM solution families to each other.\footnote{The two QNM frequency families are denoted by $\omega^{\pm}_{\ell{m}n}$ and are related to each other by $c\to-c^*$ and through Eqs.~(\ref{eqn:swSF mxc idents}) and (\ref{eqn:swSF cc idents}).}  It is clear from Eqs.~(\ref{eqn:swSF S phase}) that some consistent indexing choice must be made to fix $\ell$, thus ordering each eigensolution in the set of all solutions for fixed $s$, $m$, and $c$.  

The conditions given in Eqs.~(\ref{eqn:swSF S phase}) do not fully fix the phase freedom in SWSFs in the same way that Eqs.~(\ref{eqn:swY S idents}) do not fully fix the phase of the spin-weighted spherical harmonics.  And there is additional freedom because the SWSFs are complex.  Our goal is to find a phase-fixing scheme which is compatible with Eqs.~(\ref{eqn:swSF S phase}) and, if possible, with standard phase-fixing schemes in the spherical and scalar limits.  One approach to guarantee that Eqs.~(\ref{eqn:swSF S phase}) are satisfied is to always solve the angular Teukolsky equation (\ref{eqn:Angular Teukolsky Equation}) in a single quadrant of the complex-$c$ plane and only for $s\ge0$.  Equations~(\ref{eqn:swSF S phase}) can then be used to obtain solutions in any quadrant and for $s<0$.  While this is always possible, we prefer to seek phase-fixing schemes which allow the angular Teukolsky equation to be solved for any values of $s$, $m$, and $c$, and which are compatible with Eqs.~(\ref{eqn:swSF S phase}).

With these goals in mind, how are we to determine, from numerical solutions, if a given phase-fixing scheme is compatible with Eqs.~(\ref{eqn:swSF S phase})?  With the SWSFs defined by Eq.~(\ref{eqn:swSF-expansion}) as a linear combination of the spin-weighted spherical functions, Eqs.~(\ref{eqn:swSF S phase}) can be expressed as conditions on the expansion coefficients $\YSH{s}{\hat\ell\ell{m}}{c}$.  We find, using Eq.~(\ref{eqn:swSF-expansion}), that Eqs.~(\ref{eqn:swSF S phase}) become, respectively,
\begin{subequations}\label{eqn:swSF EC phase}
\begin{align}
\label{eqn:swSF sx EC phase}
	\YSH{-s}{\hat\ell\ell{m}}{c} &= (-1)^{\hat\ell-\ell}\YSH{s}{\hat\ell\ell{m}}{c}, \\
\label{eqn:swSF mx EC phase}
	\YSH{s}{\hat\ell\ell{(-m)}}{c} &= (-1)^{\hat\ell-\ell}\YSH{s}{\hat\ell\ell{m}}{-c}, \\
\label{eqn:swSF cc EC phase}
	\YSH{s}{\hat\ell\ell{m}}{c^*} &= \YSH[*]{s}{\hat\ell\ell{m}}{c}.
\end{align}
\end{subequations}

To test whether SWSFs satisfy these conditions, before being phase fixed or after, we choose some specific values for $s$, $m$ and a range of values for $c$.  We then solve the angular Teukolsky equation to obtain a set of solutions, for example as represented by $|\scA{0}{\ell1}{c}|$ for the first 11 eigensolutions for $s=0$ and $m=1$ in Fig.~\ref{fig:s0m1AbsA}.  This gives the base set of solutions for $\swS{s}{\ell{m}}{x}{c}$.  We also solve the angular equation 3 additional times to generate $\swS{-s}{\ell{m}}{x}{c}$, $\swS{s}{\ell{(-m)}}{x}{-c}$, and $\swS{s}{\ell{m}}{x}{c^*}$.  At each value of $c$ we choose an ordering of the solutions.  This is assumed to be a zero-based ordering with a value represented by $L$, and the corresponding value of $\ell$ is given by Eq.~(\ref{eqn:sph L def}).  Each SWSF is then represented by a set of expansion coefficients $\YSH{s}{\hat\ell\ell{m}}{c}$, where $\max(|m|,|s|)\le\hat\ell\le\hat\ell_{\rm max}$, and $\hat\ell_{\rm max}$ is large enough to accurately represent the SWSF.  Then, consistency with Eqs.~(\ref{eqn:swSF S phase}) can be verified for any computed value of $\ell$ by confirming that the expansion coefficients from the 4 solutions satisfy Eqs.~(\ref{eqn:swSF EC phase}) for all values of $\hat\ell$.

For the examples in this section, the angular Teukolsky equation (\ref{eqn:Angular Teukolsky Equation}) was solved as a discrete eigenvalue problem based on the spectral method developed by Cook and Zalutskiy\cite{cook-zalutskiy-2014}.  For additional details, see Appendix~\ref{sec:SWSpheroidal}.  Each solution returns both the eigenvalue $\scA{s}{\ell{m}}{c}$ and the set of expansion coefficients $\YSH{s}{\hat\ell\ell{m}}{c}$ needed to represent $\swS{s}{\ell{m}}{x}{c}$.  {\em Mathematica}'s {\tt Eigensystem} routine returns normalized eigenvectors so that $\sum_{\hat\ell}{|\YSH{s}{\hat\ell\ell{m}}{c}|^2}=1$ which guarantees that the SWSFs are normalized as in Eq.~(\ref{eqn:normalized SWSF}).  Furthermore, the eigenvectors returned by {\tt Eigensystem} are chosen so that the component with the maximum magnitude is real, but not necessarily positive.  So, the solutions returned by the spectral solver have a default phase choice fixed by the {\tt Eigensystem} routine in {\em Mathematica}.

We will represent a specific phase fixing scheme as $\Phase{a}$  Imposing $\Phase{a}$ on a given $\swS{s}{\ell{m}}{x}{c}$ requires determining a numerical value 
\begin{align}\label{eqn:phase-arg def}
	\Pfactor{a}(c)=e^{i\phi_a(c)},
\end{align}
which achieves the criteria specified by $\Phase{a}$ given current values for $\YSH{s}{\hat\ell\ell{m}}{c}$, and is implemented simply by
\begin{align}\label{eqn:phase change}
	\YSH{s}{\hat\ell\ell{m}}{c}\to\Pfactor{a}(c)\YSH{s}{\hat\ell\ell{m}}{c}.
\end{align}
We will refer to the default phase choice supplied by {\em Mathematica} as $\Phase{Math}$.

Interestingly, the $\Phase{Math}$ phase choice is not guaranteed to satisfy the basic symmetry conditions of Eqs.~(\ref{eqn:swSF S phase}).  This can be seen from the form of Eqs.~(\ref{eqn:swSF sx EC phase}) and (\ref{eqn:swSF mx EC phase}) which both contain a factor of $(-1)^{\hat\ell-\ell}$.  When checking to see if a particular solution satisfies the symmetries, $\ell$ is fixed and $\hat\ell$ denotes the eigenvector component.  If we fix $\hat\ell$ to the component which $\Phase{Math}$ fixes to be real, and assuming the sign of this real component is chosen consistently, then an odd value of $\hat\ell-\ell$ will cause Eqs.~(\ref{eqn:swSF sx EC phase}) and (\ref{eqn:swSF mx EC phase}) to not be satisfied.  In this case Eq.~(\ref{eqn:swSF cc EC phase}) would still be satisfied.  On the other hand, if the sign of the real component is not chosen consistently, any of the $3$ tests could fail.

\subsection{The CZ phase choice}
\label{sec:CZ-phase}
In Ref.~\cite{cook-zalutskiy-2014}, Cook and Zalutskiy (CZ) made a seemingly simple phase choice: fix the phase so that the expansion coefficient $\YSH{s}{\hat\ell\ell{m}}{c}$ with $\hat\ell=\ell$ in Eq.~(\ref{eqn:swSF-expansion}) is real and positive.  We will refer to this phase fixing scheme as $\Phase{CZ}$, and the condition is achieve by setting
\begin{align}\label{eqn:CZ phase factor}
	\Pfactor{CZ}(c)=\exp(-i\arg[\YSH{s}{\ell\ell{m}}{c}]).
\end{align}
The first step in implementing $\Phase{CZ}$ is to associate a value of $\ell$ with each SWSF.  Given values for $s$, $m$, and $c$, a specific SWSF is one eigenfunction within a set of eigensolutions which we choose to order in terms of increasing values of $|\scA{s}{\ell{m}}{c}+s|$.  One option is to let $L$ represent the zero-based index into this ordered list of solutions and fix $\ell$ via Eq.~(\ref{eqn:sph L def}).  We will refer to this as an indexed (Ind) choice for $\ell$, and designate this version of $\Phase{CZ}$ as $\Phase{CZ-Ind}$.  An alternative option is to construct a sequences of solutions all based on the same value for $\ell$.  If the sequence of solutions extends to $c=0$, for example as in Fig.~\ref{fig:s0m1AbsA}, then a natural choice is to let $L$ be the zero-based index of the eigensolution at $c=0$ and, again, fix $\ell$ for the entire sequence via Eq.~(\ref{eqn:sph L def}).  We will refer to this as a spherical-limit (SL) choice for $\ell$, and designate this version of $\Phase{CZ}$ as $\Phase{CZ-SL}$.  In the context of QNM solutions where $\Phase{CZ}$ was introduced $\Phase{CZ-SL}$ was implicitly chosen\cite{cook-zalutskiy-2014}.  Implementing this choice was straightforward because, in most cases, there is a well defined value of $\ell$ which should be associated with a given SWSF.  This is because, for QNMs of the Kerr geometry, it is natural to consider sequences of QNM solutions parameterized by the angular momentum per unit mass of the black hole $a$ and, at vanishing angular momentum, the associated solution of the angular Teukolsky equation is at $c=0$.  Even in the rare circumstances where a QNM sequence does not connect to $c=0$, there is usually enough information available to fix $\ell$.  There are also special circumstances where it is appropriate to choose $\ell$ based on an asymptotic limit where $|c|\to\infty$.\footnote{See the discussion of $\bar{L}$ and $\hat{L}$ in Sec.~\ref{sec:general sw spheroidal}.}  So long as $\ell$ is held fixed along a sequence, we will refer to the phase-fixing scheme as $\Phase{CZ-SL}$. 

Let us consider the behavior of the various phase-fixing schemes for SWSFs within the example of Fig.~\ref{fig:s0m1AbsA} where $s=0$, $m=1$, and $\arg(c)=-\frac{\pi}3$.  The first sequence of eigensolutions is labeled by $\ell=1$($L=0$) and is represented by the bottom line in the figure.  For this case, over the range $0\le|c|\le30$, $\Phase{Math}=\Phase{CZ-SL}=\Phase{CZ-Ind}$.  However, the next sequence of solutions, labeled by $\ell=2$($L=1$), is different.  In this case $\Phase{CZ-SL}=\Phase{CZ-Ind}$, but $\Phase{Math}$ differs for $|c|\gtrsim10.6$ because $\YSH{0}{221}{c}$ does not remain the largest expansion coefficient in magnitude.
\begin{figure}
	\vspace{0.15in}
	\centering
	\includegraphics[width=\linewidth,clip]{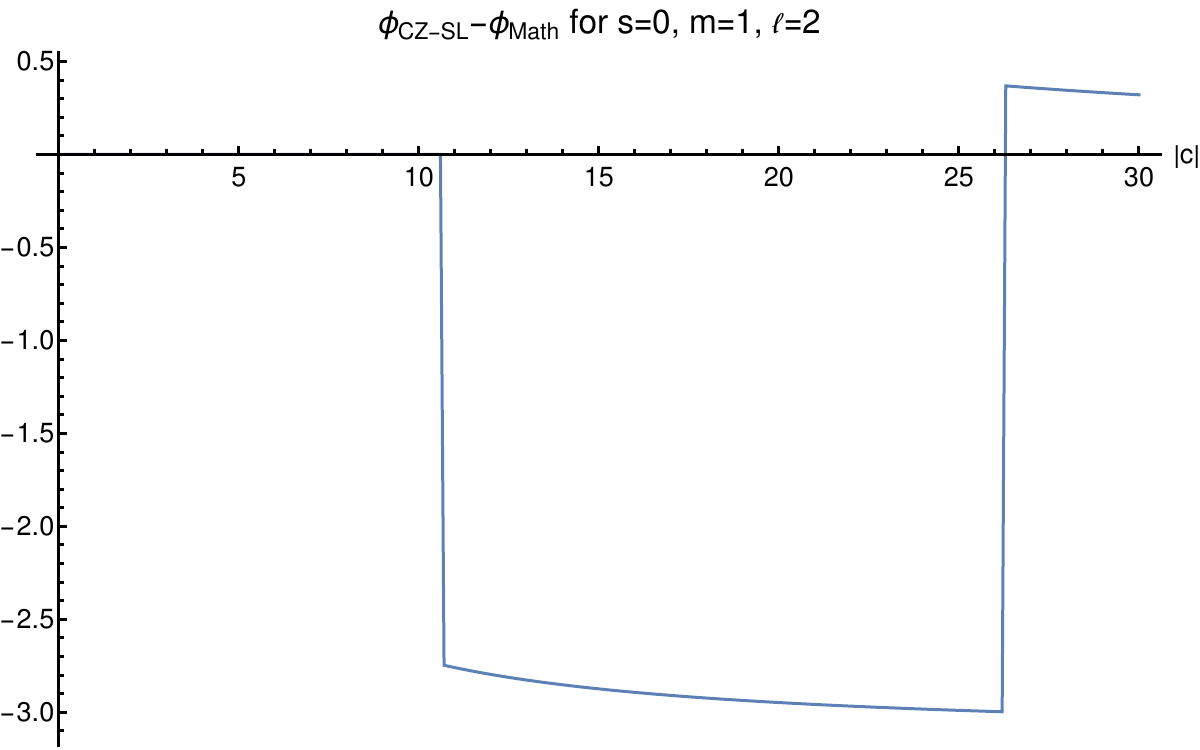}
	\caption{Phase difference between $\Phase{CZ-SL}$ and $\Phase{Math}$ imposed on $\swS{0}{21}{x}{|c|e^{-i\frac{\pi}3}}$.  The discontinuities in the phase occur as the expansion coefficient with the largest magnitude changes as $|c|$ increases.  For $|c|\lesssim10.6$, $\YSH{0}{221}{c}$ has the largest magnitude.  For $10.6\lesssim|c|\lesssim26.2$, $\YSH{0}{421}{c}$ has the largest magnitude, and for $26.2\lesssim|c|<30$, $\YSH{0}{621}{c}$ has the largest magnitude.}
	\label{fig:s0m1L1CZPhase}
\end{figure}
Figure~\ref{fig:s0m1L1CZPhase} displays the phase difference $\phi_{\rm CZ-SL}-\phi_{\rm Math}$ between the $\Phase{CZ-SL}$ phase fixed version of $\swS{0}{21}{x}{|c|e^{-i\frac{\pi}3}}$ and the version obeying $\Phase{Math}$, which clearly changes discontinuously.  When a phase difference is plotted as a function along a sequence, it is plotting the phase argument $\phi_a$ from Eq.~(\ref{eqn:phase-arg def}) needed to impose the first phase choice($\Phase{CZ-SL}$) on a sequence which currently obeys the second phase choice($\Phase{Math}$).

This example illustrates a potentially important aspect of a given phase-fixing scheme $\Phase{a}$.  The default expansion coefficients $\YSH{s}{\hat\ell\ell{m}}{c}$ automatically fixed by $\Phase{Math}$ have complex values which can be discontinuous as we vary $c$ along some sequence.  But, by applying the $\Phase{CZ-SL}$ scheme, all of the coefficients $\Pfactor{CZ-SL}(c)\YSH{s}{\hat\ell\ell{m}}{c}$ will have complex values which vary smoothly as $c$ varies, so long as $\YSH{s}{\ell\ell{m}}{c}\ne0$ along the sequence.  Of course, the $\Phase{CZ-Ind}$ scheme is not guaranteed to yield expansion coefficients which vary smoothly along a sequence because $\ell$, used to define the phase, can change discontinuously.  If, for a given application, it is necessary to interpolate through tables of expansion coefficients, then it is necessary to choose a phase-fixing scheme which guarantees smoothness.  This is a clear advantage of the $\Phase{CZ-SL}$ scheme.  But even the $\Phase{CZ-SL}$ scheme is not guaranteed to yield smooth expansion coefficients.  However, the failure mode triggered by $\YSH{s}{\ell\ell{m}}{c}=0$ is only likely to occur if a sequence extends to large values of $|c|$.  The small values of $\ell$ and overtones $n$ commonly used with QNMs keep $|c|$ reasonably small, and we have not found a case where $\YSH{s}{\ell\ell{m}}{c}$ vanishes along such QNM sequences.

\subsection{The SL phase choice}
\label{sec:SL-phase}
The $\Phase{CZ-SL}$ phase scheme is easy to implement, but its underlying principle that one specific expansion coefficient for a given $\swS{s}{\ell{m}}{x}{c}$ is always real and positive seems somewhat arbitrary, and will fail if that coefficient vanishes.  Moreover, it does not agree with the standard phase-fixing schemes used for the scalar spheroidal functions (see Eqs.~(\ref{eqn:s0 phase})).  Here, we seek to define a phase-fixing scheme which potentially has a more useful fundamental principle, and which agrees with Eqs.~(\ref{eqn:s0 phase}) and the phase choice for the spin-weighted spherical functions in Eqs.~(\ref{eqn:spherical-function def}).

The fundamental principle we wish to exploit is that there is the freedom to fix the complex function $\swS{s}{\ell{m}}{x}{c}$, or its derivative, to be real at one point which we choose to be $x=0$.  This fixes the phase only up to an overall sign choice, and we can use this additional freedom to fix either the sign of $\swS{s}{\ell{m}}{0}{c}$ or the sign of $\left.\partial_x\swS{s}{\ell{m}}{x}{c}\right|_{x=0}$.  In general, this is enough to uniquely fix the phase of $\swS{s}{\ell{m}}{x}{c}$, but there are special cases which must be dealt with. Also, we want the phase fixing to agree with existing phase choices in the limits that $s=0$ and $c=0$, and we want to ensure that the phase-fixed functions satisfy the symmetry conditions given in Eqs.~(\ref{eqn:swSF S phase}).

A necessary first step is to choose an ordering for the eigensolutions for fixed $m$, $s$, and $c$.  That is, choosing a value of the polar index $\ell$ (or $L$) to designate each solution.  With values of $\ell$, $m$, $s$, and $c$, we fix the remaining phase freedom so that either the value of the function or its derivative at $x=0$ agrees with the phase choice in the spherical limit $c=0$.  We refer to this phase-fixing scheme as the spherical limit (SL) phase choice $\Phase{SL}$.  

If $\swsS{s}{\ell{m}}{0}\ne0$ and $\swS{s}{\ell{m}}{0}{c}\ne0$, then the phase is based on choosing $\swS{s}{\ell{m}}{0}{c}$ to be real.  In this case, we choose
\begin{subequations}\label{eqn:SL non-zero}
\begin{align}\label{eqn:SL non-zero phase factor}
	&\Pfactor{SL${}_1$}(c) = \pm \exp\bigl(-i\arg[\swS{s}{\ell{m}}{0}{c}]\bigr), \\
	\intertext{with the sign chosen so that}
	\label{eqn:SL non-zero sign}
	&\sign[\Pfactor{SL${}_1$}(c)\swS{s}{\ell{m}}{0}{c}]=\sign[\swsS{s}{\ell{m}}{0}].
\end{align}
\end{subequations}
But, if either $\swsS{s}{\ell{m}}{0}=0$ or $\swS{s}{\ell{m}}{0}{c}=0$, then if $\left.\partial_x\swS{s}{\ell{m}}{x}{c}\right|_{x=0}\ne0$, the phase is based on choosing $\left.\partial_x\swS{s}{\ell{m}}{x}{c}\right|_{x=0}$ to be real.  In this case, we choose
\begin{subequations}\label{eqn:SL zero}
\begin{align}\label{eqn:SL zero phase factor}
	&\Pfactor{SL${}_2$}(c) = \pm \exp\bigl(-i\arg[\left.\partial_x\swS{s}{\ell{m}}{x}{c}\right|_0]\bigr), \\
	\intertext{with the sign chosen so that}
	\label{eqn:SL zero sign}
	&\sign[\Pfactor{SL${}_2$}(c)\left.\partial_x\swS{s}{\ell{m}}{x}{c}\right|_0]=\sign[\left.\partial_x\swsS{s}{\ell{m}}{x}\right|_0].
\end{align}
\end{subequations}
In both cases, Eqs.~(\ref{eqn:SL non-zero sign}) and (\ref{eqn:SL zero sign}) guarantee that the expansion coefficients $\YSH{s}{\hat\ell\ell{m}}{c}$ are smooth along a given sequences unless an anomalous condition arises.

The first anomalous condition that could arise is if $\swsS{s}{\ell{m}}{0}=0$ and $\left.\partial_x\swS{s}{\ell{m}}{x}{c}\right|_{x=0}=0$.  In this case, there is no way to match the phase of $\swS{s}{\ell{m}}{x}{c}$ to the phase of $\swsS{s}{\ell{m}}{x}$. Since $\swS{s}{\ell{m}}{0}{c}\ne0$ in this case\footnote{A nontrivial solution of a homogeneous 2nd order ODE and its derivative cannot simultaneously vanish at any point.}, we set the phase by Eq.~(\ref{eqn:SL non-zero phase factor}) so that $\swS{s}{\ell{m}}{0}{c}$ is real, and fix the sign freedom by demanding
\begin{align}\label{eqn:SL anomalous 1}
	\sign[\Pfactor{SL${}_1$}(c)\swS{s}{\ell{m}}{0}{c}]=(-1)^{\lfloor(\ell+m)/2\rfloor}.
\end{align}
Here, $\lfloor n\rfloor$ means the floor of $n$, and the sign choice is chosen to agree with Eq.~(\ref{eqn:s0 phase val}).

Two additional anomalous conditions can occur if $\swS{s}{\ell{m}}{0}{c}\approx0$ and either $\left.\partial_x\swS{s}{\ell{m}}{x}{c}\right|_{x=0}\approx0$ or $\left.\partial_x\swsS{s}{\ell{m}}{x}\right|_0=0$.  Analytically, both the SWSF and its derivative cannot vanish simultaneously, but this can occur with finite precision numerical results. When this condition does occur, or if the spheroidal function and the derivative of the spherical function are both zero at $x=0$, no sign comparison is possible at $x=0$.\footnote{We expect this to happen for some solutions with asymptotic leading behavior $-c^2$ as $|c|\to\infty$ where $\swS{s}{\ell{m}}{x}{c}$ is exponentially suppressed near $x=0$.}  We know of no systematic way of handling these cases and simply revert to using either $\Pfactor{CZ-SL}$ or $\Pfactor{CZ-Ind}$.  

As with $\Phase{CZ}$, there are two natural ways in which the polar index $\ell$($L$) can be chosen.  If, at each value of $c$, we let $L$ represent the zero-based index into the list of solutions ordered by $|\scA{s}{\ell{m}}{c}+s|$ and fix $\ell$ via Eq.~(\ref{eqn:sph L def}), then we have the phase-fixing scheme denoted by $\Phase{SL-Ind}$.  As with $\Phase{CZ-Ind}$, this scheme can lead to discontinuities in the expansion coefficients if we consider sequences of solutions.  The second approach for fixing the polar index $\ell$ is to choose it to be a constant along the sequence.  In most cases, $\ell$ should be chosen to agree with the spherical limit where $|c|\to0$, but there are unusual circumstances where it is appropriate to choose $\ell$ based on an asymptotic limit where $|c|\to\infty$.\footnote{See the discussion of $\bar{L}$ and $\hat{L}$ in Sec.~\ref{sec:general sw spheroidal}.}  Examples of both approaches will be explored in Sec.~\ref{sec:examples}.  In either case, however, we will denote this phase-fixing scheme as $\Phase{SL-C}$, with the ``C'' indicating that the phase choice will be continuous along sequences.

\subsection{Summary of phase choices}
\label{sec:phase summary}

When solutions of the angular Teukolsky equation, Eq.~(\ref{eqn:Angular Teukolsky Equation}), are found by any scheme, the SWSF will be given with respect to some phase choice determined simply by the particular scheme used to construct the solutions.  This is also true for all of the standard special functions, but in the most commonly used examples, the special functions are real.  If the special functions are to remain real, then any phase choice is simply an overall sign choice.  But, the SWSFs are inherently complex functions and the choice of phase is more complicated, and potentially more impactful.

The {\em Mathematica} routine {\tt Eigensystem}, when used to construct the numerical solutions of Eq.~(\ref{eqn:Angular Teukolsky Equation}), returns normalized eigenvectors with the phase chosen so that the eigenvector component with the largest magnitude is real, but not necessarily positive.  When this phase-fixing scheme is used to define the SWSFs through Eq.~(\ref{eqn:swSF-expansion}), we designate the phase choice as $\Phase{Math}$.  It is a perfectly functional phase-fixing scheme which requires no information about how a given eigensolution is labeled.  But, it has the disadvantage that the shape of $\swS{s}{\ell{m}}{x}{c}$ will change discontinuously along continuous sequences of $c$ because the position of the eigenvector component with the largest magnitude may change. Also, the phase choice $\Phase{Math}$ will not always satisfy the basic symmetries given by Eqs.~(\ref{eqn:swSF S phase}).  Thus, an alternative phase-fixing scheme should be used.

The Cook-Zalutskiy phase-fixing schemes $\Phase{CZ}$ are a small modification of $\Phase{Math}$ in which one expansion coefficient in $\YSH{s}{\hat\ell\ell{m}}{c}$ is again fixed to be real, but this coefficient is no longer required to be the coefficient with the largest magnitude.  The $\Phase{CZ}$ schemes are closely tied to the spectral expansion given in Eq.~(\ref{eqn:swSF-expansion}) and, because of this, may not be the most natural choice when other approaches, such as Leaver's method\cite{leaver-1985} or the confluent Heun method\cite{Fiziev-2010,Chen-etal-2025}, are used to find the eigensolutions of Eq.~(\ref{eqn:Angular Teukolsky Equation}).  The indices $s$ and $m$, and the value of $c$ are fixed values that are required to fully specify the angular Teukolsky equation.  The polar index $\ell$, on the other hand, is simply a value which labels each eigensolution within the infinite set of solutions.  In Eq.~(\ref{eqn:swSF-expansion}), the index $\hat\ell$ specifies the polar index of the spin-weighted spherical function $\swsS{s}{\hat\ell{m}}{x}$.  In the $\Phase{CZ}$ schemes, the expansion coefficients with $\hat\ell=\ell$ are chosen to be real and positive.  This requirement leads to one potential failure mode for the $\Phase{CZ}$ schemes.  If $\YSH{s}{\ell\ell{m}}{c}$ vanishes for some value of $c$ along a sequence, then the values of $\YSH{s}{\hat\ell\ell{m}}{c}$ will change discontinuously at this point along the sequence.

We have specified two methods for fixing the value of $\ell$ associated with each eigensolution.  For the $\Phase{CZ-SL}$ phase-fixing scheme $\ell$ takes on a constant value along a sequence of solutions.  This value is usually chosen to agree with the value for the solution with $c=0$ which smoothly connects to the given solution along a sequence of values of $c$, but other choices may be specified.  For the $\Phase{CZ-Ind}$ phase-fixing scheme, the value of $\ell$ is fixed by its sorted position within the set of all eigensolutions for the given value of $c$.  Sequences of eigensolutions parameterized by a smooth sequence of values of $c$ will, in most cases, have expansion coefficients $\YSH{s}{\hat\ell\ell{m}}{c}$ which vary smoothly along the sequence when the phase is fixed by $\Phase{CZ-SL}$, but they may not vary smoothly if fixed by $\Phase{CZ-Ind}$.  See Sec.~\ref{sec:CZ-phase} for more details.

We have defined the spherical-limit phase-fixing schemes $\Phase{SL}$ as those which fix either $\swS{s}{\ell{m}}{x}{c}$ or its derivative to be real at the equator $x=0$, and to fix the remaining phase freedom by choosing the sign of either the function or its derivative at $x=0$ based on the corresponding spherical function $\swsS{s}{\ell{m}}{x}$.  As these schemes are not fixing the phase based on a specific set of expansion functions, they should be easily applicable to SWSFs constructed by a variety of approaches.  It is also hoped that fixing the SWSFs(or their derivatives) to be real consistently at $x=0$ will facilitate extracting information from the phase, as well as the amplitude, when black-hole ringdown signals are analyzed in terms of QNMs.

As with the $\Phase{CZ}$ schemes, it is necessary to choose which value of $\ell$ is associated with each eigensolution.  We have again specified the same two methods for choosing $\ell$ as for $\Phase{CZ}$.  In the continuous SL scheme $\Phase{SL-C}$ (similar to $\Phase{CZ-SL}$), $\ell$ is chosen to be a constant along a sequence of values of $c$, and this value is usually the value associated with a solution at $c=0$.  For the $\Phase{SL-Ind}$ phase-fixing scheme, the value of $\ell$ is fixed by its sorted position within the set of all eigensolutions for the given value of $c$.  For the $\Phase{SL-C}$ scheme, sequences of eigensolutions parameterized by a smooth sequence of values of $c$ can be constructed to have expansion coefficients $\YSH{s}{\hat\ell\ell{m}}{c}$ which vary smoothly along the sequence, but smoothness can be lost when certain anomalous conditions occur (see Sec.~\ref{sec:SL-phase} for details).  As with the $\Phase{CZ-Ind}$ scheme, the $\Phase{SL-Ind}$ scheme may not have smooth sequences of expansion coefficients.

\section{Examples, public data files, and public codes}
\label{sec:examples}
In this section, we will explore the effects of the various $\Phase{CZ}$ and $\Phase{SL}$ phase fixing schemes on several QNM and TTM sequences.  A large number of sequences of QNMs and TTMs have been constructed using the methods described in Refs.~\cite{cook-zalutskiy-2014,cook-zalutskiy-2016b,cook-et-al-2018,CookLu2023}, and implemented in the {\tt KerrModes} suite of publicly available {\em Mathematica} Paclets\cite{KerrModes-1.0.6} which include full {\em Mathematica}-style documentation.

\subsection{Quasinormal modes}
\label{sec:QNM examples}
The gravitational QNMs of the Kerr geometry quantify the natural linear gravitational ringing modes of a rotating black hole\cite{berti-QNM-2009,cook-zalutskiy-2014}.  The complex mode frequencies $\omega^\pm_{\ell{m}n}$ are indexed by the standard polar index $\ell$, azimuthal index $m$, and an additional overtone index $n$.  The $\pm$ superscript differentiates between two families of solutions which obey $\omega^-_{\ell{m}n}=-(\omega^+_{\ell(-m)n})^*$ and all QNMs are evaluated with spin weight $s=-2$.  The corresponding mode function includes a product of the solutions of the radial and angular Teukolsky equations\cite{teukolsky-1973} where the oblateness parameter $c$ of the angular Teukolsky equation Eq.~(\ref{eqn:Angular Teukolsky Equation}) is given by $c=a\omega^\pm_{\ell{m}n}$ where $Ma$ is the angular momentum of the Kerr black hole.   The rotation parameter $a$ takes on values $0\le a<M$, where $M$ is the mass of the Kerr black hole, with $a=0$ corresponding to the nonrotating Schwarzschild limit and $a=M$ the extremal Kerr limit.  In the examples below, we will often use the notation $\{\ell,m,n\}$ to denote specific QNMs and TTMs.

\subsubsection*{QNM: $\ell=2$, $m=2$, $n=0$}
\label{sec:QNM 220}
As a simple example case, we first consider the $\{2,2,0\}$ mode, which is the dominant QNM in the ringdown signal of binary black hole mergers.  As we will see, all phase choices will be well behaved for this important case, but it provides a clean and clear introduction to the various details of the eigensolutions that must be considered, and how they impact the phase fixing schemes.

\begin{figure}
	\centering
	\includegraphics[width=\linewidth,clip]{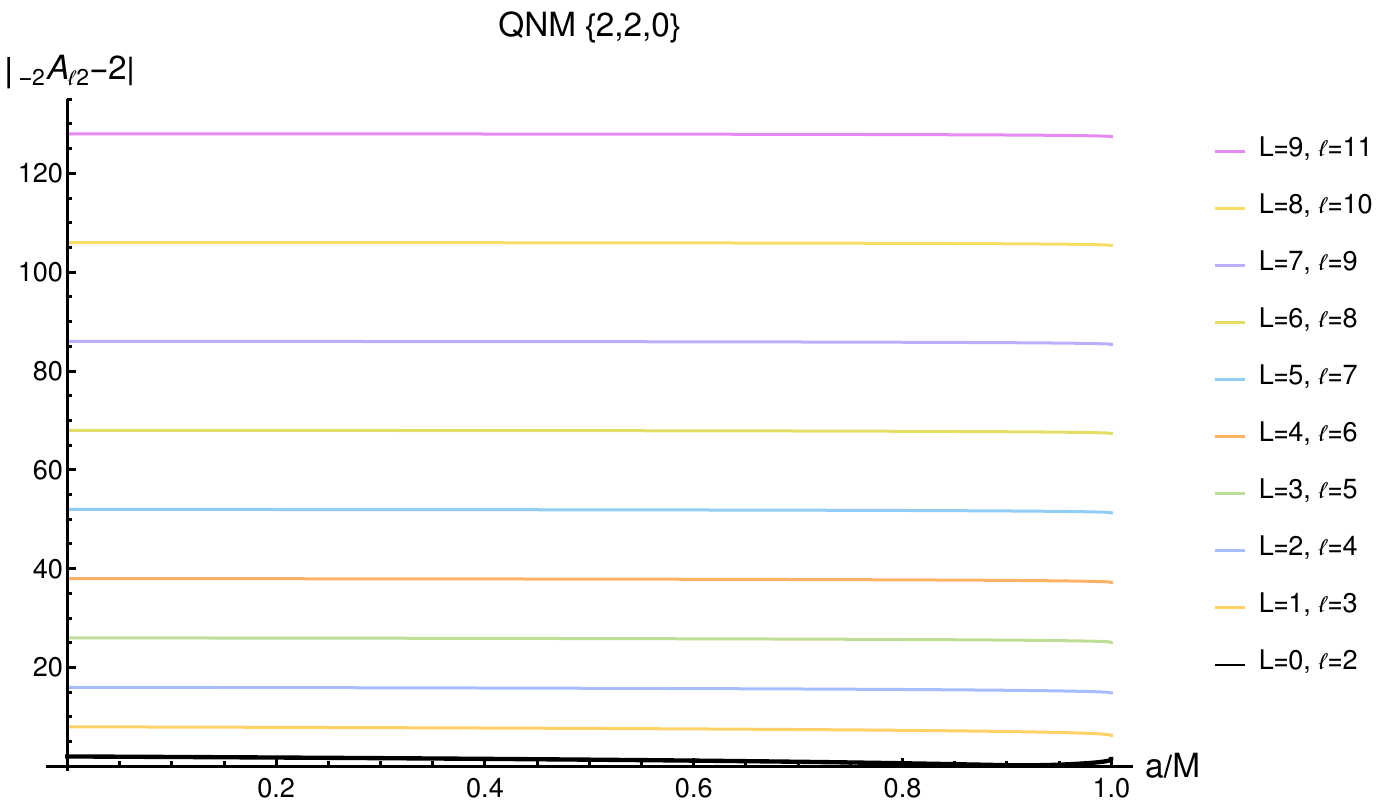}
	\caption{The first 10 eigenvalues for $s=-2$ and $m=2$ along a sequence of values of $c$ obtained from the QNM mode sequence $\{2,2,0\}$.  The plot displays $|\scA{-2}{\ell2}{c(a)}-2|$ so that the eigenvalues appear in their sorted order.  The values are displayed as functions of the dimensionless angular momentum $a/M$ which is related to the oblateness parameter by $c(a)=a\omega^+_{220}(a)$, where $\omega^+_{220}(a)$ is the complex mode frequency along the QNM sequence.  The eigenvalue sequence corresponding to the actual $\{2,2,0\}$ mode sequences is displayed as the thick black line.}
	\label{fig:QNM220Alm_a}
\end{figure}

The sequence of mode frequencies $\omega^+_{220}(a)$, evaluated as a function of the rotation parameter $a$, defines a path thought the space of the complex oblateness parameter $c(a)=a\omega^+_{220}(a)$.  Figure~\ref{fig:QNM220Alm_a} plots the magnitude $|\scA{-2}{\ell2}{c}-2|$ from the first 10 eigensolutions of the angular Teukolsky equation Eq.~(\ref{eqn:Angular Teukolsky Equation}) with $s=-2$ and $m=2$.  As discussed in Sec.~\ref{sec:general sw spheroidal}, the eigensolutions of Eq.~(\ref{eqn:Angular Teukolsky Equation}) are sorted in order of ascending values of $|\scA{s}{\ell{m}}{c}+s|$, so this plot displays the solutions in this order.  The sequence labeled $\ell=2(L=0)$ is at the bottom of the figure and is the eigenvalue sequence which pairs with the $\omega^+_{220}(a)$ sequence of QNM mode frequencies.  For clarity, we emphasize that the other eigensolutions are not paired with other QNM sequences.  For example, the sequence labeled by $\ell=3$ is not paired with the $\{3,2,0\}$ QNM sequence.  These mode frequencies define a different path through the space of the complex oblateness parameter.

\begin{figure}
	\centering
	\includegraphics[width=\linewidth,clip]{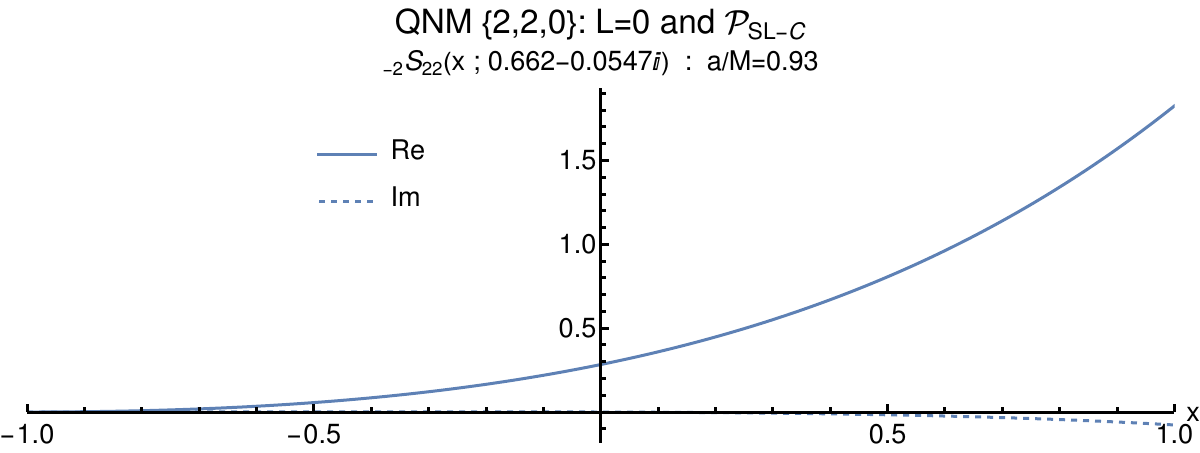}
	\includegraphics[width=\linewidth,clip]{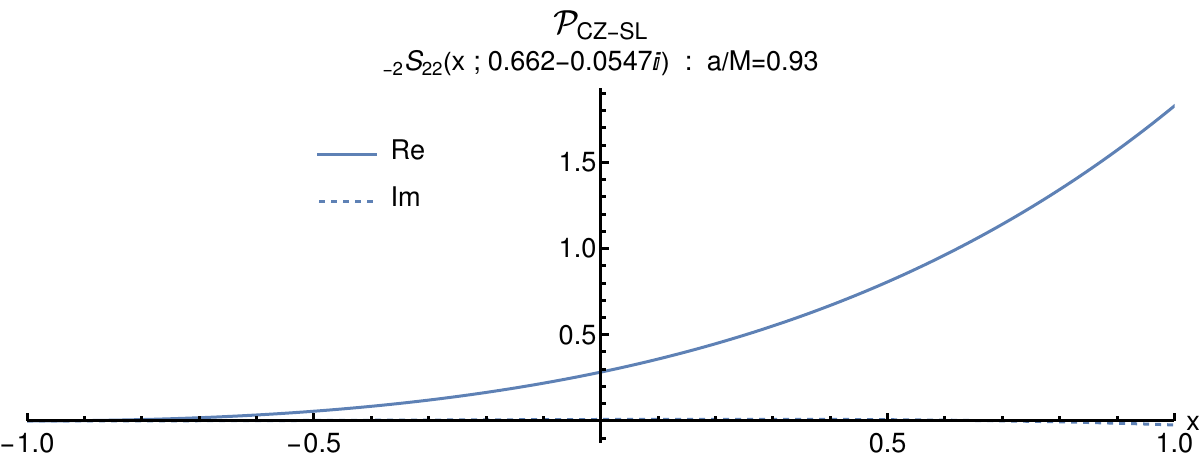}
	\caption{The function $\swS{-2}{22}{x}{0.662-0.0547i}$ where the upper plot has been phase fixed using $\Phase{SL-C}$ and the bottom using $\Phase{CZ-SL}$.  Note that there is very little difference between the two phase fixing schemes in this case.}
	\label{fig:QNM220L0S}
\end{figure}

\begin{figure}
	\centering
	\includegraphics[width=\linewidth,clip]{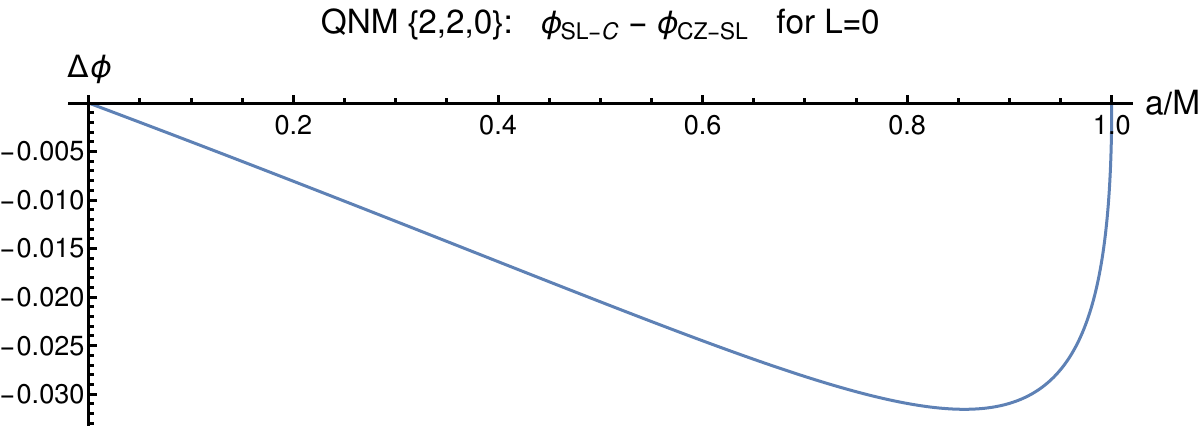}
	\includegraphics[width=\linewidth,clip]{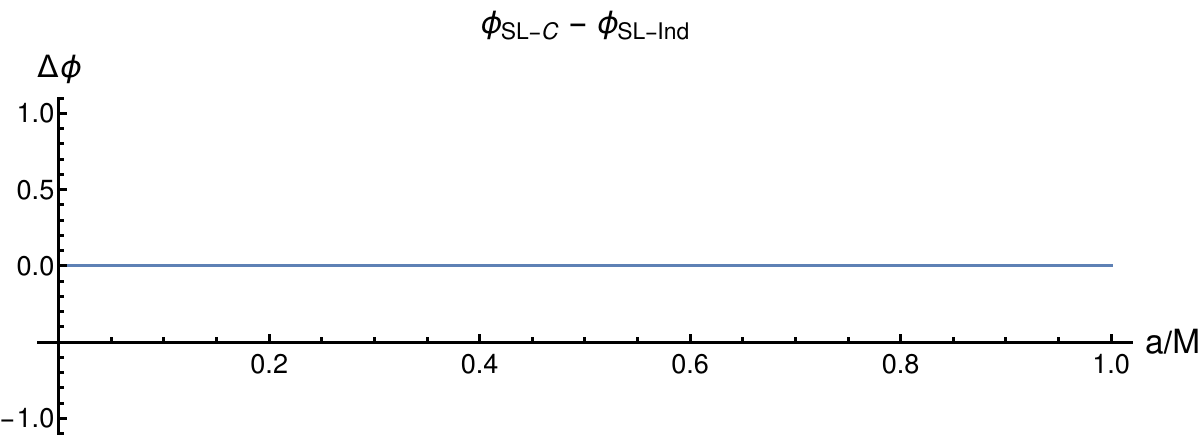}
	\caption{The phase difference between the functions $\swS{-2}{22}{x}{a\omega^+_{220}(a)}$ which have been phase fixed using different schemes.  In the top plot, the functions have been phase fixed using $\Phase{SL-C}$ and $\Phase{CZ-SL}$.  In the bottom plot, the functions have been phase fixed using $\Phase{SL-C}$ and $\Phase{SL-Ind}$.}
	\label{fig:QNM220L0phasediff}
\end{figure}

In Fig.~\ref{fig:QNM220Alm_a}, we find a simple structure to the set of eigenvalues.  For any QNMs with small values of $n$ and $\ell$, the magnitude of the mode frequencies $|\omega_{\ell{m}n}(a)|$ remain reasonably small so that $|c(a)|$ also never gets very large, in which case the values of $\scA{-2}{\ell{m}}{c}$ do not deviate enough from their values at $c=0$ to cause crossings in the eigenvalues.  Also, so long as $|c|$ remains relatively small, there tends to be only small phase differences between the $\Phase{CZ}$ and $\Phase{SL}$ phase fixing schemes as illustrated in Figs~\ref{fig:QNM220L0S} and \ref{fig:QNM220L0phasediff}.  In Fig.~\ref{fig:QNM220L0S}, we display the behavior of the eigenfunction $\swS{-2}{22}{x}{0.662-0.0547i}$ for both the $\Phase{SL-C}$ and $\Phase{CZ-SL}$ phase fixing schemes.  These solutions correspond to a point along the sequence where $a\approx0.93M$, which is near the point with the largest deviation between the solutions.  This example explores the case where the $\Phase{SL-C}$ scheme fixes the phase so that $\Im[\swS{-2}{22}{0}{c}]=0$.  While it is difficult to see in this figure, the function is entirely real at $x=0$ in the top plot where the phase has been fixed using $\Phase{SL-C}$.  In the bottom plot the function has a small imaginary component at $x=0$.
The difference between the two functions is more easily seen in Fig.~\ref{fig:QNM220L0phasediff} which plots actual phase differences.  In the upper plot, we display the phase difference between $\Phase{SL-C}$ and $\Phase{CZ-SL}$ as a function of $a$, and find that the largest phase difference has a magnitude of about $0.03$.
In the lower plot, we display the phase difference between $\Phase{SL-C}$ and $\Phase{SL-Ind}$.  We find no phase difference, which is expected since the $\ell=2$ sequence in Fig.~\ref{fig:QNM220Alm_a} exhibits no crossings.

\begin{figure}
	\centering
	\includegraphics[width=\linewidth,clip]{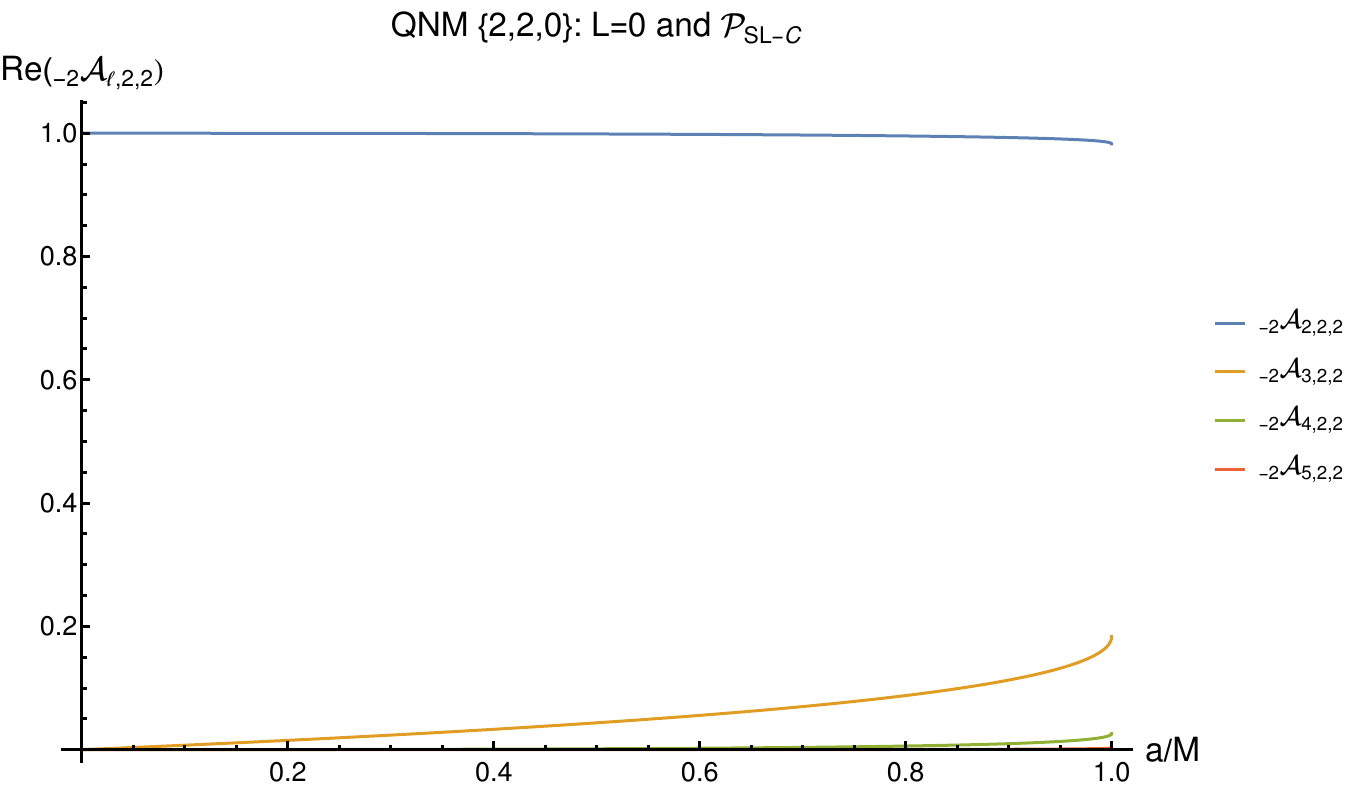}
	\caption{The real part of the first $4$ expansion coefficients $\YSH{-2}{\ell22}{a\omega^+_{220}(a)}$ are plotted to demonstrate that the expansion coefficients are smooth functions of $a$ for the $\Phase{SL-C}$ phase choice.}
	\label{fig:QNM220L0ExCoefRe}
\end{figure}

In this simple example, all phase fixing schemes yield eigenfunctions which vary smoothly along the sequence $0\le a<M$.  We illustrate this in Fig.~\ref{fig:QNM220L0ExCoefRe} which displays the real part of the first $4$ expansion coefficients $\YSH{-2}{\ell22}{c(a)}$ from Eq.~(\ref{eqn:swSF-expansion}).
These $4$ coefficients contain most of the power in the expansion of the SWSF $\swS{-2}{22}{x}{c(a)}$ as a linear combination of spin-weighted spherical functions $\swsS{-2}{\ell2}{x}$.  As illustrated by the first $4$ expansion coefficients in Fig.~\ref{fig:QNM220L0ExCoefRe}, all of the expansion coefficients $\YSH{-2}{\ell22}{c(a)}$ are smooth functions of $a$ along the $\{2,2,0\}$ QNM sequence.  In fact, for this simple example, all phase choices, including $\Phase{Math}$, yield smooth expansion coefficients.  This is not always the case.

We have also carefully explored whether or not our phase-fixed eigenfunctions satisfy the basic symmetries of the SWSFs given in Eqs.~(\ref{eqn:swSF S phase}) by checking that the expansion coefficients satisfy Eqs.~(\ref{eqn:swSF EC phase}) for all $10$ eigensolutions represented in Fig.~\ref{fig:QNM220Alm_a} at each computed value along the sequences.  This involves constructing $4$ full sets of eigensolutions for $\swS{-2}{22}{x}{c(a)}$, $\swS{2}{22}{x}{c(a)}$, $\swS{-2}{2(-2)}{x}{-c(a)}$, and $\swS{-2}{22}{x}{c^*(a)}$, as well as the phase corrections $\Pfactor{CZ-SL}(a)$, $\Pfactor{CZ-Ind}(a)$, $\Pfactor{SL-C}(a)$, and $\Pfactor{SL-Ind}(a)$ for each solution.  The phase fixed versions of each expansion coefficient for each solution are compared using Eqs.~(\ref{eqn:swSF EC phase}) to verify that each condition is satisfied.  For this simple example, all phase choices, including $\Phase{Math}$, satisfy the basic symmetries given in Eqs.~(\ref{eqn:swSF S phase}).

\subsubsection*{QNM: $\ell=3$, $m=0$, $n=0$}
\label{sec:QNM 300}
Now consider the example case of the $\{3,0,0\}$ mode, where there are two main points on which to focus.  First, in this example it is the derivative $\partial_x\swsS{-2}{30}{x}|_{x=0}$ that is fixed to be real for the $\Phase{SL-C}$ phase choice, as opposed to the function itself as in the case of $\{2,2,0\}$.  Second, it is the $\ell=3(L=1)$ eigenvalue sequence which pairs with the $\omega^+_{300}(a)$ sequence of QNM mode frequencies.  Figure~\ref{fig:QNM300Alm_a}, found in Appendix~\ref{sec:additional_figures}, plots the magnitudes $|\scA{-2}{\ell0}{c}-2|$ of the first $10$ eigenvalues of the angular Teukolsky equation Eq.~(\ref{eqn:Angular Teukolsky Equation}) with $s=-2$ and $m=0$.  While Figs.~\ref{fig:QNM220Alm_a} and \ref{fig:QNM300Alm_a} look qualitatively similar, we emphasize that the two figures correspond to different paths through the space of the complex oblateness parameter, agreeing only at $a=0$ where $c=0$.

\begin{figure}
	\centering
	\includegraphics[width=\linewidth,clip]{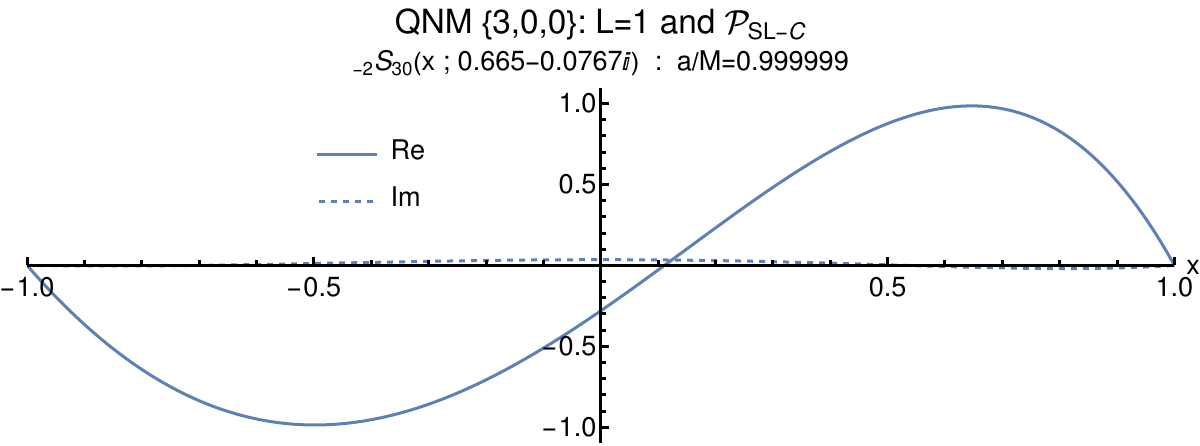}
	\includegraphics[width=\linewidth,clip]{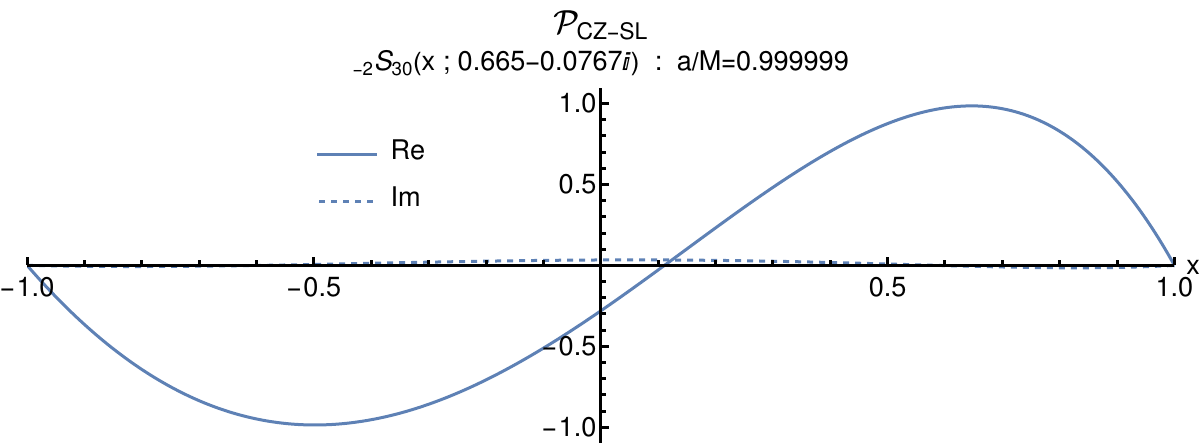}
	\caption{The functions $\swS{-2}{30}{x}{0.665-0.0767i}$ where the upper plot has been phase fixed using $\Phase{SL-C}$ and the bottom using $\Phase{CZ-SL}$.  Note that there is very little difference between the two phase fixing schemes in this case.}
	\label{fig:QNM300L1S}
\end{figure}

In Fig.~\ref{fig:QNM300L1S}, we display the behavior of the eigenfunction $\swS{-2}{30}{x}{0.665-0.0767i}$ for both the $\Phase{SL-C}$ and $\Phase{CZ-SL}$ phase fixing schemes.  These solutions correspond to a point along the sequence where $a\approx M$, which is at the point with the largest deviation between the solutions.  It is again difficult to see in this figure, but the slope of the imaginary part of the function vanishes at $x=0$ in the top plot where the phase has been fixed using $\Phase{SL-C}$.  In the bottom plot the slope of the function has a small imaginary component at $x=0$.

The difference between the two phase-fixing schemes is more easily seen in Fig.~\ref{fig:QNM300L1phasediff}, found in Appendix~\ref{sec:additional_figures}, which plots actual phase differences.  The phase difference between $\Phase{SL-C}$ and $\Phase{CZ-SL}$ as a function of $a$ has a maximum phase difference of about $0.006$.  There is, again, no phase difference between $\Phase{SL-C}$ and $\Phase{SL-Ind}$, which is expected since the $\ell=3$ sequence in Fig.~\ref{fig:QNM300Alm_a} exhibits no crossings.

Again, all phase fixing schemes yield eigenfunctions which vary smoothly along the sequence $0\le a<M$.  We display this in Fig.~\ref{fig:QNM300L1ExCoefRe}, found in Appendix~\ref{sec:additional_figures}, which displays the real part of the first $5$ expansion coefficients $\YSH{-2}{\ell30}{c(a)}$ from Eq.~(\ref{eqn:swSF-expansion}).  Again, for this simple example, all phase choices, including $\Phase{Math}$, yield smooth expansion coefficients.  This is not always the case.

As with the $\{2,2,0\}$ example, we compared the phase fixed versions of each expansion coefficient for each solution using Eqs.~(\ref{eqn:swSF EC phase}) to verify that each condition is satisfied.  For this simple example, all phase choices, including $\Phase{Math}$, again satisfy the basic symmetries given in Eqs.~(\ref{eqn:swSF S phase}).

\subsubsection*{QNM: $\ell=2$, $m=-2$, $n=13$}
\label{sec:QNM 2m213}
\begin{figure}
	\centering
	\includegraphics[width=\linewidth,clip]{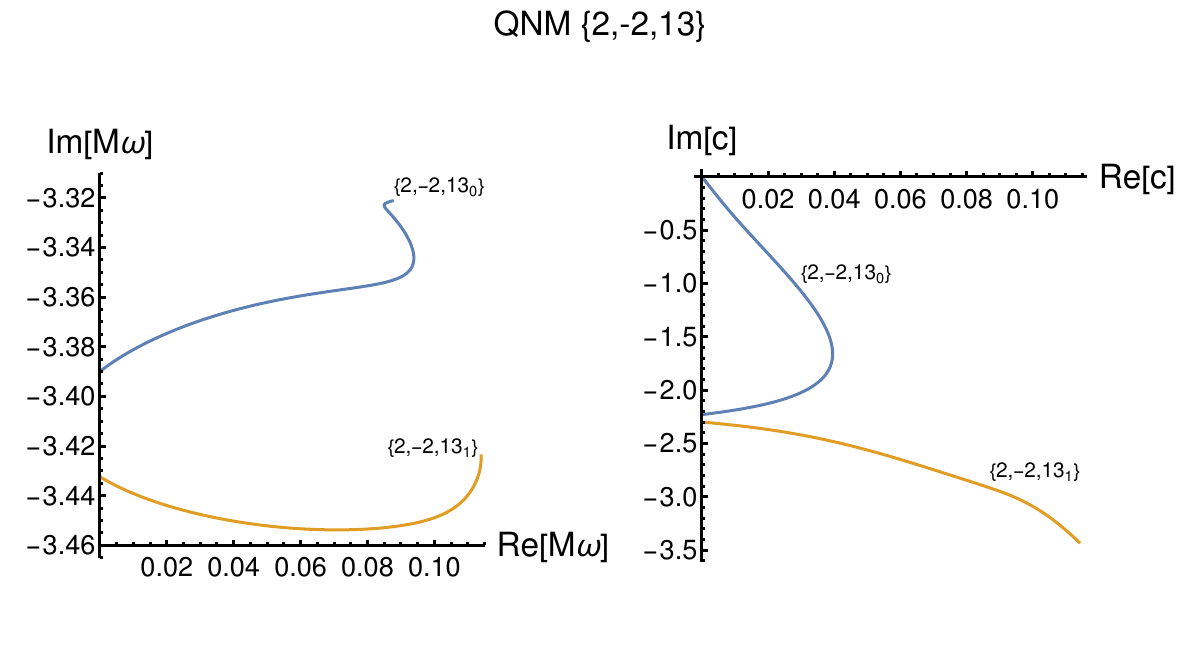}
	\caption{The QNM $\{2,-2,13\}$ sequence is composed of two parts. The first part, $\{2,-2,13_0\}$, begins at $a=0$ and extends to $a\approx0.657472M$.  No further QNMs exist for the sequence until $a\gtrsim0.669473M$ when the second part, $\{2,-2,13_1\}$, begins.  The left plot displays the path of $\omega^+_{2(-2)13}(a)$ through the complex frequency plane for the two overtone multiplets.  The right plot displays the path of corresponding oblateness parameter $c(a)=a\omega^+_{2(-2)13}(a)$ through the complex $c$ plane.}
	\label{fig:QNM2m213omegac}
\end{figure}

In this example, we consider a QNM with a higher overtone and having the unusual structure of consisting of two distinct parts, one of which does not connect directly to $a=0$.  The $\{2,-2,13\}$ QNM is an overtone multiplet.  The first portion of the sequence is labeled as $\{2,-2,13_0\}$, begins at $a=0$, but extends only to $a\approx0.657472M$ where it terminates with a mode frequency in the neighborhood of the negative imaginary axis (NIA).\footnote{Ref.~\cite{cook-zalutskiy-2016b} argued that a QNM on this sequence cannot exist exactly on the NIA. However, recent work\cite{Chen-etal-2025} claims to have found a method to extend this, and similar sequences, through the NIA.}  The second portion of the sequence is labeled as $\{2,-2,13_1\}$ and emerges from the NIA with $a\approx0.669473M$.  The left plot of Fig.~\ref{fig:QNM2m213omegac} shows the paths of the two overtone multiplet sequences $\omega^+_{2(-2)13_0}(a)$ and $\omega^+_{2(-2)13_1}(a)$ through the complex frequency plane.
The right plot of the figure shows the corresponding paths of the complex oblateness parameter $c(a)=a\omega^+_{2(-2)13}(a)$ for each overtone multiplet.

As discussed in Secs.~\ref{sec:CZ-phase} and \ref{sec:SL-phase}, when imposing the $\Phase{SL-C}$ and $\Phase{CZ-SL}$ phase choices, it is necessary to select a value of $\ell$ for the entire sequence.  In general, we take this value to be the same as the solution at $c=0$ along a smoothly connected sequence.  Clearly all of the solutions along the $\{2,-2,13_0\}$ multiplet are smoothly connected to $c=0$ and we choose $\ell=2$.  For the $\{2,-2,13_1\}$ multiplet, the QNM sequence itself has no direct connection to $c=0$ and we must find some other way to choose the value of $\ell$ needed to impose a phase choice.  In Ref.~\cite{cook-zalutskiy-2016b}, the two QNM sequences $\{2,-2,13_0\}$ and $\{2,-2,13_1\}$ were grouped together as an overtone multiplet with $\ell=2$ based on the behavior of nearby QNMs.  From Fig.~\ref{fig:QNM2m213omegac}, it is clear that we could choose a smooth path between the two multiplets to connect $\{2,-2,13_1\}$ to $c=0$, so the choice of $\ell=2$ is also justified for the $\{2,-2,13_1\}$ multiplet.

\begin{figure}
	\centering
	\includegraphics[width=\linewidth,clip]{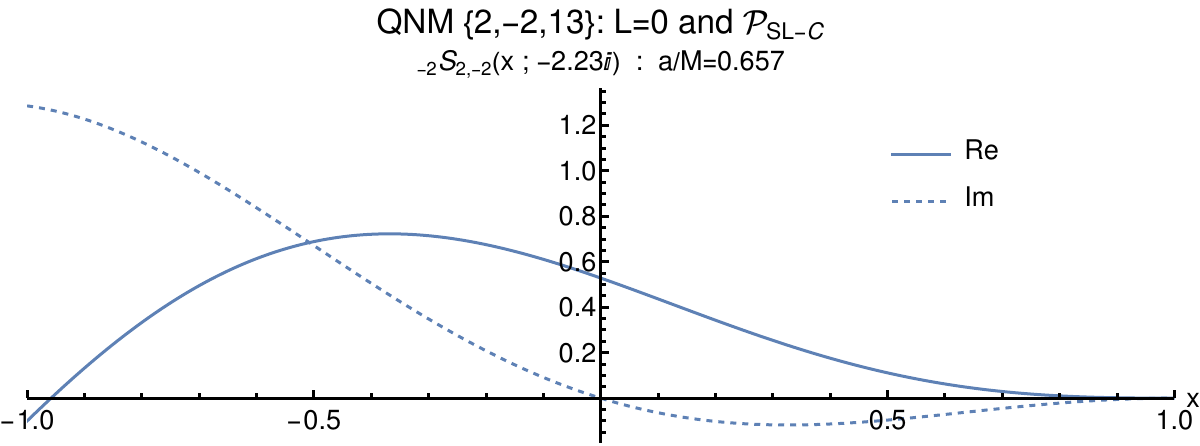}
	\includegraphics[width=\linewidth,clip]{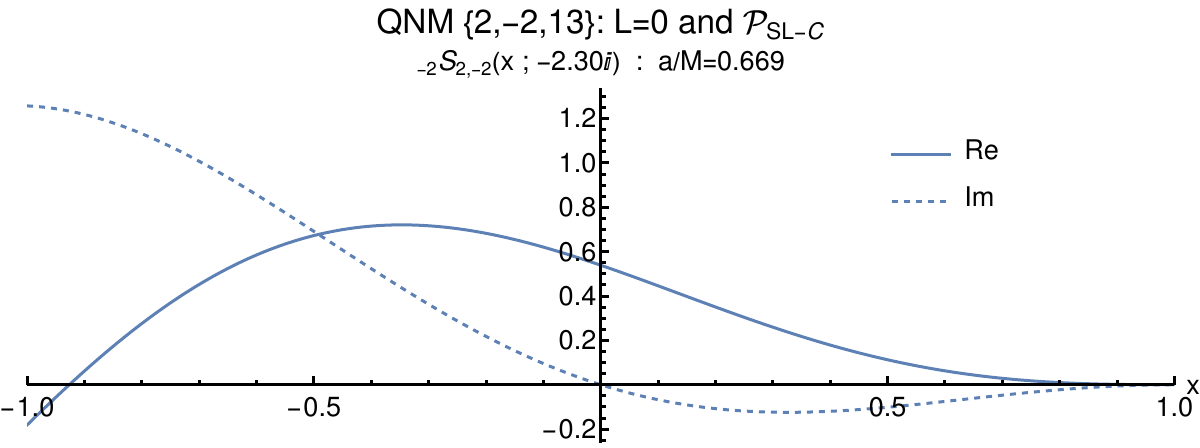}
	\caption{The functions $\swS{-2}{2,-2}{x}{-2.23i}$ and $\swS{-2}{2,-2}{x}{-2.30i}$ are displayed respectively in the upper and lower plots, with both phase fixed using $\Phase{SL-C}$.  The upper plot corresponds to $a=0.657$ at the beginning of the gap, and the lower plot to $a=0.669$ at the end of the gap.}
	\label{fig:QNM2m213L0S}
\end{figure}

Figure~\ref{fig:QNM2m213Alm_a}, found in Appendix~\ref{sec:additional_figures}, plots the magnitudes $|\scA{-2}{\ell(-2)}{c}-2|$ of the first 10 eigensolutions of the angular Teukolsky equation Eq.~(\ref{eqn:Angular Teukolsky Equation}) with $s=-2$ and $m=-2$, and no eigenvalue crossings are seen.
In Fig.~\ref{fig:QNM2m213L0S}, we display the behavior of the eigenfunction $\swS{-2}{2(-2)}{x}{c}$ for the $\Phase{SL-C}$ phase fixing scheme only.  In the upper plot, the function is plotted at $c\approx-2.23i$ corresponding to termination of the $\{2,-2,13_0\}$ multiplet where $a\approx0.657M$.  In the lower plot, the function is plotted at $c\approx-2.30i$ corresponding to the beginning of the $\{2,-2,13_1\}$ multiplet where $a\approx0.669M$.  Even though the two sequences are not directly connected, we see that the eigenfunction has a very similar shape on both sides of the solution gap.  Also, in this example, the value of $|c|$ is large enough that we can clearly see the phase-fixing condition that the function is real at $x=0$.

\begin{figure}
	\centering
	\includegraphics[width=\linewidth,clip]{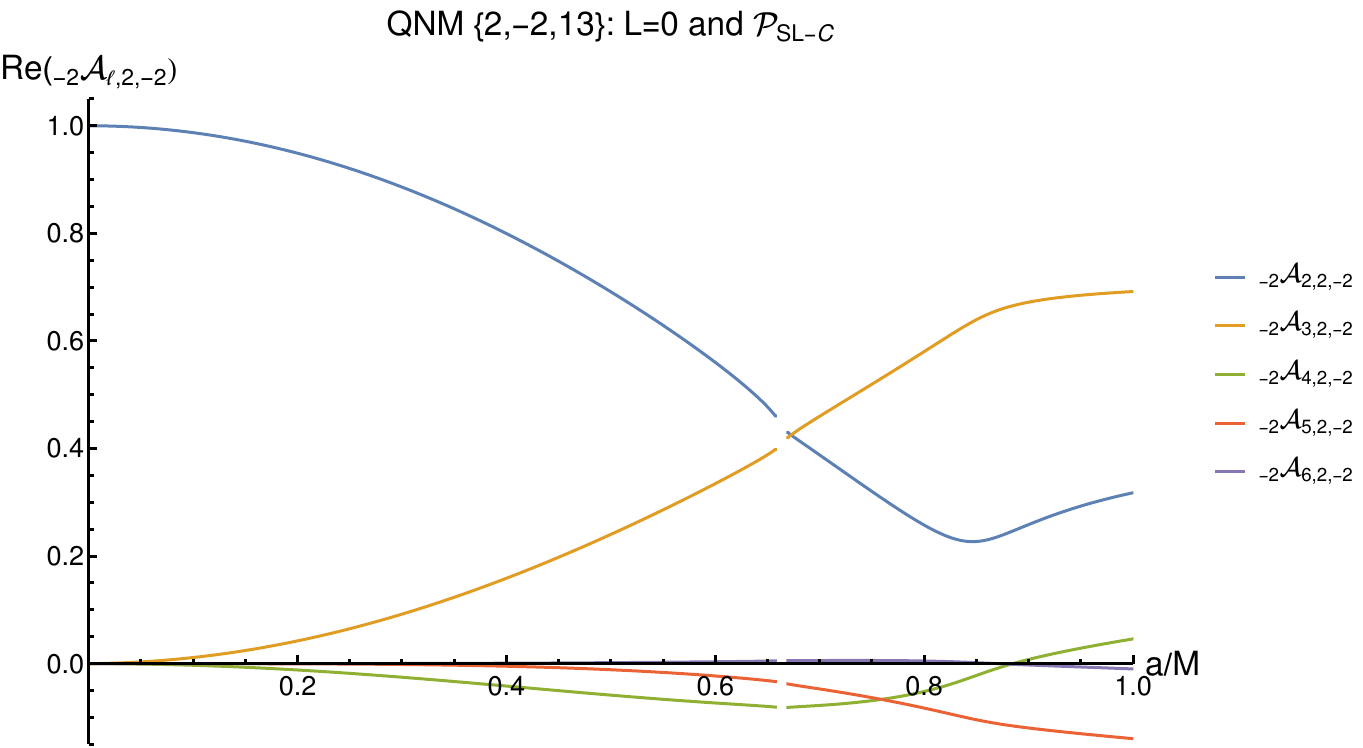}
	\caption{The real part of the first $5$ expansion coefficients $\YSH{-2}{\ell2(-2)}{a\omega^+_{2(-2)13}(a)}$ are plotted to demonstrate that the expansion coefficients are smooth functions of $a/M$ along each overtone multiplet for the $\Phase{SL-C}$ phase choice.}
	\label{fig:QNM2m213L0ExCoefRe}
\end{figure}

The difference between the $\Phase{SL-C}$ and $\Phase{CZ-SL}$ phase fixing schemes now becomes significant for the $\{2,-2,13\}$ QNM sequence with a maximum phase difference of order 1 which can be seen in Fig.~\ref{fig:QNM2m213L0phasediff}.
Again, all phase fixing schemes yield eigenfunctions which vary smoothly along each overtone multiplet.  We display this in Fig.~\ref{fig:QNM2m213L0ExCoefRe} which displays the real part of the first $5$ expansion coefficients $\YSH{-2}{\ell2(-2)}{c(a)}$ from Eq.~(\ref{eqn:swSF-expansion}).  We see that the different expansion coefficients are consistent across the gap of missing QNMs.
We include this figure to emphasize that, even though the expansion coefficients of the two branches of the sequences are smooth, care must be taken with QNMs which are overtone multiplets.  Interpolation between the branches of the sequence should not be attempted.

We, again, compared the phase fixed versions of each expansion coefficient for each solution using Eqs.~(\ref{eqn:swSF EC phase}) to verify that each condition is satisfied.  For this simple example, all phase choices, including $\Phase{Math}$, again satisfy the basic symmetries given in Eqs.~(\ref{eqn:swSF S phase}) along the $n=13_0$ segment of the multiplet.  But the situation is slightly more complicated along the $n=13_1$ segment.  Near $a\approx0.882M$, the magnitudes $|\YSH{-2}{\ell2(-2)}{a\omega^+_{2(-2)13}(a)}|$ for $\ell=2$ and $\ell=3$ cross and the $\Phase{Math}$ phase choice changes discontinuously.\footnote{Figure~\ref{fig:QNM2m213L0ExCoefRe} shows the real part of $\YSH{-2}{\ell2(-2)}{a\omega^+_{2(-2)13}(a)}$, not its magnitude.}  Furthermore, the two basic symmetries in Eqs.~(\ref{eqn:swSF sx S phase}) and (\ref{eqn:swSF mx S phase}) fail to be satisfied for $\Phase{Math}$ after the discontinuous phase change.  However, all of the $\Phase{SL}$ or $\Phase{CZ}$ phase choices yield expansion coefficients which satisfy the basic symmetries everywhere.

\subsubsection*{QNM: $\ell=2$, $m=2$, $n=31$}
\label{sec:QNM 2231}
\begin{figure}
	\centering
	\includegraphics[width=\linewidth,clip]{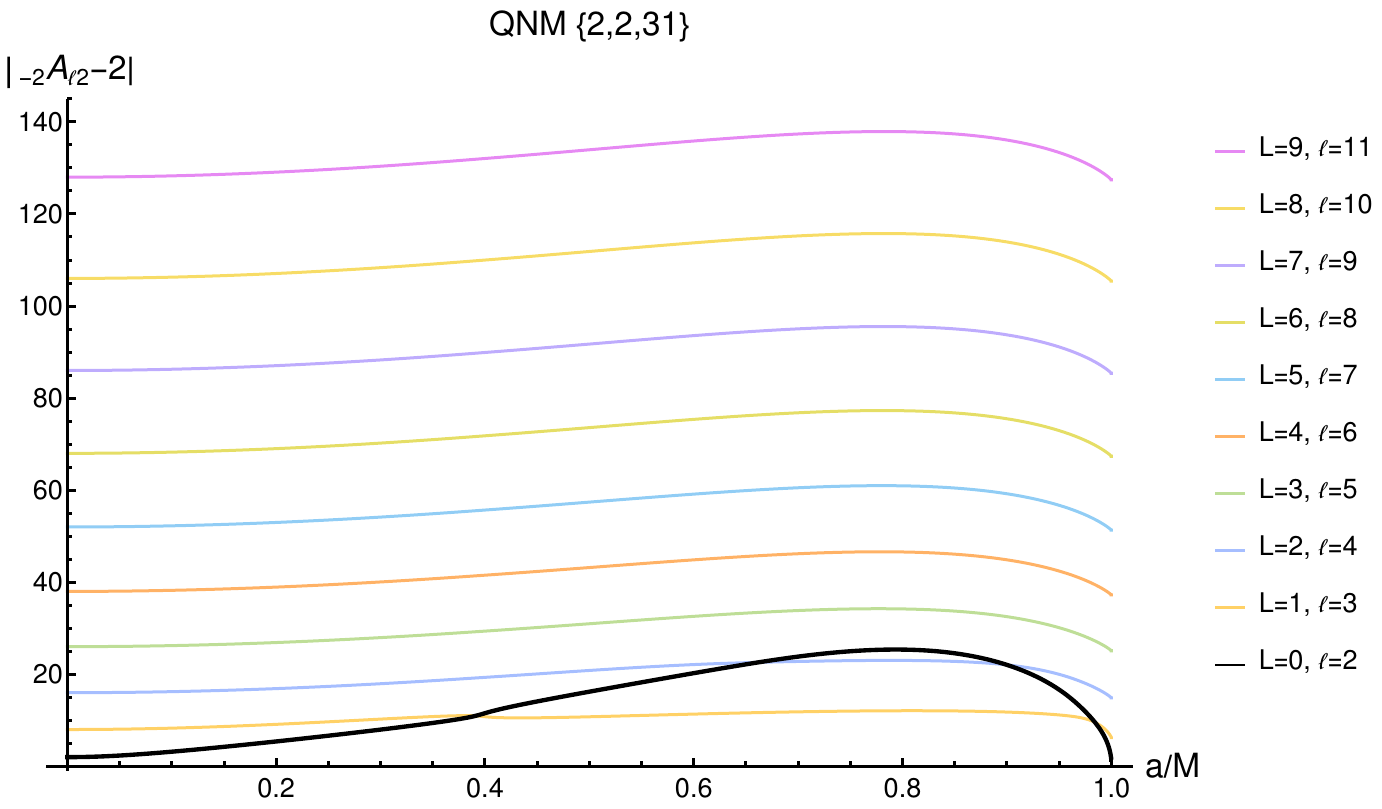}
	\caption{The first 10 eigenvalues for $s=-2$ and $m=2$ along a sequence of values of $c$ obtained from the QNM mode sequence $\{2,2,31\}$.  The plot displays $|\scA{-2}{\ell2}{c}-2|$ so that the eigenvalues appear in their sorted order.  The values are displayed as functions of the dimensionless angular momentum $a/M$ which is related to the oblateness parameter by $c=a\omega^+_{2,2,31}$, where $\omega^+_{2,2,31}$ is the complex mode frequency along the QNM sequence.  The eigenvalue sequence corresponding to the actual $\{2,2,31\}$ mode sequences is displayed as the thick black line.}
	\label{fig:QNM2231Alm_a}
\end{figure}

As a final example from the QNMs, we consider a QNM with a very high overtone.  The $\{2,2,31\}$ QNM presents an example where the sequences of $|\scA{-2}{\ell2}{a\omega^+_{2,2,31}(a)}-2|$ cross.  Figure~\ref{fig:QNM2231Alm_a} plots the magnitudes $|\scA{-2}{\ell2}{a\omega^+_{2,2,31}(a)}-2|$ of the first $10$ eigensolutions of the angular Teukolsky equation.  The sequence labeled $\ell=2(L=0)$ is the eigenvalue sequence which pairs with the $\omega^+_{2,2,31}(a)$ sequence of QNM mode frequencies.  For small values of $a/M$, it begins as the first eigenvalue, but near $a\approx0.4M$ it becomes the second sorted eigenvalue, and then the third near $a\approx0.67M$.  As $a/M$ increases further, it exhibits two more crossings as it transitions back to being the first eigenvalue.
This reordering of the eigenvalues along the sequence affects the smoothness of some of the phase choices.

\begin{figure}
	\centering
	\includegraphics[width=\linewidth,clip]{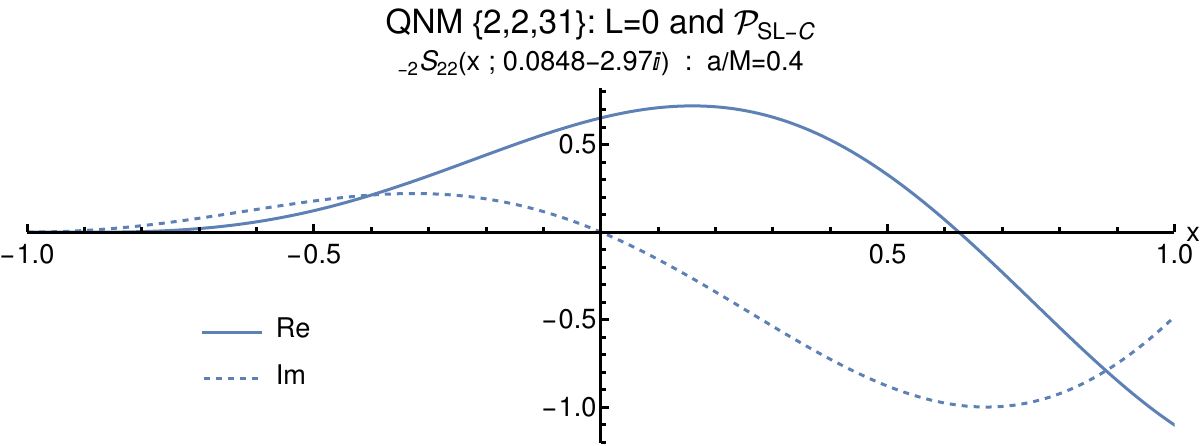}
	\includegraphics[width=\linewidth,clip]{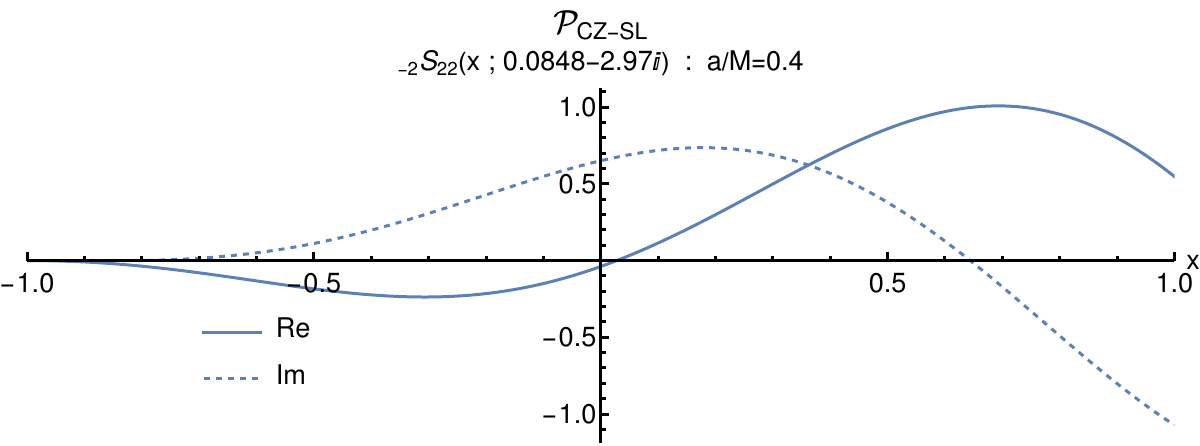}
	\caption{The functions $\swS{-2}{22}{x}{0.0848-2.97i}$ where the upper plot has been phase fixed using $\Phase{SL-C}$ and the bottom using $\Phase{CZ-SL}$.  Note that the two phase fixing schemes yield very different results in this case.}
	\label{fig:QNM2231L0S}
\end{figure}

\begin{figure}
	\centering
	\includegraphics[width=\linewidth,clip]{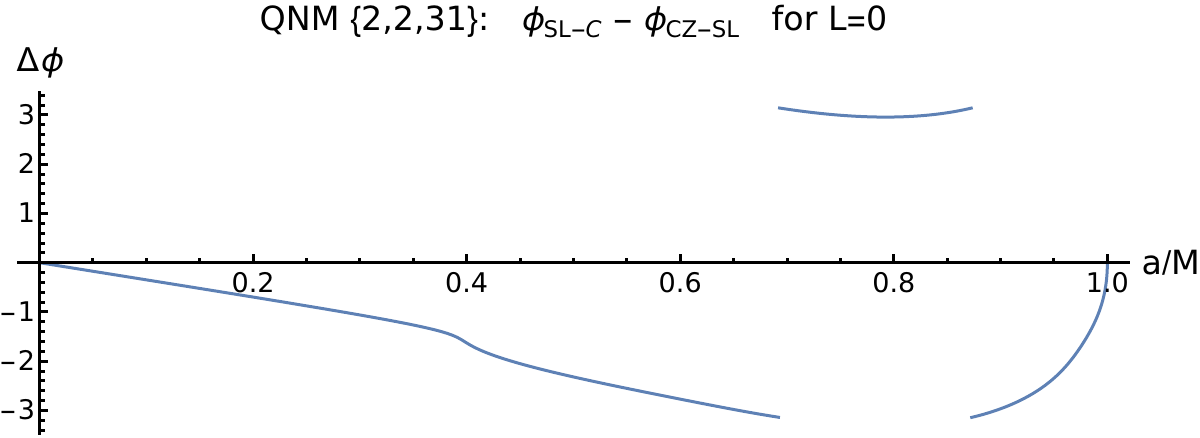}
	\includegraphics[width=\linewidth,clip]{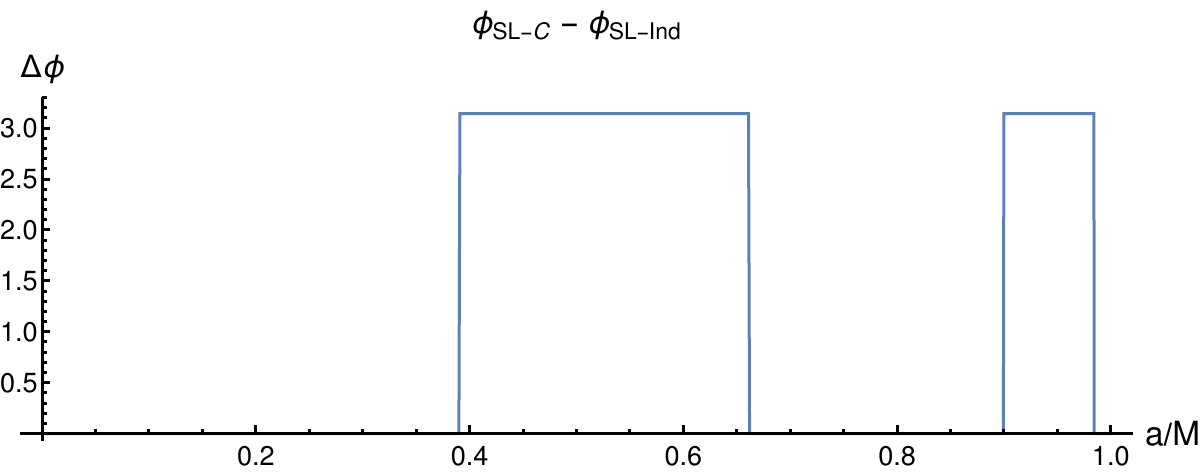}
	\caption{The phase difference between the functions $\swS{-2}{22}{x}{a\omega^+_{2,2,31}(a)}$ which have been phase fixed using different schemes.  In the top plot, the functions have been phase fixed using $\Phase{SL-C}$ and $\Phase{CZ-SL}$.  Both phase choices are continuous.  The discontinuities are simply the result of restricting phases to the range $(-\pi,\pi]$.  Such discontinuities will always be displayed as a gap.  In the bottom plot, the functions have been phase fixed using $\Phase{SL-C}$ and $\Phase{SL-Ind}$.  The discontinuities in the bottom plot are the result of eigenvalue crossings which lead to discontinuities in the $\Phase{SL-Ind}$ scheme.  Actual phase discontinuities will always be displayed with a continuous line.}
	\label{fig:QNM2231L0phasediff}
\end{figure}

In Fig.~\ref{fig:QNM2231L0S}, we display the behavior of the eigenfunction $\swS{-2}{22}{x}{0.0848-2.97i}$ for both the $\Phase{SL-C}$ and $\Phase{CZ-SL}$ phase fixing schemes.  We can clearly see that the $\Phase{SL-C}$ phase fixing scheme has fixed the function to be real and positive at $x=0$, while the $\Phase{CZ-SL}$ scheme yields a complex value there.
The top portion of Fig.~\ref{fig:QNM2231L0phasediff} displays the phase difference between these two phase choices along the entire sequence.  Both choices yield smoothly varying eigenfunctions along the entire sequence.  The bottom portion of Fig.~\ref{fig:QNM2231L0phasediff} displays the phase difference between the $\Phase{SL-C}$ and $\Phase{SL-Ind}$ schemes.  The discontinuities are the result of the value for $\ell$ used by $\Phase{SL-Ind}$ changing as we move along the sequence.  A careful comparison with Fig.~\ref{fig:QNM2231Alm_a} shows that the discontinuities occur precisely at values of $a/M$ where crossings occur in $|\scA{-2}{\ell2}{a\omega^+_{2,2,31}(a)}-2|$.
Because of these crossings, only the $\Phase{SL-C}$ and $\Phase{CZ-SL}$ phase fixing schemes allow $\swS{-2}{22}{x}{a\omega^+_{2,2,31}(a)}$ to vary smoothly along the sequence.

\begin{figure}
	\centering
	\includegraphics[width=\linewidth,clip]{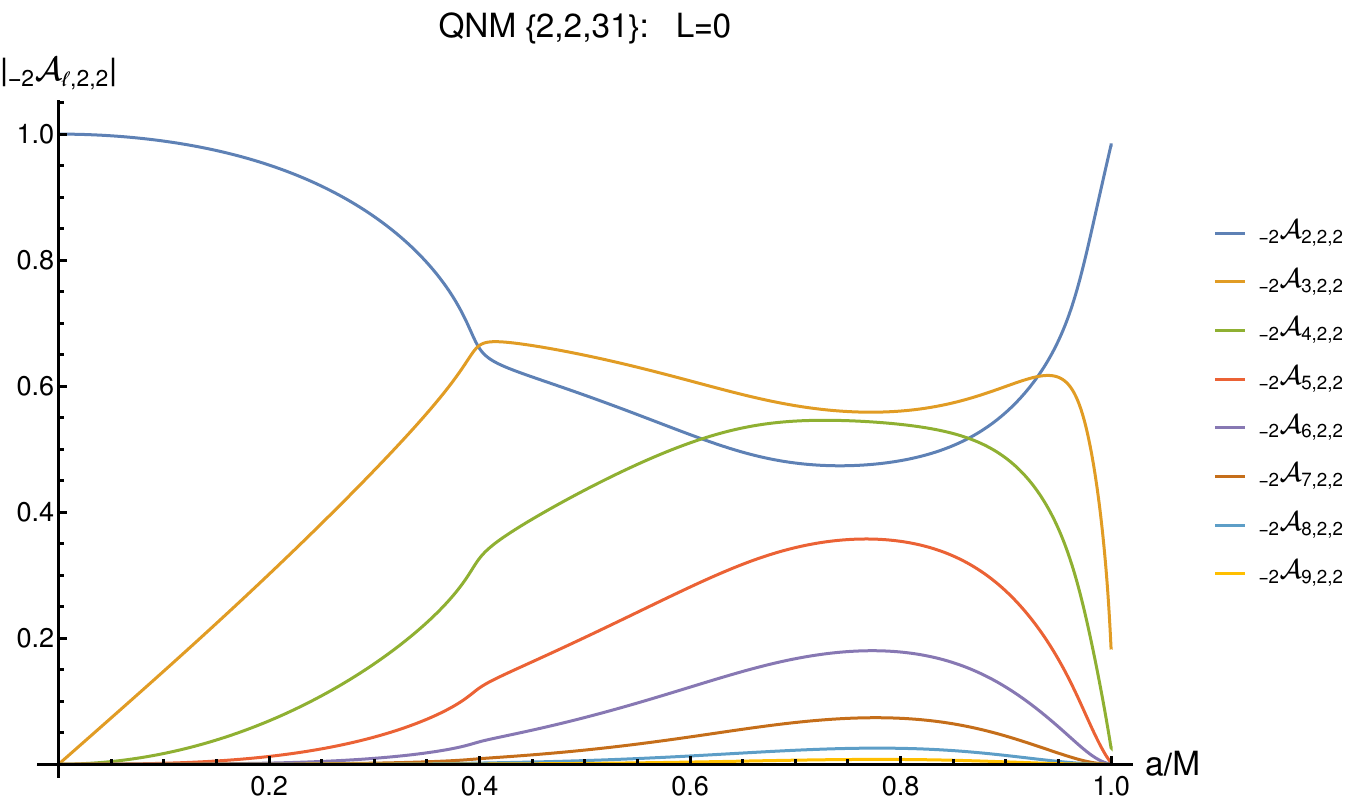}
	\caption{The magnitudes of the first $8$ expansion coefficients $\YSH{-2}{\ell22}{a\omega^+_{2,2,31}(a)}$ are plotted to show how the coefficient with the largest magnitude can change as $a/M$ varies the sequence.  This plot is independent of phase choice.}
	\label{fig:QNM2231L0ExCoefAbs}
\end{figure}

In Fig.~\ref{fig:QNM2231L0ExCoefAbs} we display $|\YSH{-2}{\ell22}{a\omega^+_{2,2,31}(a)}|$ for the first $8$ values of $\ell$ instead of the real part of these coefficients as shown in similar previous plots.
We clearly see that the $\YSH{-2}{222}{a\omega^+_{2,2,31}(a)}$ expansion coefficient is dominant for small values of $a/M$, but between $0.4\lesssim a/M \lesssim 0.93$ it becomes subdominant.

These changes in the dominant expansion coefficient cause the $\Phase{Math}$ phase choice to change discontinuously.  This also leads to a failure of $\Phase{Math}$ to satisfy the two basic symmetries in Eqs.~(\ref{eqn:swSF sx S phase}) and (\ref{eqn:swSF mx S phase}) along portions of the sequence.  However, all the $\Phase{SL}$ or $\Phase{CZ}$ phase choices yield expansion coefficients which satisfy the basic symmetries everywhere.

\subsection{Total-transmission modes}
\label{sec:TTM examples}
The gravitational TTMs of the Kerr geometry are less well known than the QNMs and have not yet found a significant application in gravitational physics (see however, Ref.~\cite{tuncer-2026}).  Nevertheless, they are a part of the full mode structure of the Kerr geometry, and their angular mode solutions offer interesting and more extreme examples to consider.  There are two types of TTMs: the left-TTMs (TTM${}_{\rm L}$s) and the right-TTMs (TTM${}_{\rm R}$s).  TTM${}_{\rm L}$s represent modes which propagate away from the black-hole horizon and toward spatial infinity without experiencing any net scattering back into the black hole.  TTM${}_{\rm R}$s propagate in the opposite direction, without any net scattering back toward spatial infinity.  

Both types of TTMs obey the same symmetries which lead to the two families of QNMs differentiated by the $\pm$ superscript on the mode frequency (see Sec.~\ref{sec:QNM examples}).  They are also indexed in the same way, except that they do not have the same structure of an infinite number of overtones for each mode.  Only a single TTM was thought to exist for each harmonic mode\cite{chandra-1984} until recently when two additional families of solutions were found\cite{CookLu2023}.  It is convenient to differentiate between these $3$ families using the third(overtone) index.  The original set of TTMs is designated by $n=0$, and the mode frequencies in the Schwarzschild limit($a=0$) are imaginary and finite, with values $M\omega^\pm_{\ell{m}0}(0)=-\frac{i}{12}(\ell-1)\ell(\ell+1)(\ell+2)$.  The two new sets of solutions are distinguished by having Schwarzschild-limit frequencies existing at complex infinity.  The second set, designated by $n=1$, has mode frequencies with leading order behavior $M\omega^+_{\ell{m}1}(a)=-i2^{2/3}3^{1/3}(M/a)^{4/3}$ in the Schwarzschild limit.  The third set, designated by $n=2$, has mode frequencies with leading order behavior $M\omega^+_{\ell{m}2}(a)=-i(-1)^{1/3}2^{2/3}3^{1/3}(M/a)^{4/3}$.  

Below, we will explore two examples from the $n=2$ set of TTM${}_{\rm R}$s which allow us to explore the behavior of SWSFs for large values of $|c|$ as $a\to0$.  Because the $n=2$ TTMs do not connect to $c=0$, we must find another way to fix a value of $\ell$ for the both the $\Phase{SL-C}$ and $\Phase{CZ-SL}$ phase-fixing schemes.  As discussed in Sec.~\ref{sec:general sw spheroidal}, this can be accomplished by examining the asymptotic behavior of $\scA{s}{\ell{m}}{c}$ as $|c|\to\infty$.

\subsubsection*{TTM${}_R$: $\ell=2$, $m=2$, $n=2$}
\label{sec:TTMR 222}
\begin{figure}
	\centering
	\includegraphics[width=\linewidth,clip]{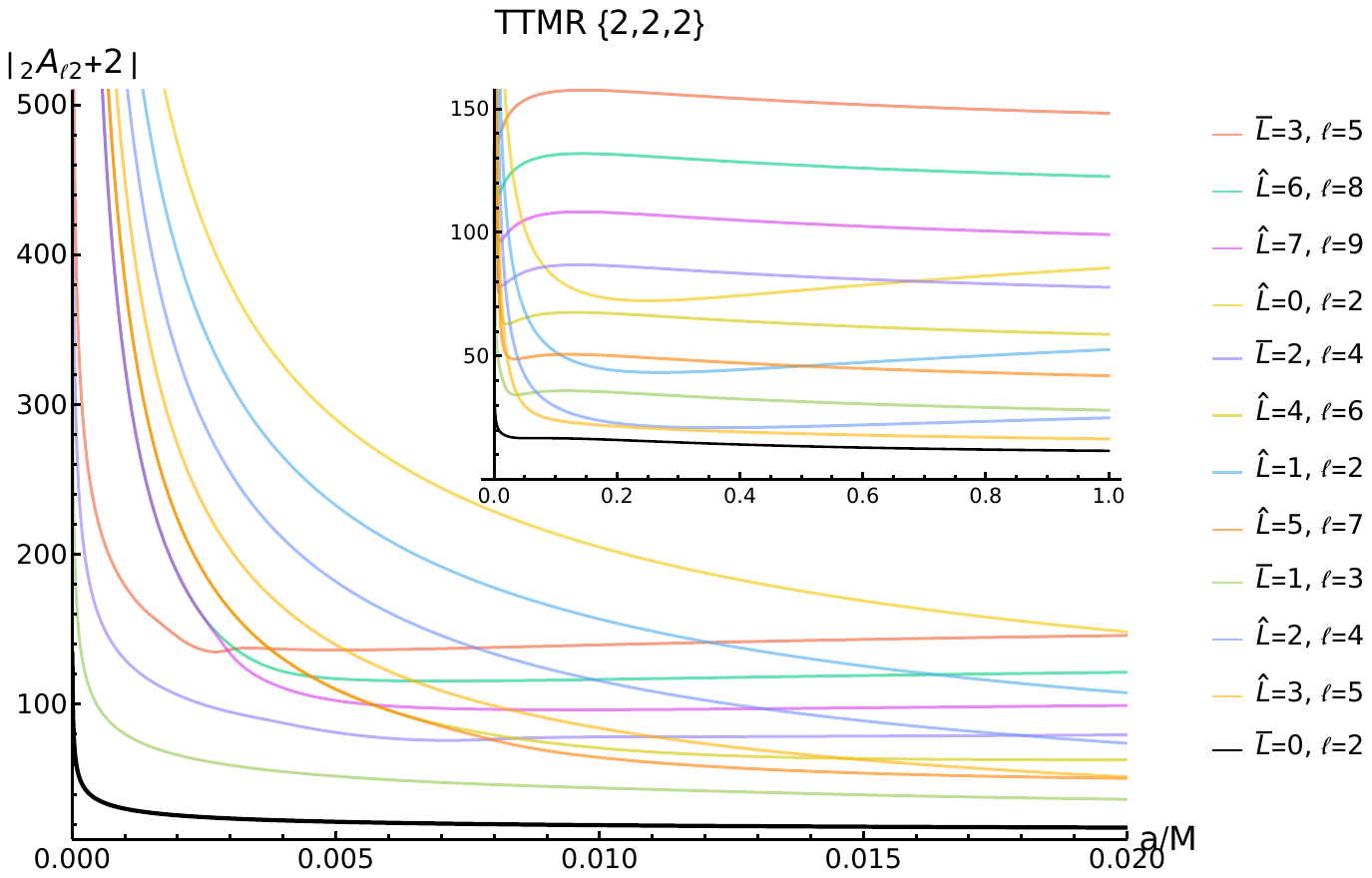}
	\caption{The first 12 eigenvalues for $s=2$ and $m=2$ along a sequence of values of $c$ obtained from the TTM${}_{\rm R}$ mode sequence $\{2,2,2\}$.  The plot displays $|\scA{2}{\ell2}{c}+2|$ so that the eigenvalues appear in their sorted order.  The values are display as functions of the dimensionless angular momentum $a/M$ which is related to the oblateness parameter by $c=a\omega^+_{222}$, where $\omega^+_{222}$ is the complex mode frequency along the TTM${}_{\rm R}$ sequence.  The eigenvalue sequence corresponding to the actual $\{2,2,2\}$ mode sequences is displayed as the thick black line.  The inset figure show the plot along the full range of $a$, while the full plot show the behavior for small values of $a$.}
	\label{fig:TTMR222Alm_a}
\end{figure}

The $\{2,2,2\}$ TTM${}_{\rm R}$ presents an example with $s=2$ where the sequences of $|\scA{2}{\ell2}{a\omega^+_{2,2,2}(a)}+2|$ display a complicated structure.  Figure~\ref{fig:TTMR222Alm_a} plots the magnitude $|\scA{2}{\ell2}{a\omega^+_{2,2,2}(a)}+2|$ from the first $12$ eigensolutions of the angular Teukolsky equation.  The sequence labeled $\bar{L}=0$ is the eigenvalue sequence which pairs with the $\omega^+_{2,2,2}(a)$ sequence of TTM${}_{\rm R}$ mode frequencies, and is seen as the first(lowest) eigenvalue for all values of $a$.  The remaining eigenvalue sequences appear labeled by various values of $\bar{L}$ or $\hat{L}$ which seem to have no obvious ordering.  The ordering of sequences in the plot legend corresponds to the ordering of the eigenvalue sequences at $a/M=1$.  As $a$ decreases, there are numerous crossings, but as $a\to0$, the sequences display two distinct behaviors.

\begin{figure}
	\centering
	\includegraphics[width=\linewidth,clip]{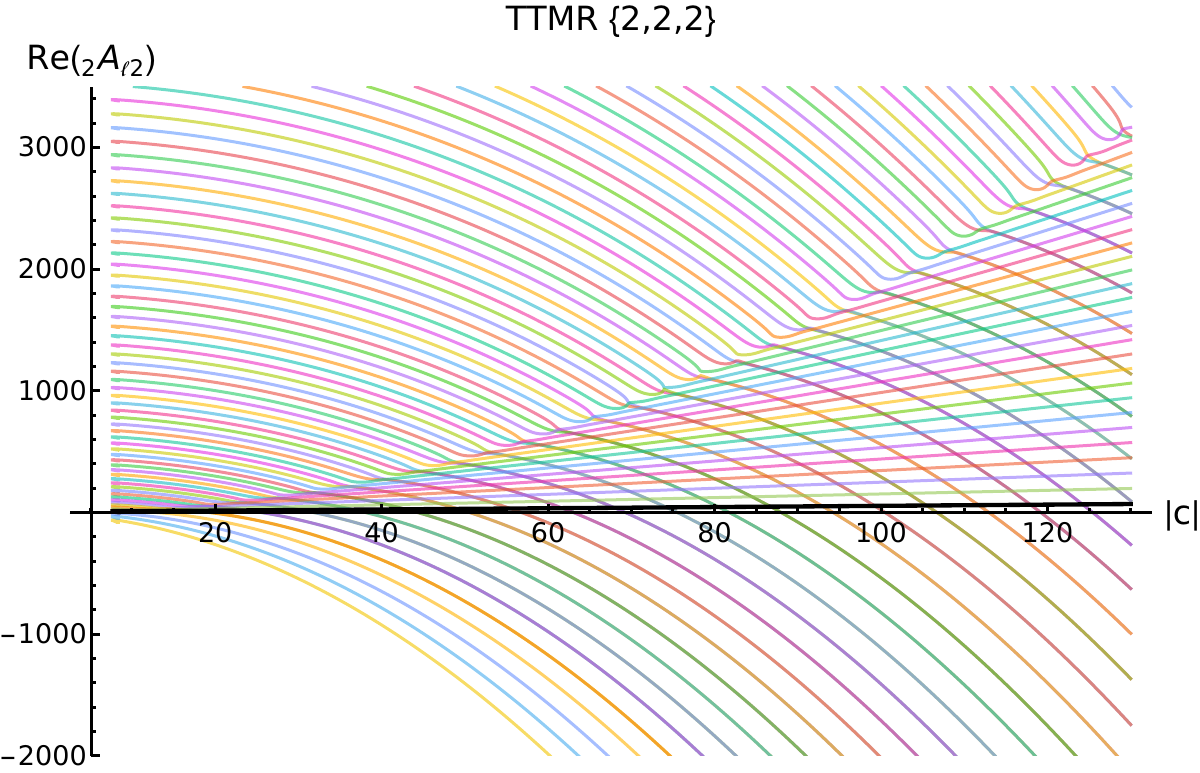}
	\caption{A large number of eigenvalues for $s=2$ and $m=2$ along a sequence of values of $c$ obtained from the TTM${}_{\rm R}$ mode sequence $\{2,2,2\}$.  The plot displays sequences of $\Re(\scA{2}{\ell2}{c})$ as functions of $|c|$ where the oblateness parameter $c=a\omega^+_{222}(a)$, and $\omega^+_{222}(a)$ is the complex mode frequency along the TTM${}_{\rm R}$ sequence.  The eigenvalue sequence corresponding to the actual $\{2,2,2\}$ mode sequences is displayed as the thick black line.}
	\label{fig:TTMR222ReAlm_cAll}
\end{figure}

\begin{figure}
	\centering
	\includegraphics[width=\linewidth,clip]{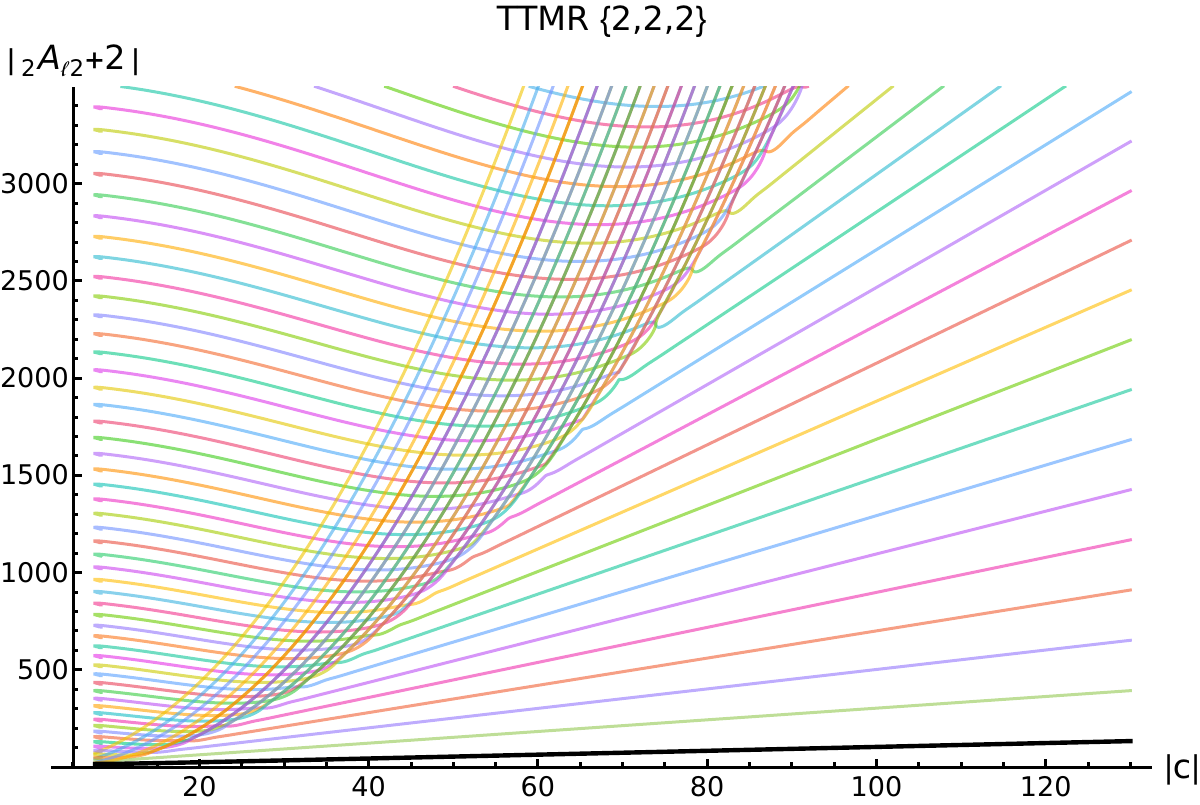}
	\caption{A large number of eigenvalues for $s=2$ and $m=2$ along a sequence of values of $c$ obtained from the TTM${}_{\rm R}$ mode sequence $\{2,2,2\}$.  The plot displays sequences of $|\scA{2}{\ell2}{c}+2|$ so that the eigenvalues appear in their sorted order.  The values are display as functions of $|c|$ where the oblateness parameter $c=a\omega^+_{222}(a)$, and $\omega^+_{222}(a)$ is the complex mode frequency along the TTM${}_{\rm R}$ sequence.  The eigenvalue sequence corresponding to the actual $\{2,2,2\}$ mode sequences is displayed as the thick black line.}
	\label{fig:TTMR222AbsAlm_cAll}
\end{figure}

\begin{figure}
	\centering
	\includegraphics[width=\linewidth,clip]{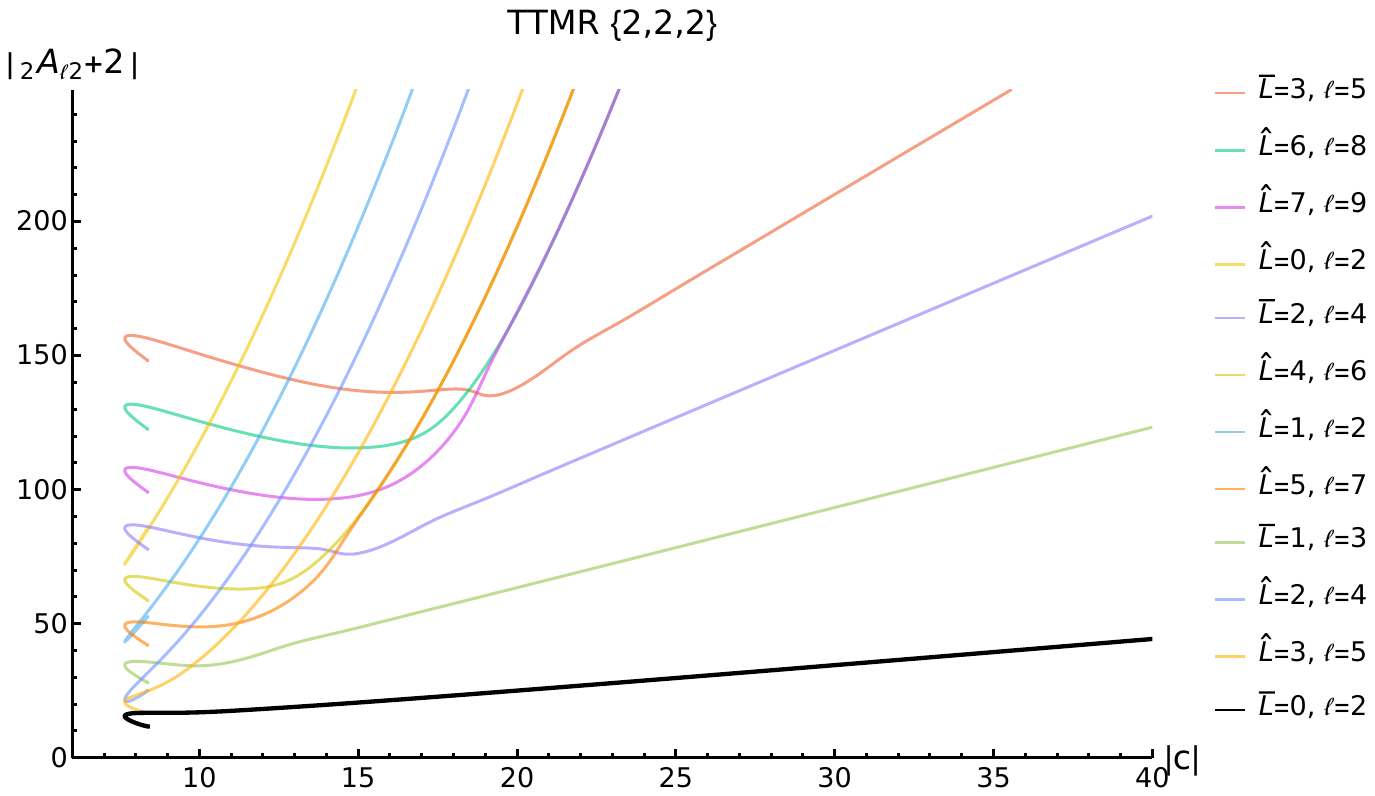}
	\caption{$12$ sequences from the large set of sequences in Fig.~\ref{fig:TTMR222AbsAlm_cAll}.  See the caption for this figure for additional details.  The sequences labeled by $\bar{L}$ have leading asymptotic behavior $ic(2\bar{L}+1)$.  Those labeled by $\hat{L}$ have leading asymptotic behavior $-c^2$.}
	\label{fig:TTMR222AbsAlm_c12}
\end{figure}

The asymptotic behavior is more easily seen in Fig.~\ref{fig:TTMR222ReAlm_cAll} which displays the real part of $\scA{2}{\ell2}{c}$ as a function of $|c|$ for a large number of eigenvalues along $c(a) = a\omega^+_{222}(a)$, with asymptotic behavior $c(a)\approx-i(-1)^{1/3}2^{2/3}3^{1/3}(M/a)^{1/3}$.
The eigenvalue sequences in one subset have leading order asymptotic behavior $-c^2$, and can be clearly seen extending to large negative values.  The remaining sequences have leading order asymptotic behavior $ic(2\bar{L}+1)$, but are less clearly visible as they are repeatedly crossed by members of the first subset.  They are more easily seen in Fig.~\ref{fig:TTMR222AbsAlm_cAll} which displays $|\scA{2}{\ell2}{c}+2|$ as a function of $|c|$ for the same set of sequences.
In this figure, the subset of sequences with leading order asymptotic behavior $ic(2\bar{L}+1)$ are easily seen, while sequences in the $-c^2$ subset are repeatedly crossing other sequences.  The ordering and labeling of these sequences is more easily understood by examining a subset of these sequences over a smaller range of $|c|$ as presented in Fig.~\ref{fig:TTMR222AbsAlm_c12} which displays only $12$ sequences.
The asymptotic indexing of the sequences behaving as $ic(2\bar{L}+1)$ is easily understood in this figure.  The indexing of the sequences behaving asymptotically as $-c^2$ is still not discernible in this figure.

\begin{figure}
	\centering
	\includegraphics[width=\linewidth,clip]{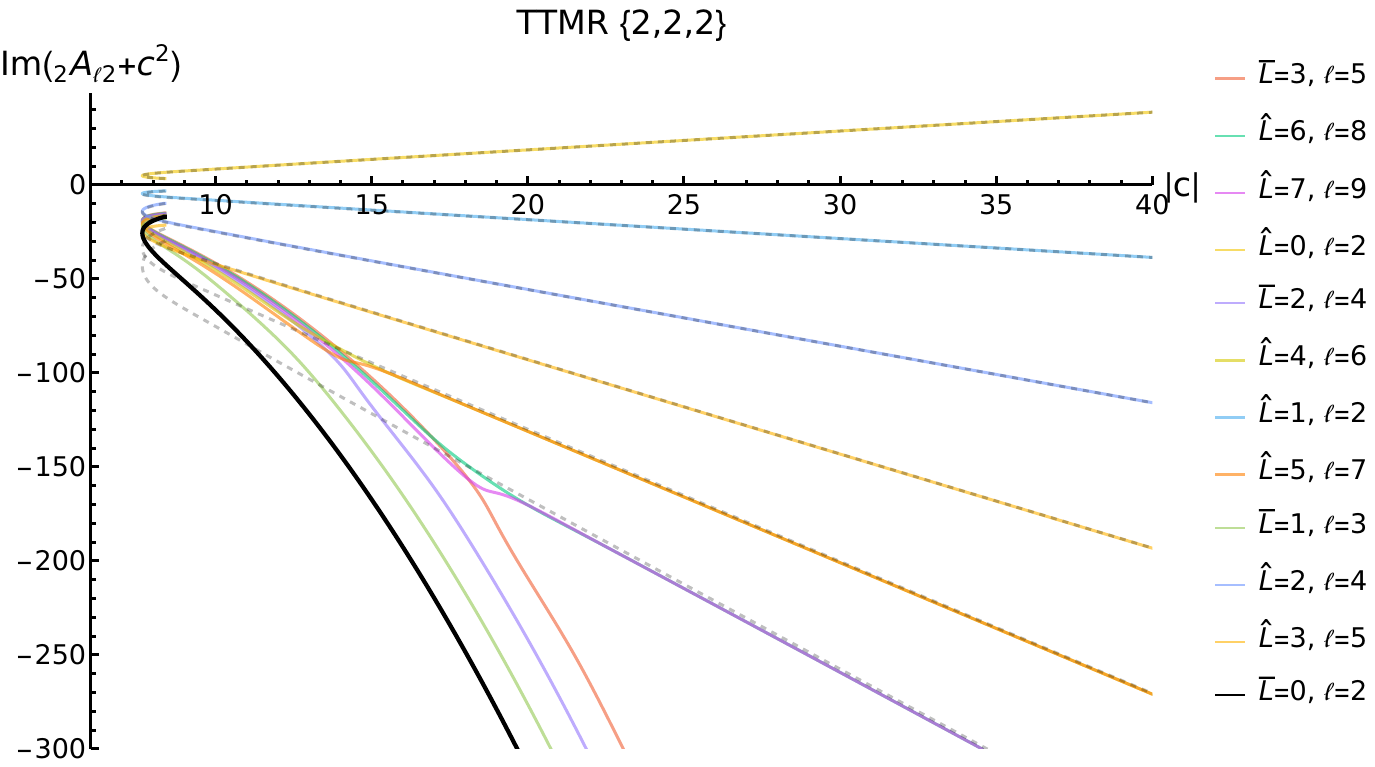}
	\caption{The same $12$ sequences displayed in Fig.~\ref{fig:TTMR222AbsAlm_c12}, but now plotting $\Im[\scA{2}{\ell2}{c}+c^2]$.  The sequences labeled by $\hat{L}$ have leading asymptotic behavior $-c^2$, so this plot displays the next-to-leading order behavior of these sequences.  The dashed lines take on values of $2(2i-1)\Im(c)$ for $0\le i\le5$, where the oblateness parameter $c=a\omega^+_{222}(a)$, and $\omega^+_{222}(a)$ is the complex mode frequency along the TTM${}_{\rm R}$ sequence.}
	\label{fig:TTMR222ImAlm_c12}
\end{figure}

The indexing of the eigenvalues which behave asymptotically as $-c^2$ can be extracted from their next-to-leading order behavior.  This behavior is well understood in the cases where $c^2$ is real\cite{Casalas-oblate-2005,VickersCook2022}, but not for general values of $c$, and is complicated by the asymptotic degeneracy of certain pairs of solutions.  In Fig.~\ref{fig:TTMR222ImAlm_c12}, we plot the imaginary part of $\scA{2}{\ell2}{c}+c^2$ to reveal the next-to-leading order behavior of these sequences.  The dashed lines take on values of $2(2i-1)\Im(c)$ for $0\le i\le5$ and we see excellent asymptotic agreement with this behavior.  The sequences labeled by $\hat{L}=0\ldots3$ corresponds to $i=0\ldots3$, but $i=4$ gives the behavior of the two asymptotically degenerate sequences labeled by $\hat{L}=4$ and $5$, while $i=5$ gives the behavior of the two degenerate sequences labeled by $\hat{L}=6$ and $7$.  We do not have any rigorous justification for which of the degenerate sequences receives which of the two possible values of $\hat{L}$.

\begin{figure}
	\centering
	\includegraphics[width=\linewidth,clip]{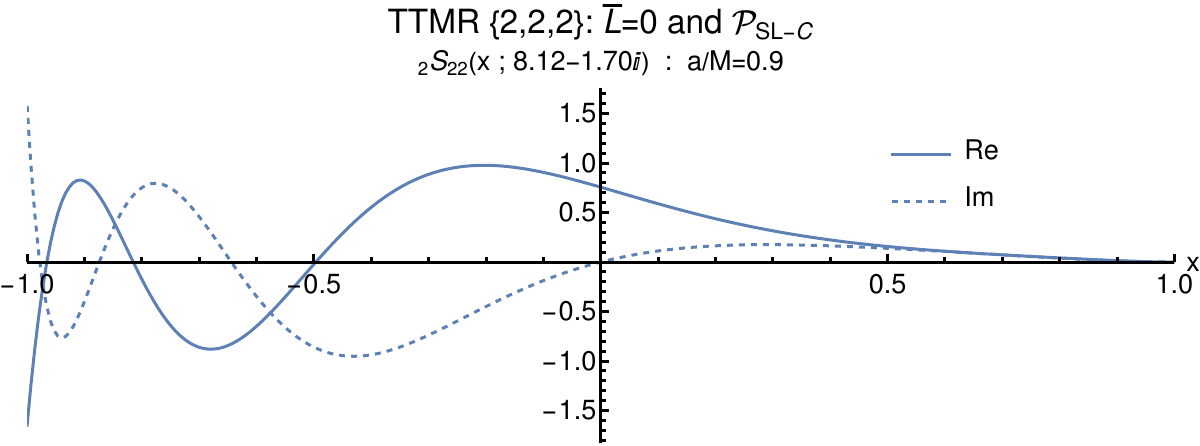}
	\includegraphics[width=\linewidth,clip]{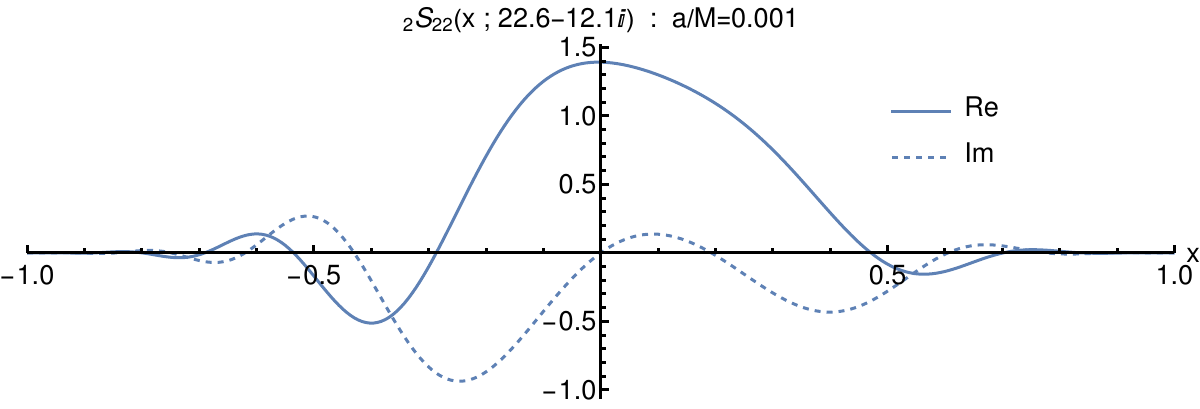}
	\includegraphics[width=\linewidth,clip]{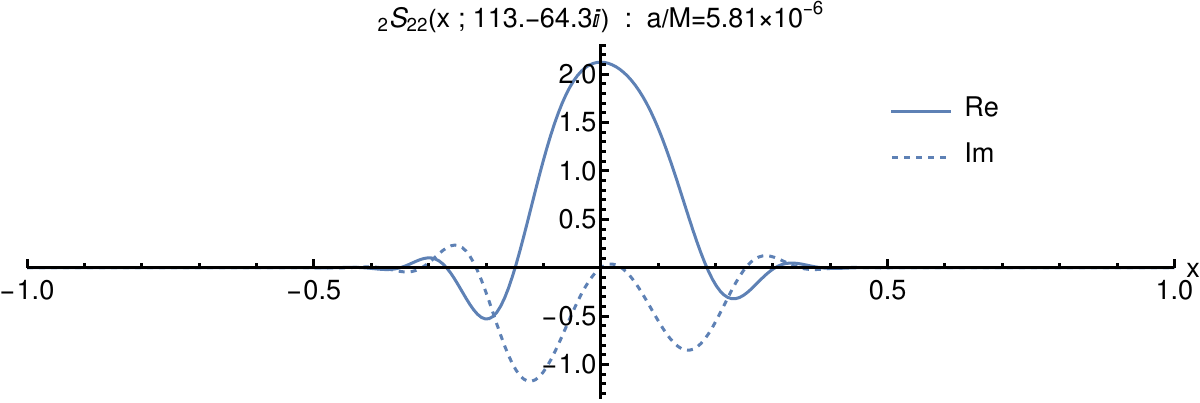}
	\caption{The functions $\swS{2}{22}{x}{c}$ for $c=8.12-1.70i$, $22.8-12.1i$, and $113.-64.3i$ are shown, respectively, in the upper, middle, and lower plots of the figure.  All satisfy $\Phase{SL-C}$.}
	\label{fig:TTMR222Lb0S}
\end{figure}

Now consider the behavior of the eigenfunctions along two of the sequences displayed in Figs.~\ref{fig:TTMR222AbsAlm_cAll} and \ref{fig:TTMR222AbsAlm_c12}.  In Fig.~\ref{fig:TTMR222Lb0S} we display the behavior of the eigenfunction $\swS{2}{22}{x}{c}$ at $3$ different values of $c$ along the $\bar{L}=0$ sequence.  Because this sequence does not extend to $c=0$, but does extend to asymptotically large values of $|c|$, we use the $\bar{L}=0$ asymptotic behavior to fix $\ell=2$ everywhere on this sequence for the $\Phase{SL-C}$ and $\Phase{CZ-SL}$ phase-fixing schemes.  The top plot in Fig.~\ref{fig:TTMR222Lb0S} shows the eigenfunction for a small value of $|c|$ along the sequence corresponding to $a=0.9M$.  The middle and lower plots show the eigenfunction for successively larger values of $|c|$ and smaller values of $a$.  All are phase fixed using $\Phase{SL-C}$ and we can clearly see that $\swS{2}{22}{0}{c}$ is real and positive in each plot.
The smoothness of the eigenfunction as it varies along the sequence can be verified in Fig.~\ref{fig:TTMR222Lb0ExCoefRe}, found in Appendix~\ref{sec:additional_figures}, which plots the real part of the first $10$ expansion coefficients $\YSH{2}{\ell22}{a\omega^+_{222}(a)}$.  Each coefficient varies smoothly as a function of $a$, showing that the $\Phase{SL-C}$ yields a smooth sequence of eigenfunctions.
The phase difference between $\Phase{SL-C}$ and $\Phase{CZ-SL}$ is smooth and rather large for all values of $a$ along the TTM${}_{\rm R}$ $\{2,2,2\}$ sequences, as can be seen in Fig.~\ref{fig:TTMR222Lb0phasediff} of Appendix~\ref{sec:additional_figures}.  We also see that the $\Phase{SL-Ind}$ phase-fixing scheme also yields a smooth sequence, as expected since there are no crossings for the $\bar{L}=0$ sequences (see Fig.~\ref{fig:TTMR222AbsAlm_c12}).

\begin{figure}
	\centering
	\includegraphics[width=\linewidth,clip]{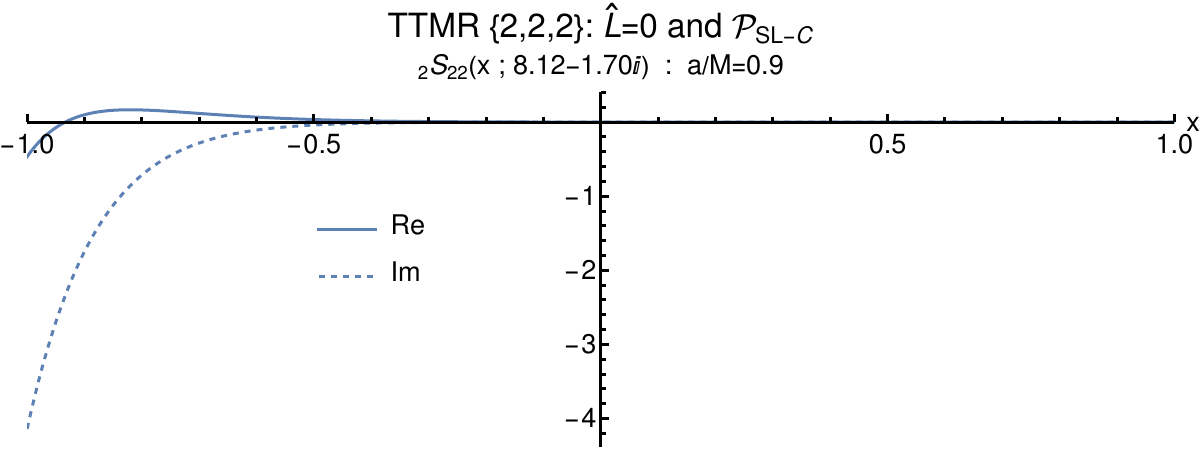}
	\includegraphics[width=\linewidth,clip]{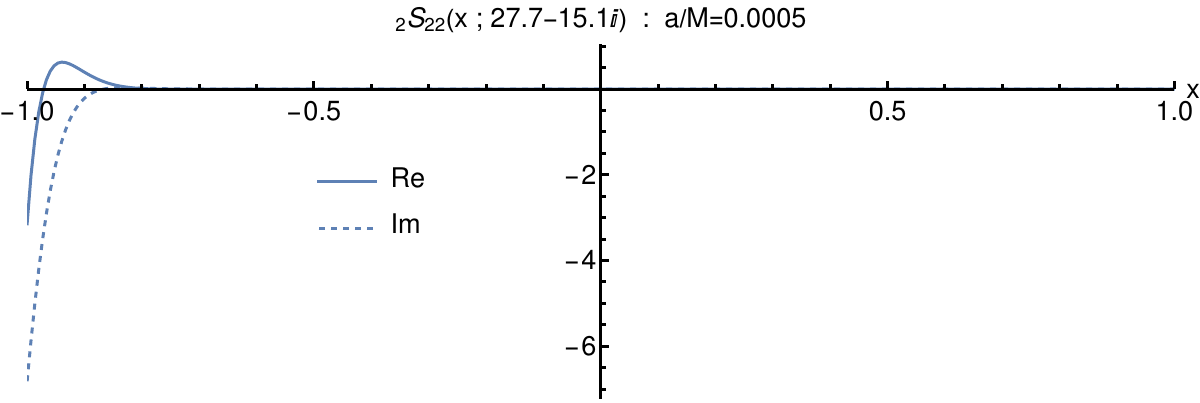}
	\caption{The functions $\swS{2}{22}{x}{c}$ for $c=9.04-4.16i$ and $27.7-15.1i$ are shown, respectively, in the upper and lower plots of the figure.  All satisfy $\Phase{SL-C}$.}
	\label{fig:TTMR222Lh0S}
\end{figure}

\begin{figure}
	\centering  
	\includegraphics[width=\linewidth,clip]{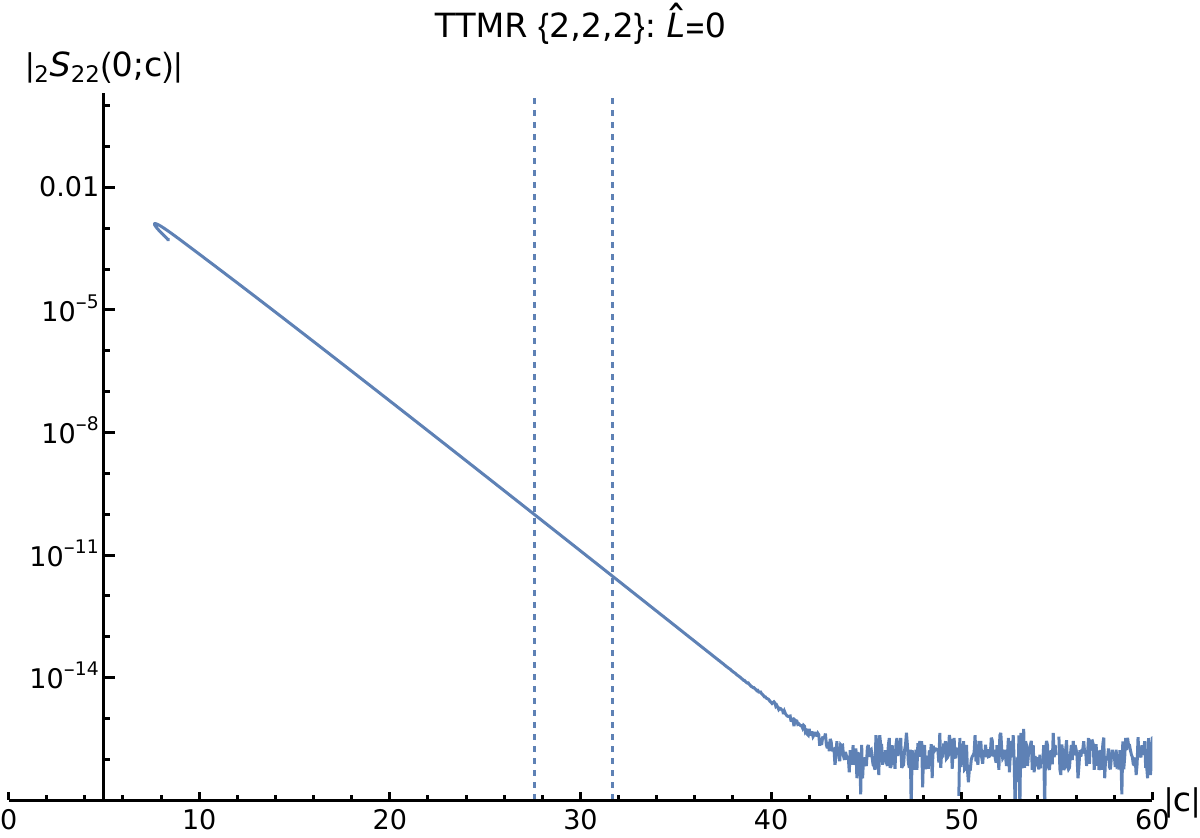}
	\caption{Log plot of the magnitude of the eigenfunction $\swS{2}{22}{x}{c}$ for $\hat{L}=0$ at $x=0$ along $c=a\omega^+_{222}(a)$, and $\omega^+_{222}(a)$ is the complex mode frequency along the TTM${}_{\rm R}$ sequence.  The left vertical dashed line marks the threshold($10^{-10}$) where $\Phase{SL-C}$ determines $\swS{2}{22}{0}{c}\approx0$, cannot fix the phase using Eqs.~(\ref{eqn:SL non-zero}), and switches to using Eqs.~(\ref{eqn:SL zero}).  The right vertical dashed line marks the threshold ($10^{-10}$) where $\Phase{SL-C}$ determines both $\swS{2}{22}{0}{c}\approx0$ and $\left.\partial_x\swS{2}{22}{x}{c}\right|_{x=0}\approx0$, and begins to fix the phase using $\Phase{CZ-SL}$.}
	\label{fig:TTMR222Lh0Sx0}
\end{figure}

For the second example, we choose the $\hat{L}=0$ sequence which is the first element in the set of eigenvalues which behaves asymptotically like $-c^2$.  This case presents a few challenges, and tests the $\Phase{SL-C}$ scheme in exceptional situations.  We emphasize that this sequence is not part of any QNM or TTM solution, but is an element of the set of eigensolutions associated with the TTM${}_{\rm R}$ $\{2,2,2\}$ sequence which takes the path $c(a)=a\omega^+_{222}(a)$ with $0<a\le1$.  Figure~\ref{fig:TTMR222Lh0S} presents the behavior of the eigenfunction $\swS{2}{22}{x}{c}$ at two different values of $c$ along the $\hat{L}=0$ sequence.  Because this sequence does not extend to $c=0$, but does extend to asymptotically large values of $|c|$, we use the $\hat{L}=0$ asymptotic behavior to fix $\ell=2$ everywhere on this sequence for the $\Phase{SL-C}$ and $\Phase{CZ-SL}$ phase-fixing schemes.
Notice that we have labeled the two different sequences of eigenfunctions in Figs.~\ref{fig:TTMR222Lb0S} and \ref{fig:TTMR222Lh0S} as $\swS{2}{22}{x}{c}$ using $\ell=2$ for both.  This is entirely a limitation of the conventional notation we use, and requires some extra care to avoid confusing the two solutions.  More importantly, since the solutions along $\hat{L}=0$ also use $\ell=2$, the $\Phase{SL-C}$ scheme will also fix the phase so that $\swS{2}{22}{0}{c}$ is real and positive.  Imposing this phase choice will become problematic for large values of $|c|$ as shown in Fig.~\ref{fig:TTMR222Lh0Sx0} which plots the magnitude of the eigenfunction at $x=0$.
The magnitude decays exponentially with increasing $|c|$, so it will eventually become difficult to accurately set the phase based on the value of the function at $x=0$.  As described in Sec.~\ref{sec:SL-phase}, our approach within $\Phase{SL-C}$ in this situation is to switch from fixing the phase based the value of the function via Eqs.~(\ref{eqn:SL non-zero}) to fixing the phase based on the derivative of the function via Eqs.~(\ref{eqn:SL zero}).  This transition is controlled by a threshold currently set to $10^{-10}$.  This transition is marked in Fig.~\ref{fig:TTMR222Lh0Sx0} by the left vertical dashed line.

\begin{figure}
	\centering
	\includegraphics[width=\linewidth,clip]{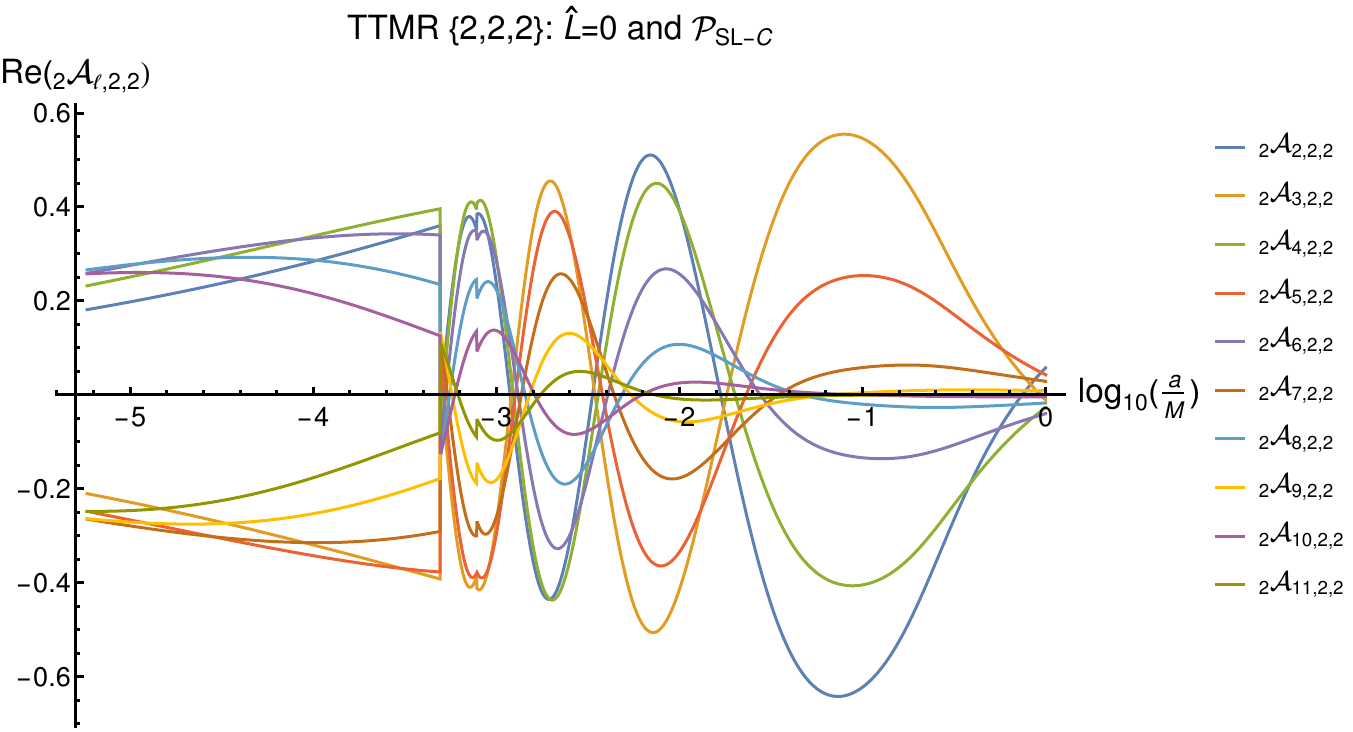}
	\caption{The real part of the first $10$ expansion coefficients $\YSH{2}{\ell22}{a\omega^+_{222}(a)}$ for $\hat{L}=0$ are plotted to demonstrate how the $\Phase{SL-C}$ scheme can lead to nonsmooth behavior along the sequence when the eigenfunction becomes very small at $x=0$.  Note the discontinuous behavior at $\log_{10}(a/M)\approx-3.1$ and $\log_{10}(a/M)\approx-3.3$ where, first the value of the function, and then its derivative become too small to accurately set the phase.}
	\label{fig:TTMR222Lh0ExCoefRe}
\end{figure}

Discontinuously changing the phase fixing scheme in this way will lead to a discontinuity in the behavior of the eigenfunction as we move along the sequence.  This can be seen in Fig.~\ref{fig:TTMR222Lh0ExCoefRe} which plots the real part of the first $10$ expansion coefficients $\YSH{2}{\ell22}{a\omega^+_{222}(a)}$ for $\hat{L}=0$.  Each coefficient varies smoothly as a function of $a$ for $\log_{10}(a/M)\gtrsim-3.1$, but changes discontinuously near $-3.1$.
Soon, the derivative also falls below a similar threshold and, as described in Sec.~\ref{sec:SL-phase} the $\Phase{SL-C}$ scheme abandons trying to use the behavior at $x=0$ to fix the phase and falls back to using the $\Phase{CZ}$ approach.  This transition is marked in Fig.~\ref{fig:TTMR222Lh0Sx0} by the right vertical dashed line, and can be seen in Fig.~\ref{fig:TTMR222Lh0ExCoefRe} near $\log_{10}(a/M)\approx-3.3$.

\begin{figure}
	\centering
	\includegraphics[width=\linewidth,clip]{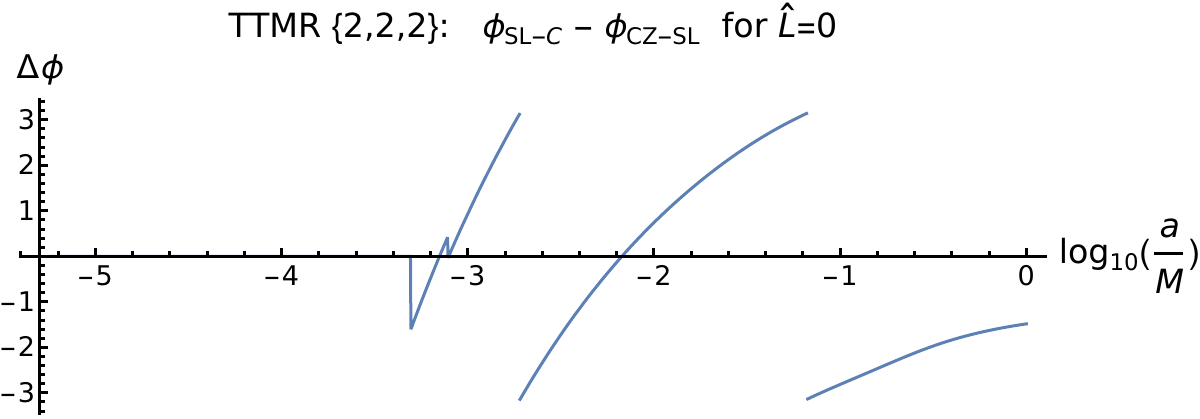}
	\includegraphics[width=\linewidth,clip]{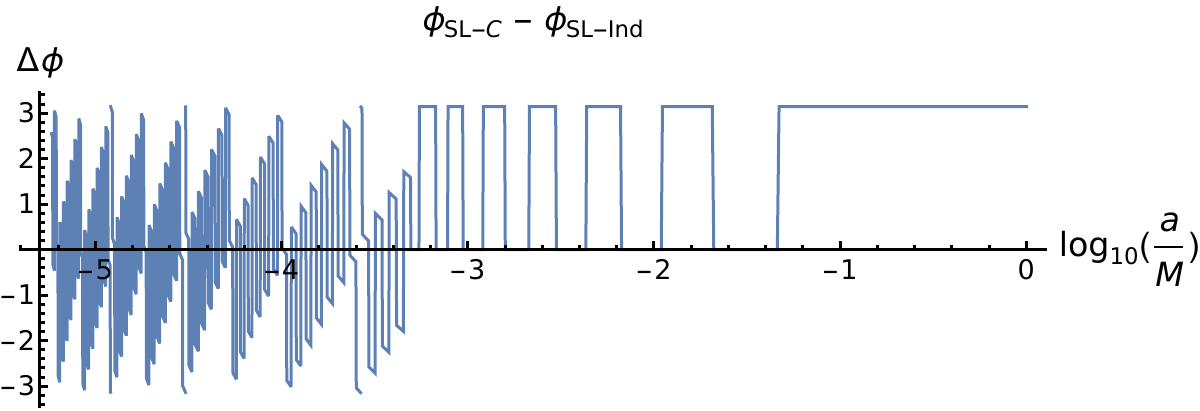}
	\caption{The phase difference between the functions $\swS{2}{22}{x}{a\omega^+_{222}(a)}$ along $\hat{L}=0$ which have been phase fixed using different schemes.  In the top plot, the functions have been phase fixed using $\Phase{SL-C}$ and $\Phase{CZ-SL}$.  In the bottom plot, the functions have been phase fixed using $\Phase{SL-C}$ and $\Phase{SL-Ind}$.}
	\label{fig:TTMR222Lh0phasediff}
\end{figure}

These discontinuities can also be seen in phase differences between the $\Phase{SL-C}$ and other phase choices.  Figure~\ref{fig:TTMR222Lh0phasediff} displays two phase differences along the $\hat{L}=0$ sequence.  In the upper plot, we display the difference between the $\Phase{SL-C}$ and $\Phase{CZ-SL}$ phase choices as a function of $a$ along the sequences.  The phase difference varies smoothly for large values of $a$ down to $\log_{10}(a/M)\approx-3.1$ where the $\Phase{SL-C}$ scheme discontinuously changes its behavior as described above.  Below $\log_{10}(a/M)\approx-3.3$, the function becomes so small at $x=0$ that the $\Phase{SL-C}$ scheme discontinuously changes again to its fallback behavior of using the $\Phase{CZ}$ scheme.
In the lower plot, we display the difference between the $\Phase{SL-C}$ and $\Phase{SL-Ind}$ phase choices.  Note the large number of discontinuous changes that occur due to the many crossings that can be seen in Fig.~\ref{fig:TTMR222AbsAlm_cAll}.  The $\hat{L}=0$ sequence is the left most sequence growing as $|c|^2$ in the figure.  At each crossing, the value of $\ell$ for the $\Phase{SL-Ind}$ scheme changes discontinuously.  While gaps in both plots are due to restricting the phase difference to $(-\pi,\pi]$, the other discontinuities are real.

Most of the sequences we have examined in this example satisfy the basic symmetries given in Eqs.~(\ref{eqn:swSF S phase}), but verifying this is challenging for sequences in the set of eigenvalues which behave asymptotically like $-c^2$ and which are also degenerate.  As before, we compared the phase fixed versions of each expansion coefficient for each solution on many sequence using Eqs.~(\ref{eqn:swSF EC phase}) to verify that each condition is satisfied.  For sequences with leading asymptotic behavior $ic(2\bar{L}+1)$, the examined sequences ($\bar{L}=0,1,2$) all satisfy the basic symmetries when phase fixed using any of the $\Phase{SL}$ or $\Phase{CZ}$ phase choices.  But, the $\Phase{Math}$ phase choice fails to satisfy the symmetries in Eqs.~(\ref{eqn:swSF sx S phase}) and (\ref{eqn:swSF mx S phase}) for the reasons described in Sec.~\ref{sec:Phase choices}.  For the sequences with leading asymptotic behavior $-c^2$, the nondegenerate sequences ($\hat{L}=0,1,2,3$) have exactly the same behavior.

However, for the degenerate sequences we examined ($\hat{L}=4,5,6,7$), we also find that all phase fixing schemes fail the symmetry in Eq.~(\ref{eqn:swSF sx S phase}) for sufficiently large values of $|c|$ along the sequences.  This failure is ultimately due to the current algorithm which combines results from eigensystem solutions at each value of $c$ into continuous sequences.  Consider 2 sequences $A$ and $B$ which become degenerate for sufficiently large values of $|c|$.  Our current approach to create a sequence of solutions looks for the first eigenvalue which matches a predicted value based on continuity and quadratic extrapolation.  Our sequences reach sufficiently large values of $|c|$ that the difference between the eigenvalues for sequences $A$ and $B$ become smaller than the precision of the calculations ($1$ part in $10^{24}$ in this example).  When this happens, the sequencing algorithm will fail to properly assign solutions to each sequence.  Two of the $3$ symmetries in Eqs.~(\ref{eqn:swSF S phase}) yield the same eigenvalue, but the symmetry manifest in Eq.~(\ref{eqn:swSF sx S phase}) causes the eigenvalues to differ by $2s$.  This small difference in sequences where the eigenvalue is behaving asymptotically as $-c^2$ eventually causes the sequencing algorithm to inconsistently choose between the $A$ and $B$ solutions for this symmetry test.  A more sophisticated approach to construct smooth sequences of eigensolutions would be necessary to deal with this situation.  It should be possible to use smoothness of the expansion coefficients $\YSH{s}{\acute\ell\ell{m}}{c}$, in addition to the eigenvalues $\scA{s}{\ell{m}}{c}$, to properly assign solutions to sequences as the eigenvalues become degenerate, but we have not attempted to implement this.

\subsubsection*{TTM${}_R$: $\ell=3$, $m=0$, $n=2$}
\label{sec:TTMR 302}

\begin{figure}
	\centering
	\includegraphics[width=\linewidth,clip]{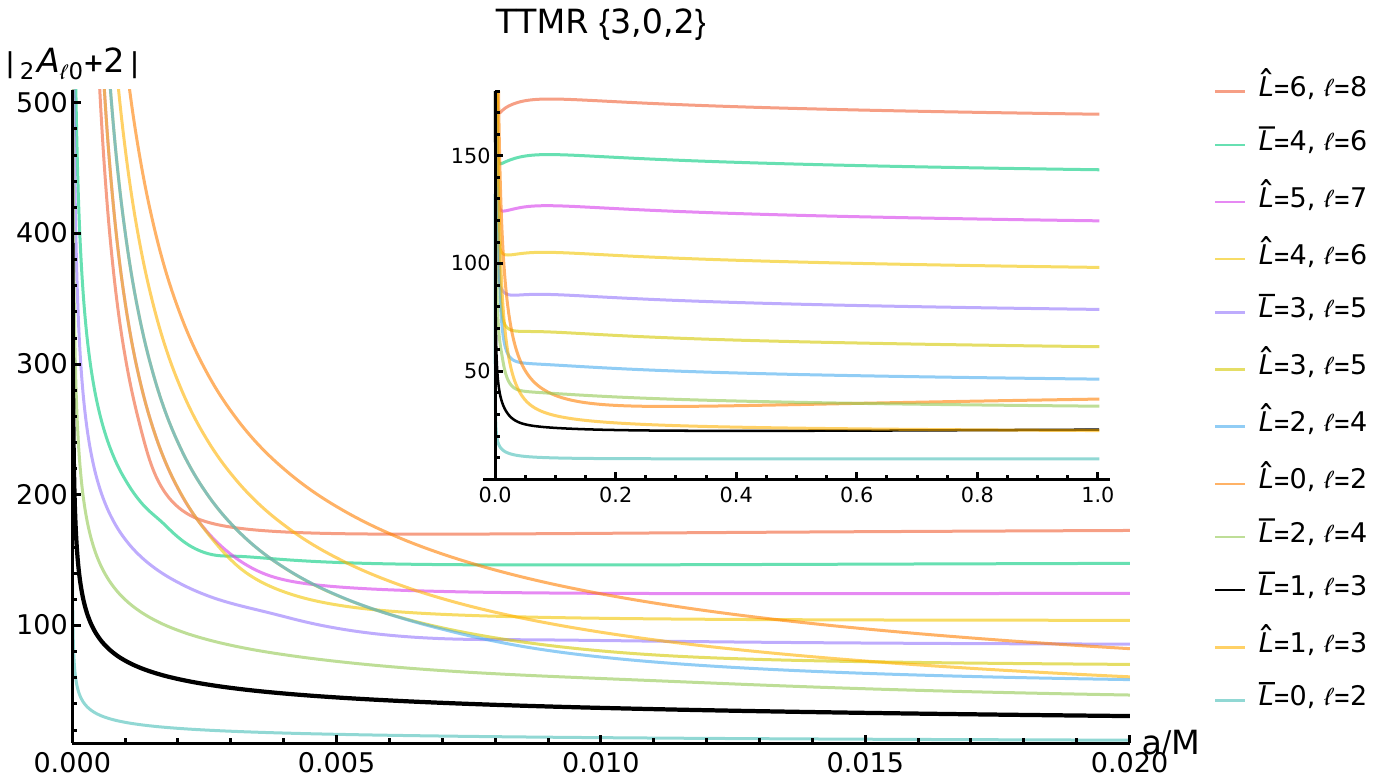}
	\caption{The first 12 eigenvalues for $s=2$ and $m=0$ along a sequence of values of $c$ obtained from the TTM${}_{\rm R}$ mode sequence $\{3,0,2\}$.  The plot displays $|\scA{2}{\ell0}{c}+2|$ [see Eq.~(\ref{eqn:swSF sx A ident})] so that the eigenvalues appear in their sorted order.  The values are display as functions of the dimensionless angular momentum $a/M$ which is related to the oblateness parameter by $c=a\omega^+_{302}$, where $\omega^+_{302}$ is the complex mode frequency along the TTM${}_{\rm R}$ sequence.  The eigenvalue sequence corresponding to the actual $\{3,0,2\}$ mode sequences is displayed as the thick black line.  The inset figure show the plot along the full range of $a$, while the full plot show the behavior for small values of $a$.}
	\label{fig:TTMR302Alm_a}
\end{figure}

The $\{3,0,2\}$ TTM${}_{\rm R}$ presents another example with $s=2$ where the sequences of $|\scA{2}{\ell0}{a\omega^+_{3,0,2}(a)}+2|$ display a complicated structure, and which further test the $\Phase{SL-C}$ phase-fixing scheme.  Figure~\ref{fig:TTMR302Alm_a} plots the magnitude $|\scA{2}{\ell0}{a\omega^+_{3,0,2}(a)}+2|$ from the first $12$ eigensolutions of the angular Teukolsky equation.  The sequence labeled $\bar{L}=1$ is the eigenvalue sequence which pairs with the $\omega^+_{3,0,2}(a)$ sequence of TTM${}_{\rm R}$ mode frequencies, and is seen as the second eigenvalue for most values of $a$.  The remaining eigenvalue sequences appear labeled by various values of $\bar{L}$ or $\hat{L}$ based on their asymptotic behaviors.

\begin{figure}
	\centering
	\includegraphics[width=\linewidth,clip]{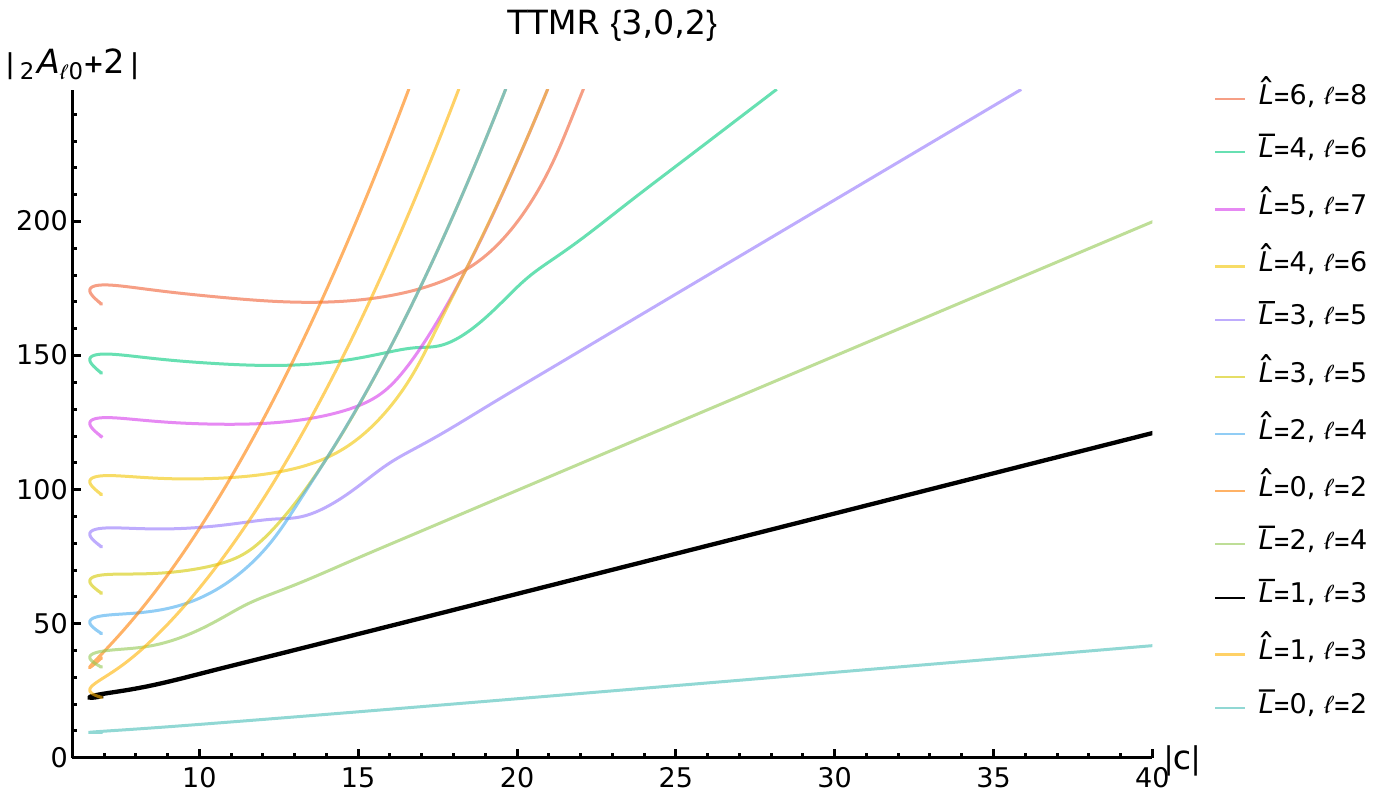}
	\caption{$12$ sequences from the large set of sequences in Fig.~\ref{fig:TTMR302AbsAlm_cAll}.  See the caption for this figure for additional details.  The sequences labeled by $\bar{L}$ have leading asymptotic behavior $ic(2\bar{L}+1)$.  Those labeled by $\hat{L}$ have leading asymptotic behavior $-c^2$.}
	\label{fig:TTMR302AbsAlm_c12}
\end{figure}

\begin{figure}
	\centering
	\includegraphics[width=\linewidth,clip]{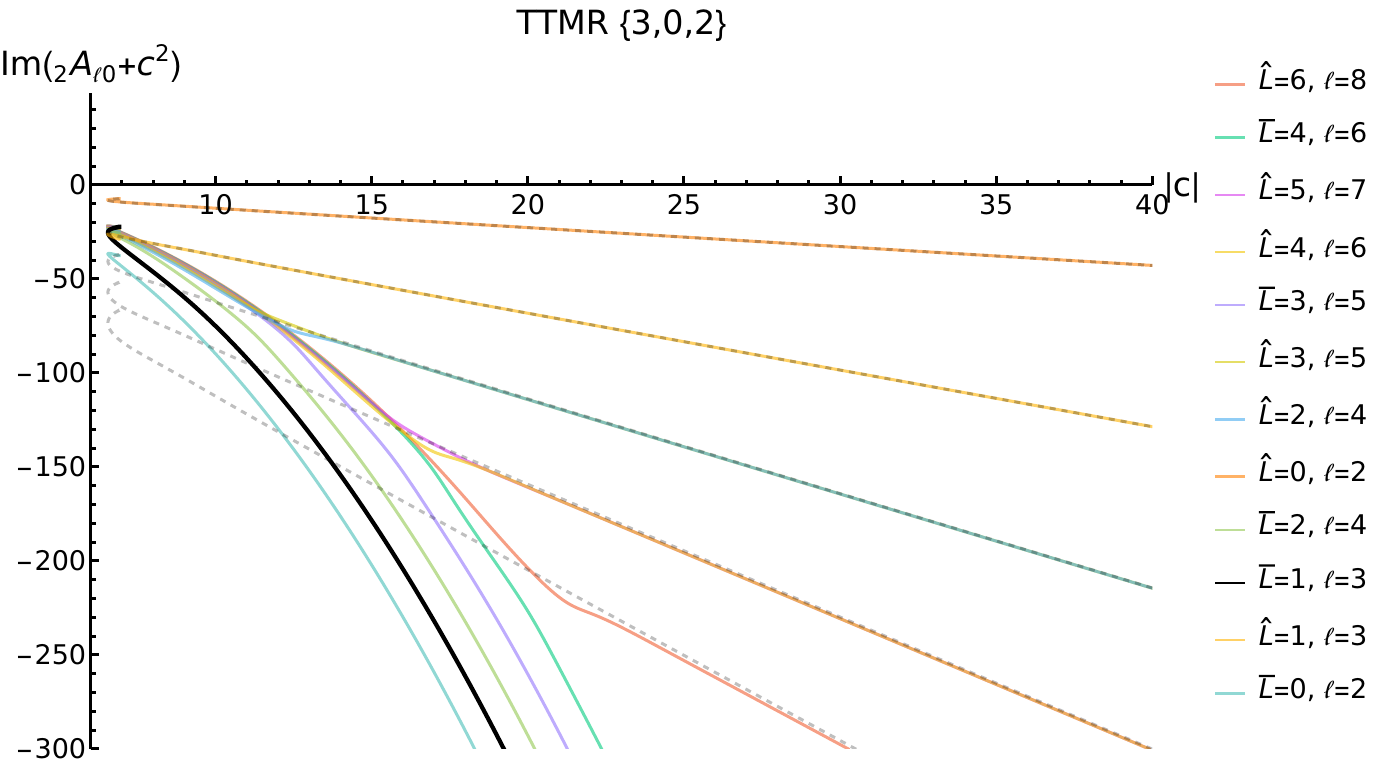}
	\caption{The same $12$ sequences displayed in Fig.~\ref{fig:TTMR302AbsAlm_c12}, but now plotting $\Im[\scA{2}{\ell0}{c}+c^2]$.  The sequences labeled by $\hat{L}$ have leading asymptotic behavior $-c^2$, so this plot displays the next-to-leading order behavior of these sequences.  The dashed lines take on values of $2(2i+1)\Im(c)$ for $0\le i\le4$, where the oblateness parameter $c=a\omega^+_{302}(a)$, and $\omega^+_{302}(a)$ is the complex mode frequency along the TTM${}_{\rm R}$ sequence.}
	\label{fig:TTMR302ImAlm_c12}
\end{figure}

The asymptotic behavior is more easily seen if Figs.~\ref{fig:TTMR302ReAlm_cAll} and \ref{fig:TTMR302AbsAlm_cAll}, found in Appendix~\ref{sec:additional_figures}, which, respectively, display the real part of $\scA{2}{\ell0}{c}$ and $|\scA{2}{\ell0}{c}+2|$ as functions of $|c|$ for a large number of eigenvalues along $c(a) = a\omega^+_{302}(a)$, with asymptotic behavior $c(a)\approx-i(-1)^{1/3}2^{2/3}3^{1/3}(M/a)^{1/3}$.  These figures are qualitatively similar to Figs.~\ref{fig:TTMR222ReAlm_cAll} and \ref{fig:TTMR222AbsAlm_cAll}, but are quantitatively different as the solutions are along a different path through the complex $c$ space.
The behavior and labeling of the various sequences is more easily understood from Figs.~\ref{fig:TTMR302AbsAlm_c12} and \ref{fig:TTMR302ImAlm_c12}
In Fig.~\ref{fig:TTMR302AbsAlm_c12}, the values of $\bar{L}$ are again obtained from the asymptotic behavior $ic(2\bar{L}+1)$.  As in the $\{2,2,2\}$ TTM${}_{\rm R}$ example, the indexing of the eigenvalues which behave asymptotically as $-c^2$ can be extracted from their next-to-leading order behavior which is not well understood for general values of $c$.  In Fig.~\ref{fig:TTMR302ImAlm_c12}, we plot the imaginary part of $\scA{2}{\ell0}{c}+c^2$ and the dashed lines take on values of $2(2i+1)\Im(c)$ for $0\le i\le4$.  The sequences labeled by $\hat{L}=0$ and $1$ correspond respectively to $i=0$ and $1$, but $i=2$ gives the behavior of the two asymptotically degenerate sequences labeled by $\hat{L}=2$ and $3$, while $i=3$ gives the behavior of the two degenerate sequences labeled by $\hat{L}=4$ and $5$.  For $i=4$, the figure only displays one of the two degenerate sequences.

\begin{figure}
	\centering
	\includegraphics[width=\linewidth,clip]{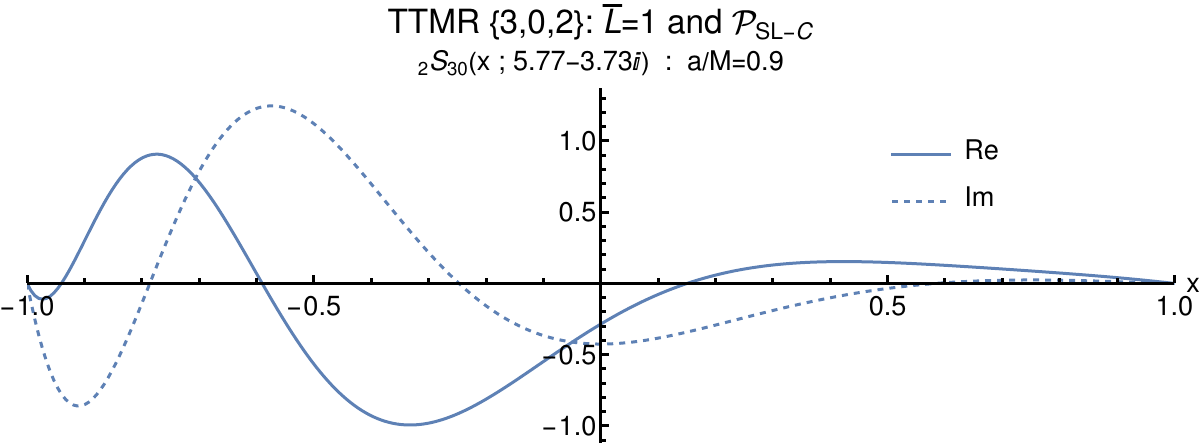}
	\includegraphics[width=\linewidth,clip]{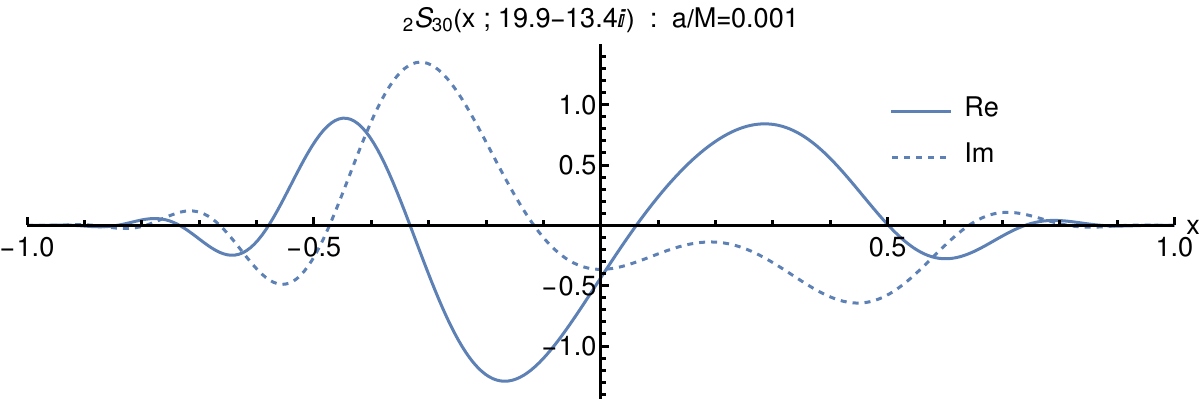}
	\includegraphics[width=\linewidth,clip]{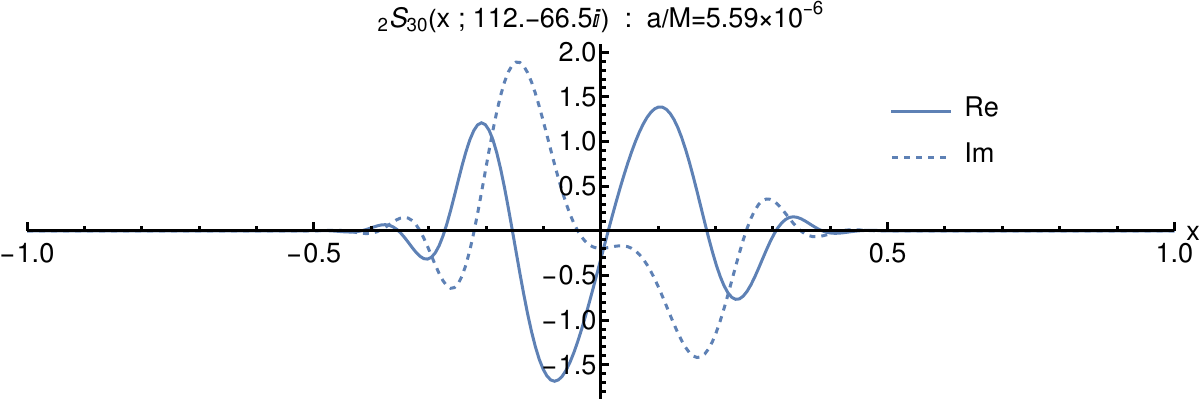}
	\caption{The functions $\swS{2}{30}{x}{c}$ for $c=5.77-3.73i$, $19.9-13.4i$, and $112.-66.5i$ are shown, respectively, in the upper, middle, and lower plots of the figure.  All satisfy $\Phase{SL-C}$.}
	\label{fig:TTMR302Lb1S}
\end{figure}

Now consider the behavior of the eigenfunctions along two of the sequences displayed in Figs.~\ref{fig:TTMR302AbsAlm_cAll} and \ref{fig:TTMR302AbsAlm_c12}.  In Fig.~\ref{fig:TTMR302Lb1S} we display the behavior of the eigenfunction $\swS{2}{30}{x}{c}$ at $3$ different values of $c$ along the $\bar{L}=1$ sequence.  Because this sequence does not extend to $c=0$, but does extend to asymptotically large values of $|c|$, we use the $\bar{L}=1$ asymptotic behavior to fix $\ell=3$ everywhere on this sequence for the $\Phase{SL-C}$ and $\Phase{CZ-SL}$ phase-fixing schemes.  The top plot in Fig.~\ref{fig:TTMR302Lb1S} shows the eigenfunction for a small value of $|c|$ along the sequence corresponding to $a=0.9M$.  The middle and lower plots show the eigenfunction for successively larger values of $|c|$ and smaller values of $a$.  All are phase fixed using $\Phase{SL-C}$ and we can clearly see that $\left.\partial_x\swS{2}{30}{x}{c}\right|_{x=0}$ is real and positive in each plot.
The smoothness of the eigenfunction as it varies along the sequence can be verified in Fig.~\ref{fig:TTMR302Lb1ExCoefRe}, found in Appendix~\ref{sec:additional_figures}, which plots the real part of the first $10$ expansion coefficients $\YSH{2}{\ell30}{a\omega^+_{302}(a)}$.  Each coefficient varies smoothly as a function of $a$, showing that the $\Phase{SL-C}$ yields a smooth sequence of eigenfunctions.
Finally, Fig.~\ref{fig:TTMR302Lb1phasediff} in Appendix~\ref{sec:additional_figures} illustrates that the phase difference between $\Phase{SL-C}$ and $\Phase{CZ-SL}$ is smooth and rather large for all values of $a$ along the TTM${}_{\rm R}$ $\{3,0,2\}$ sequences.  However, we see that the $\Phase{SL-Ind}$ phase-fixing scheme does not yield a smooth sequence, as the $\bar{L}=1$ sequences does experience a crossing near $a=M$ (see Fig.~\ref{fig:TTMR302AbsAlm_c12}).

\begin{figure}
	\centering
	\includegraphics[width=\linewidth,clip]{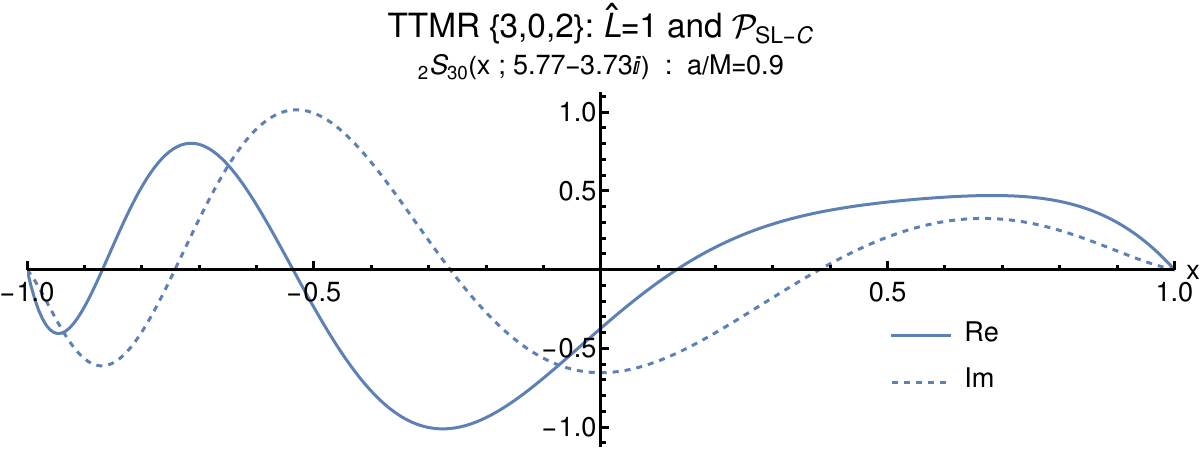}
	\includegraphics[width=\linewidth,clip]{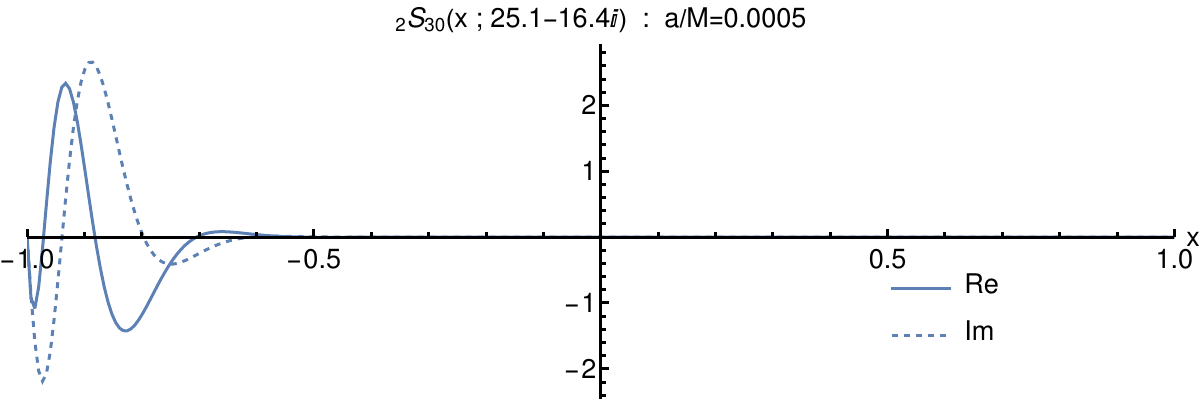}
	\caption{The functions $\swS{2}{30}{x}{c}$ for $c=5.77-3.73i$ and $25.1-16.4i$ are shown, respectively, in the upper and lower plots of the figure.  All satisfy $\Phase{SL-C}$.}
	\label{fig:TTMR302Lh1S}
\end{figure}

\begin{figure}
	\centering
	\includegraphics[width=\linewidth,clip]{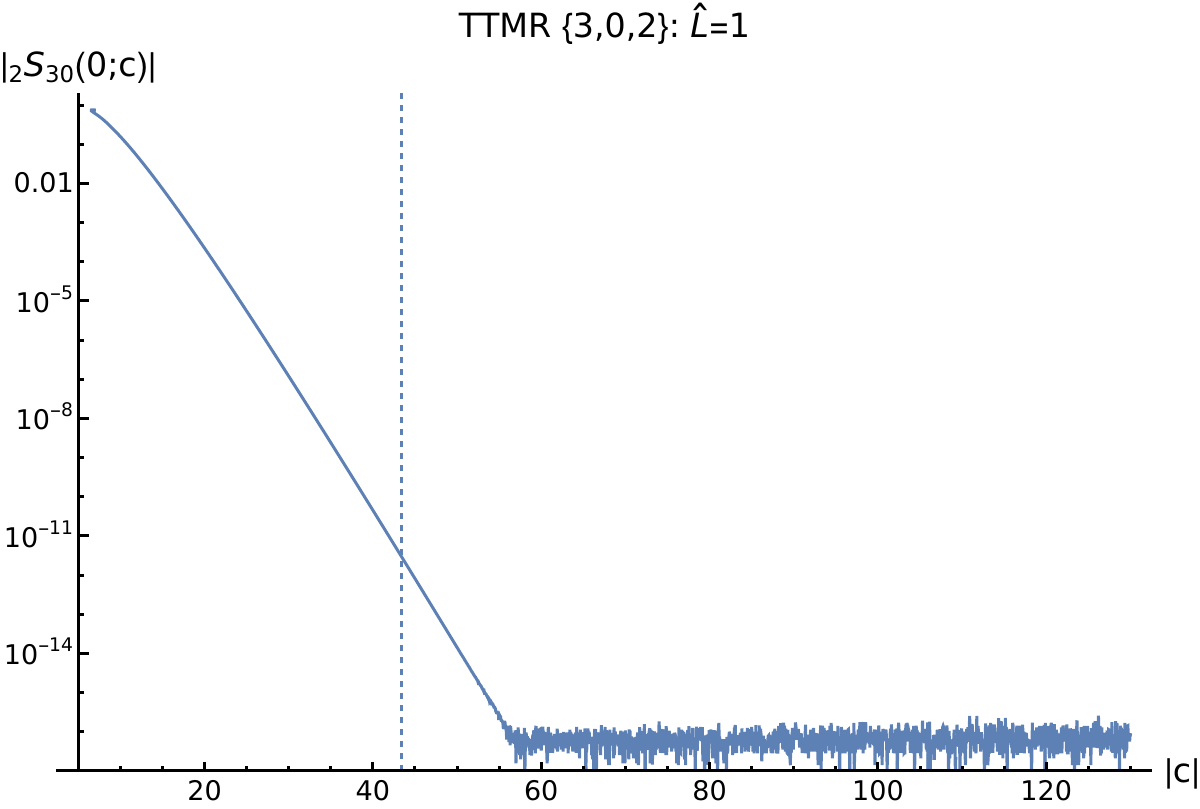}
	\caption{Log plot of the magnitude of the eigenfunction $\swS{2}{30}{x}{c}$ for $\hat{L}=1$ at $x=0$ along $c=a\omega^+_{302}(a)$, and $\omega^+_{302}(a)$ is the complex mode frequency along the TTM${}_{\rm R}$ sequence.  The vertical dashed line marks the threshold($10^{-10}$) where $\Phase{SL-C}$ determines $\left.\partial_x\swS{2}{30}{x}{c}\right|_{x=0}\approx0$, cannot fix the phase using Eqs.~(\ref{eqn:SL zero}), and begins to fix the phase using $\Phase{CZ-SL}$.}
	\label{fig:TTMR302Lh1Sx0}
\end{figure}

For the second example, we choose the $\hat{L}=1$ sequence which is the second element in the set of eigenvalues which behaves asymptotically like $-c^2$.  This case presents similar challenges to those encountered in the $\hat{L}=0$ sequence from TTM${}_{\rm R}$ $\{2,2,2\}$.  Again, this sequence is not part of any QNM or TTM solution, but is an element of the set of eigensolutions associated with the TTM${}_{\rm R}$ $\{3,0,2\}$ sequence which takes the path $c(a)=a\omega^+_{302}(a)$ with $0<a\le1$.  Figure~\ref{fig:TTMR302Lh1S} presents the behavior of the eigenfunction $\swS{2}{30}{x}{c}$ at two different values of $c$ along the $\hat{L}=1$ sequence.  Because this sequence does not extend to $c=0$, but does extend to asymptotically large values of $|c|$, we use the $\hat{L}=1$ asymptotic behavior to fix $\ell=3$ everywhere on this sequence for the $\Phase{SL-C}$ and $\Phase{CZ-SL}$ phase-fixing schemes.
We have labeled the two different sequences of eigenfunctions in Figs.~\ref{fig:TTMR302Lb1S} and \ref{fig:TTMR302Lh1S} as $\swS{2}{30}{x}{c}$ using $\ell=3$ for both, and we must be careful to avoid confusing the two solutions.  Also, since the solutions along $\hat{L}=1$ also use $\ell=3$, the $\Phase{SL-C}$ scheme will also fix the phase so that $\left.\partial_x\swS{2}{30}{x}{c}\right|_{x=0}$ is real and positive.  Imposing this phase choice will become problematic for large values of $|c|$ as shown in Fig.~\ref{fig:TTMR302Lh1Sx0} which plots the magnitude of the eigenfunction at $x=0$.
The magnitude decays exponentially with increasing $|c|$, so it will eventually become difficult to accurately set the phase based on the value of the derivative of the function at $x=0$.  As described in Sec.~\ref{sec:SL-phase}, our approach within $\Phase{SL-C}$ in this situation is to abandon trying to use the behavior at $x=0$ to fix the phase and fall back to using the $\Phase{CZ}$ approach.  This transition is controlled by a threshold currently set to $10^{-10}$, and is marked in Fig.~\ref{fig:TTMR302Lh1Sx0} by the vertical dashed line.

Discontinuously changing the phase fixing scheme in this way will lead to a discontinuity in the behavior of the eigenfunction as we move along the sequence.  This can be seen in Fig.~\ref{fig:TTMR302Lh1ExCoefRe}, found in Appendix~\ref{sec:additional_figures}, which plots the real part of the first $5$ expansion coefficients $\YSH{2}{\ell30}{a\omega^+_{302}(a)}$ for $\hat{L}=1$.  Each coefficient varies smoothly as a function of $a$ for $\log_{10}(a/M)\gtrsim-3.8$, but changes discontinuously near $-3.8$.

This discontinuity can also be seen in phase differences between the $\Phase{SL-C}$ and other phase choices.  Figure~\ref{fig:TTMR302Lh1phasediff} in Appendix~\ref{sec:additional_figures} displays two phase differences along the $\hat{L}=1$ sequence.  In the upper plot, we display the difference between the $\Phase{SL-C}$ and $\Phase{CZ-SL}$ phase choices as a function of $a$ along the sequences.  The phase difference varies smoothly for large values of $a$ down to $\log_{10}(a/M)\approx-3.8$ where the $\Phase{SL-C}$ scheme discontinuously changes its behavior as described above to its fallback behavior of using the $\Phase{CZ}$ scheme.
In the lower plot, we display the difference between the $\Phase{SL-C}$ and $\Phase{SL-Ind}$ phase choices.  Note the large number of discontinuous changes that occur due to the many crossing that can be seen in Fig.~\ref{fig:TTMR302AbsAlm_cAll}.  The $\hat{L}=1$ sequence is second from the left among those growing as $|c|^2$ in the figure.  At each crossing, the value of $\ell$ for the $\Phase{SL-Ind}$ scheme changes discontinuously.  While gaps in both plots are due to restricting the phase difference to $(-\pi,\pi]$, the other discontinuities are real.

As before, we compared the phase fixed versions of each expansion coefficient for each solution on many sequence using Eqs.~(\ref{eqn:swSF EC phase}) to verify that each condition is satisfied.  For sequences with leading asymptotic behavior $ic(2\bar{L}+1)$, the examined sequences ($\bar{L}=0,1,2,3$) all satisfy the basic symmetries when phase fixed using any of the $\Phase{SL}$ or $\Phase{CZ}$ phase choices.  But, the $\Phase{Math}$ phase choice fails to satisfy the symmetries in Eqs.~(\ref{eqn:swSF sx S phase}) and (\ref{eqn:swSF mx S phase}) for the reasons described in Sec.~\ref{sec:Phase choices} for all cases except $\bar{L}=0$.  For the sequences with leading asymptotic behavior $-c^2$, the non-degenerate sequences ($\hat{L}=0,1$) have the same behavior, except that $\Phase{Math}$ fails for both sequences.

For the degenerate sequences which we examined ($\hat{L}=2,3,4,5$), we also find that all phase fixing schemes fail the symmetry in Eq.~(\ref{eqn:swSF sx S phase}) for sufficiently large values of $|c|$ along the sequences.  This failure occurs because the current algorithm which combines results from eigensystem solutions at each value of $c$ into continuous sequences is not properly choosing from the two nearly degenerate eigenvalues, as described in more detail above at the end of the TTM${}_R$ example with $\ell=2$, $m=2$, $n=2$.

\subsection{Public data files}
\label{sec:public_data_files}
A large number of gravitational QNM and TTM sequences, including all those discussed in Secs.~\ref{sec:QNM examples} and \ref{sec:TTM examples}, are publicly available for download as HDF5\cite{HDF5} files through Zenodo\cite{KerrModeData-cook-2025}.  By default, all of the sequences have been phase fixed base on the $\Phase{SL-C}$ scheme, but each sequence includes all of the information needed to easily use any of the $\Phase{SL}$ or $\Phase{CZ}$ schemes.  Each sequence extends over all allowed values of $a$ with a maximum step size of $0.001M$, but an adaptive stepping algorithm was used to resolve arbitrarily complex paths through the complex frequency space.  Finally, none of the sequences in these HDF5 files are affected by the uncertainty associated with asymptotically degenerate eigenvalues as explored in Sec.~\ref{sec:TTM examples}.

Data files exist corresponding to each available overtone for the QNMs, TTM${}_{\rm L}$s, and TTM${}_{\rm R}$s.  The names are ${\tt KerrQNM\_NN.h5}$, ${\tt KerrTTML\_NN.h5}$, and ${\tt KerrTTMR\_NN.h5}$, where ${\tt NN}$ designates the overtone number (with a leading zero when needed).  Individual sequences are stored hierarchically using HDF5 groups which have the logical structure of folders.  In all cases, the file contains a single top level group named ${\tt nNN}$ and it contains HDF5 groups for each available azimuthal index name ${\tt m\pm{MM}}$ where ${\tt MM}$ designates the azimuthal index number (with a leading zero when needed).  Within each azimuthal subgroup are the HDF5 datasets for each available polar index $\ell$, with each named ${\tt\{\ell,m,n\}}$.  Here, $\ell$, ${\tt m}$, and ${\tt n}$ are simply index values with no leading zeros, and an explicit minus sign on ${\tt m}$ when needed.  The overtone ${\tt n}$ is usually just the index value.  However, when overtone multiplets exist, they are stored as separate HDF5 datasets, and overtone has the form ${\tt\{n,0\}}$ or ${\tt\{n,1\}}$.  For example, the $\{2,2,0\}$ QNM sequence would be accessed as ${\tt KerrQNM\_00.h5/n00/m\mbox{+}02/\{2,2,0\}}$, The $\{3,0,1_0\}$ TTM${}_{\rm R}$ multiplet would be accessed as ${\tt KerrTTMR\_01.h5/n01/m\mbox{+}00/\{3,0,\{1,0\}\}}$, and the $\{16,-10,13\}$ QNM sequence would be accessed as ${\tt KerrQNM\_13.h5/n13/m\mbox{-}10/\{16,\mbox{-}10,13\}}$.

The format of each dataset is the same for all entries.  Each row in a dataset contains all the information about a mode at a specific value of $a$ along the sequence.  The contents of each row are listed in Table~\ref{tab:hdf}.
\begin{table}[ht!]
\begin{tabular}{c|c}
\hline
Column(s) & Value(s) \\
\hline
1 & $a/M$ \\ 
2,3 & $\Re,\Im[M\omega^+_{\ell{m}n}]$ \\ 
4,5 & $\Re,\Im[\scA{s}{\ell{m}}{a\omega^+_{\ell{m}n}}]$ \\ 
6 & $L$ \\ 
7,8 & $\Re,\Im[\Pfactor{SL-Ind}]$ \\ 
9,10 & $\Re,\Im[\YSH{s}{\ell_{\rm min}\ell{m}}{a\omega^+_{\ell{m}n}}]$ \\ 
11,12 & $\Re,\Im[\YSH{s}{(\ell_{\rm min}+1)\ell{m}}{a\omega^+_{\ell{m}n}}]$ \\
\vdots & \vdots \\ 
\hline 
\end{tabular}
\caption{The data stored in each row of an HDF5 dataset for a Kerr mode sequence.  Complex values are stored as two consecutive real values.  $L$ is the zero-based index of the given $\scA{s}{\ell{m}}{a\omega^+_{\ell{m}n}}$ in the sorted set of all eigenvalues, where the values are sorted based on $|\scA{s}{\ell{m}}{a\omega^+_{\ell{m}n}}+s|$.  $\Pfactor{SL-Ind}$ is the complex phase factor required to convert data to $\Phase{SL-Ind}$.  The remaining columns contain the complex expansion coefficients $\YSH{s}{\hat\ell\ell{m}}{a\omega^+_{\ell,m,n}}$.}
\label{tab:hdf}
\end{table}
The expansion coefficients $\YSH{s}{\hat\ell\ell{m}}{a\omega^+_{\ell,m,n}}$ require the majority of the storage space for each solution.  The stored values are phase fixed to $\Phase{SL-C}$, and the value of $\hat\ell$ for the first coefficient is $\ell_{\rm min}=\max(|s|,|m|)$.  Each row contains the same range of values for $\hat\ell$ based on the maximum number needed to represent any solution in the sequence, and values are padded with zeros as necessary.  These expansion coefficients can be easily modified for any other desired phase-fixing scheme.  The value of $\ell$ can be used to convert to $\Phase{CZ-SL}$ by ensuring that $\YSH{s}{\ell\ell{m}}{a\omega^+_{\ell,m,n}}$ is real and positive. The index $L$ is the zero based index for $\scA{s}{\ell{m}}{a\omega^+_{\ell{m}n}}$ within the set of sorted eigenvalues, where the values are sorted based on $|\scA{s}{\ell{m}}{a\omega^+_{\ell{m}n}}+s|$ as described in Sec.~\ref{sec:general sw spheroidal}.  This value of $L$ can be used to determine the correct value of $\hat\ell$ for converting to $\Phase{CZ-Ind}$.  Since the conversion to $\Phase{SL-Ind}$ is complicated, we provide this required conversion factor $\Pfactor{SL-Ind}$.

Below, we give details for the available sequences.

\subsubsection{QNM sequences}
\label{seq:qnm_sequences}
A total of 4917 phase-fixed QNM sequences are available through Zenodo\cite{KerrModeData-cook-2025} and cover overtones from $n=0$ through $n=32$.  The first 16 overtones, $n\in\{0,1,\cdots,15\}$, include all $2\ell+1$ values of the azimuthal index $m$ for polar indices $\ell\in\{2,3,\cdots,16\}$.  Each solution has been computed with a requested absolute error of $10^{-12}$, and the values for $\omega_{\ell{m}n}$ and $\scA{s}{\ell{m}}{a\omega^+_{\ell{m}n}}$ should be accurate to that level.  The next 17 overtones, $n\in\{16,17,\cdots,32\}$, include all $2\ell+1$ values of the azimuthal index $m$ for polar indices $\ell\in\{2,3,4\}$.  Each solution has been computed with a requested absolute error of $10^{-8}$, and the values for $\omega_{\ell{m}n}$ and $\scA{s}{\ell{m}}{a\omega^+_{\ell{m}n}}$ should be accurate to that level.  Details of these solutions\footnote{Except for the case $n=32$.} can be found in Refs.~\cite{cook-zalutskiy-2014,cook-zalutskiy-2016b}

Most sequences cover the full range of possible values of the angular momentum per unit mass $0\le a<M$.  Each sequence is discretized with a step size of $\Delta{a}\le0.001M$, and adaptive stepping is used to ensure that complicated trajectories through the complex $\omega$ plane are well resolved.  A small number of sequences exist which do not cover the full range of $a$.  The existence of these unusual sequences led to the introduction of overtone-multiplet notation in Ref.~\cite{cook-zalutskiy-2014} in order to group together sequences which seemed to share certain properties.  All of these sequences contain mode frequencies which are either on, or approach, the negative imaginary axis(NIA) of the complex frequency plane.  Cook and Zalutskiy\cite{cook-zalutskiy-2016b}used Pincherle's theorem to conclude that QNM solutions can only exist on the NIA in the form of polynomial solutions, and that this accounts for the incompleteness of some sequences.  Other research has explored the existence of unconventional QNMs\cite{van_den_brink-2003,Fiziev-2011}, and recently Chen, {\em et al}.\cite{Chen-etal-2025} have developed an approach for computing unconventional QNMs which seem to complete the QNM sequences that, to date, had not covered the full range of $a$.  To our knowledge, these results have not been independently verified, and doing so is beyond the scope of this work.  We, therefore, simply acknowledge this work and inform anyone using these datasets that they may be incomplete.  All of the sequences which may be incomplete can be found in Ref.~\cite{cook-zalutskiy-2016b} in the notation used in the publicly available data files.

\subsubsection{TTM sequences}
\label{seq:ttm_sequences}
A total of 161 phase-fixed TTM${}_{\rm L}$, and 161 phase fixed TTM${}_{\rm R}$ sequences are available through Zenodo\cite{KerrModeData-cook-2025}.  TTMs do not have conventional overtones in the same sense that QNMs do.  However, there are 3 known distinct families of TTMs and Cook and Lu\cite{CookLu2023} adopted the notation of using the overtone index to denote TTM families.  All overtone families of the TTMs have been computed for polar indices $\ell\in\{2,3,\cdots,8\}$.  The first 2 families of overtones, denoted by $n=0$ and $n=1$, exhibit a symmetry in which the two families of solutions $\omega^\pm$ are degenerate.  That is, $\omega^-_{\ell{m}n}=\omega^+_{\ell{m}n}$ for the $n=0$ and $n=1$ families of TTMs.  Since all modes obey the symmetry that $\omega^-_{\ell{m}n}=-(\omega^+_{\ell(-m)n})^*$, this means that the full set of modes can be fully represented by the $\ell+1$ values of $m\ge0$. The $n=2$ family does not exhibit the same degeneracy and includes all $2\ell+1$ values of the azimuthal index $m$ for each $\ell$.  Each solution has been computed with a requested absolute error of $10^{-8}$, and the values for $\omega_{\ell{m}n}$ and $\scA{s}{\ell{m}}{a\omega^+_{\ell{m}n}}$ should be accurate to that level.  Details of these solutions can be found in Refs.~\cite{CookLu2023}.

The $n=0$ family of sequences cover the full range of possible values of the angular momentum per unit mass $0\le a\le M$.  The $n=1$ and $n=2$ families of sequences cover the range of possible values of the angular momentum per unit mass $0<a\le M$.  For these 2 families, the mode frequency of the Schwarzschild limit is at complex infinity, so these sequences only approach $a=0$ asymptotically.  Each sequence is discretized with a step size of $\Delta{a}\le0.001M$, and adaptive stepping is used to ensure that complicated trajectories through the complex $\omega$ plane are well resolved.  The $m=0$ sequences of both $n=0$ and $n=1$ families are stored as overtone multiplets, however this is not because of any difficulty approaching, or crossing, the NIA.  All of these sequences contain segments in which the mode frequency is purely imaginary from $a=0$ to some critical value of $a$ where the sequence of mode frequencies discontinuously turns to take on fully complex values.  Overtone multiplets have been used in this instance to simply separate these two different portions of the full sequence.

\section{Discussion}
\label{sec:discussion}
The SWSFs and SWSHs are an important class of special functions for black-hole perturbation theory, and in other areas, but they are challenging to work with.  In general, the separation constant $\scA{s}{\ell{m}}{c}$ and its associated SWSF $\swS{s}{\ell{m}}{x}{c}$ must be computed numerically.  Most of the attention in the literature has focused on computation of the separation constant $\scA{s}{\ell{m}}{c}$, or analytic approximations of it in the large- and small-$|c|$ limits.  When working with the SWSFs, it has been generally assumed that the phase choice for each function is not important.  This assumption is true in many situations, but if phase differences are considered between quantities that depend of different SWSFs, then the choice of phase can make a difference.

In this paper, we have considered the implicit phase choice imposed by a general eigenvalue solver ($\Phase{Math}$ when {\em Mathematica}'s {\tt Eigensystem} is used) when a spectral method is employed to solve the eigensystem.  We have also looked at the Cook-Zalutskiy phase choice $\Phase{CZ}$.  Both choices will yield reasonable behavior so long as $|c|$ is not too large, but both approaches can result in discontinuous behavior for the phase along sequences of solutions.  The discontinuous behavior of $\Phase{Math}$ was explicitly illustrated in Fig.~\ref{fig:s0m1L1CZPhase}.  The behavior of the phase along sequences can be important when using discrete(tabulated) data along a sequence if interpolation to intermediate values is necessary.  It is, therefore, not advisable to rely on any implicit phase choice such as $\Phase{Math}$.  

The $\Phase{CZ-SL}$ phase choice will, in almost all cases yield a continuous phase along sequences of solutions, but there is a failure mode where either of the $\Phase{CZ}$ phase choices will simply fail. This failure mode occurs anywhere $\YSH{s}{\ell\ell{m}}{c}=0$ along a sequence.  While possible, the failure mode in $\Phase{CZ-SL}$ is highly unlikely to happen and we have not seen an explicit example of it anywhere among the sequences we have considered.  If it did occur somewhere along a sequence, then the SWSF would undergo a $180^\circ$ phase change at this point.  Such a change is relatively easy to correct, and doing so would ensure that the phase of a sequence of SWSFs remains continuous.  Thus, the $\Phase{CZ-SL}$ is a reasonable choice when phase-continuous sequences of SWSFs are required.

While the $\Phase{CZ}$ scheme is reasonable, it is largely motivated by a specific numerical solution scheme.  If the spectral approach for solving the angular Teukolsky equation is not used, then the $\Phase{CZ}$ scheme requires the accurate computation of at least one integral in order to determine $\YSH{s}{\ell\ell{m}}{c}$.  As an alternative, we have proposed the spherical limit phase choice $\Phase{SL}$, and in particular its continuous variant $\Phase{SL-C}$, as a more generally motivated phase choice.  We have demonstrated that $\Phase{SL-C}$ ensures that the phase of each SWSF $\swS{s}{\ell{m}}{x}{c}$ changes smoothly along a sequence of solutions in all but the most extreme of circumstances.  Like the $\Phase{CZ}$ scheme, there is an anomalous situation which can lead to a discontinuous phase choice when both $\swsS{s}{\ell{m}}{0}=0$ and $\left.\partial_x\swS{s}{\ell{m}}{x}{c}\right|_{x=0}=0$ for some specific value of $c$.  This failure mode is also highly unlikely to happen and we have not seen an explicit example of it anywhere among the sequences we have considered.  Just as with $\Phase{CZ-SL}$, if it did occur somewhere along a sequence, then the SWSF would again undergo a $180^\circ$ phase change at this point.  Such a change is, again, relatively easy to correct.

Because the $\Phase{SL-C}$ scheme is designed to extend to finite values of $|c|$ the same behavior at the equator ($x=0$) found in the spherical limit, we believe that the $\Phase{SL-C}$ scheme is better, and more generally well motivated, than the $\Phase{CZ}$ scheme.  Compared to the cost of solving the eigensystem, it is not computationally expensive to implement the scheme.  We, thus, propose that the $\Phase{SL-C}$ scheme should be adopted as the preferred phase-fixing scheme for SWSFs and SWSHs.  We conjecture that this new phase choice may provide an advantage when mapping ringdown fitting results to the space of progenitor systems.  Verifying this conjecture will likely require highly accurate numerical simulation waveforms, and is beyond the scope of this paper.

We have made publicly available a large number of tabulated QNM and TTM mode sequences in which the SWSFs have been phase-fixed using $\Phase{SL-C}$ and tested to ensure continuity of the phase choice\cite{KerrModeData-cook-2025}.  We have also made publicly available the {\tt SWSpheroidal} {\em Mathematica} Paclet\cite{KerrModes-1.0.6} which includes two independent methods for numerically solving the angular Teukolsky eigensystem.  These two independent methods have been tested against each other and yield results that are identical, limited only by the requested precision and accuracy of the solutions.

\acknowledgments
Some computations were performed using the Wake Forest University (WFU) High Performance Computing Facility, a centrally managed computational resource available to WFU researchers including faculty, staff, students, and collaborators\cite{DEAC-Cluster}.

\section*{Data Availability Statement}
Some data and software that support the findings of this article are openly available. The phase-fixed datasets for the QNM and TTM sequences are available through Ref.~\cite{KerrModeData-cook-2025}. The software used to generate this data is available through Ref.~\cite{KerrModes-1.0.6}. The figures in this article were generated from this data and using this software, but the notebooks which generate each figure are not publicly available.

\appendix
\section{Additional Figures Relevant to Sec.~\ref{sec:examples}}
\label{sec:additional_figures}
$13$ figures associated with several examples discussed in Secs.~\ref{sec:QNM examples} and \ref{sec:TTM examples} are included in this Appendix for completeness.  They provide additional details for some of these examples, but are qualitatively similar to certain plots in the main text.

\begin{figure}[h!]
	\centering
	\includegraphics[width=\linewidth,clip]{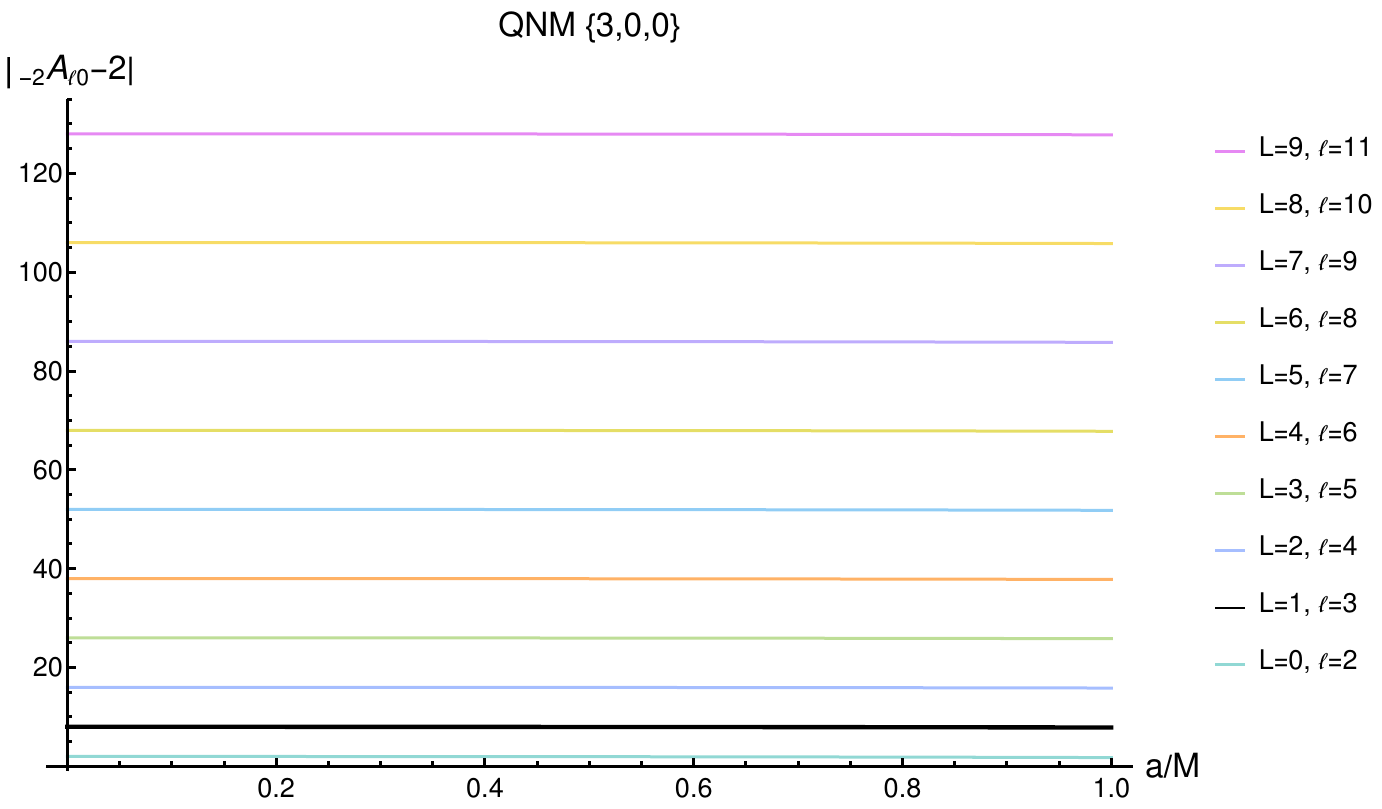}
	\caption{The first 10 eigenvalues for $s=-2$ and $m=0$ along a sequence of values of $c$ obtained from the QNM mode sequence $\{3,0,0\}$.  The plot displays $|\scA{-2}{\ell0}{c(a)}-2|$ so that the eigenvalues appear in their sorted order.  The values are display as functions of the dimensionless angular momentum $a/M$ which is related to the oblateness parameter by $c=a\omega^+_{300}$, where $\omega^+_{300}$ is the complex mode frequency along the QNM sequence.  The eigenvalue sequence corresponding to the actual $\{3,0,0\}$ mode sequences is displayed as the thick black line.}
	\label{fig:QNM300Alm_a}
\end{figure}

\begin{figure}[h!]
	\centering
	\includegraphics[width=\linewidth,clip]{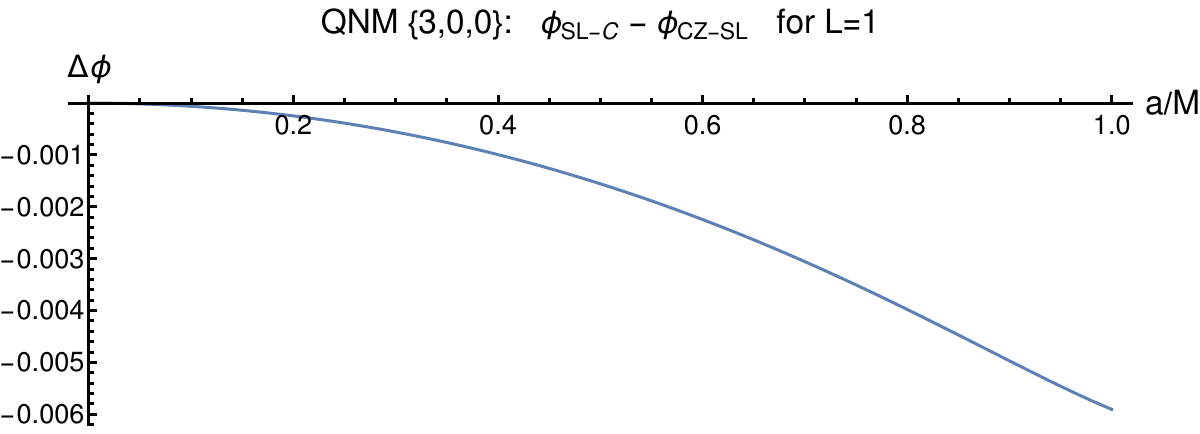}
	\includegraphics[width=\linewidth,clip]{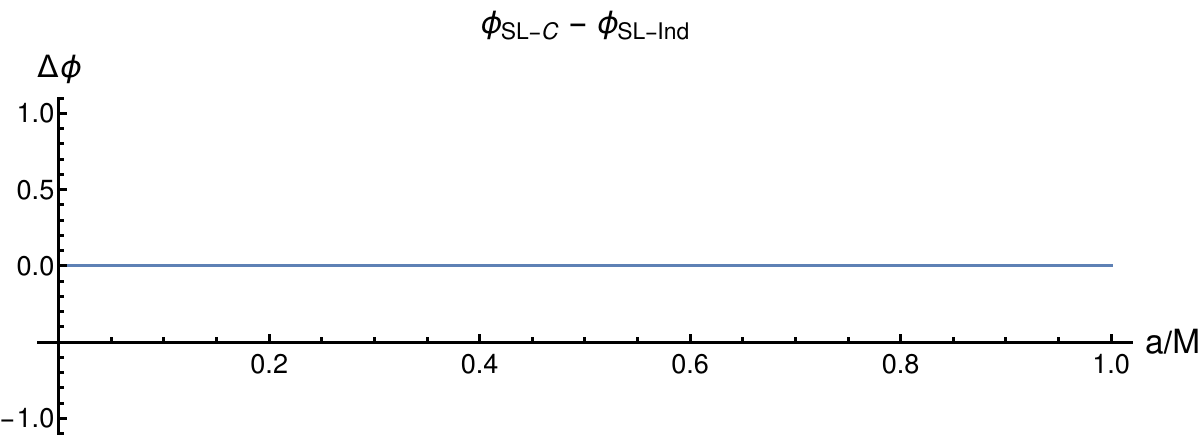}
	\caption{The phase difference between the functions $\swS{-2}{30}{x}{a\omega^+_{300}(a)}$ which have been phase fixed using different schemes.  In the top plot, the functions have been phase fixed using $\Phase{SL-C}$ and $\Phase{CZ-SL}$.  In the bottom plot, the functions have been phase fixed using $\Phase{SL-C}$ and $\Phase{SL-Ind}$.}
	\label{fig:QNM300L1phasediff}
\end{figure}

\begin{figure}[h!]
	\centering
	\includegraphics[width=\linewidth,clip]{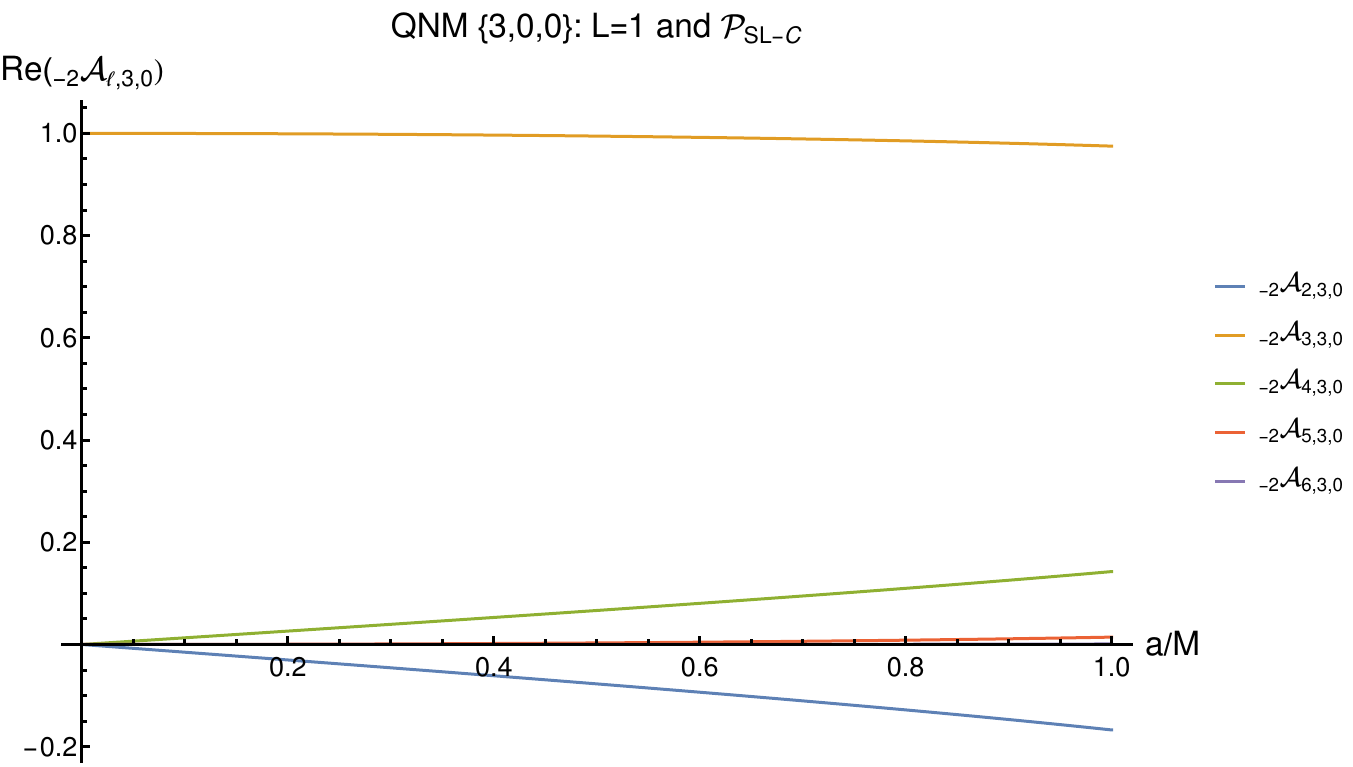}
	\caption{The real part of the first $5$ expansion coefficients $\YSH{-2}{\ell30}{a\omega^+_{300}(a)}$ are plotted to demonstrate that the expansion coefficients are smooth functions of $a$ for the $\Phase{SL-C}$ phase choice.}
	\label{fig:QNM300L1ExCoefRe}
\end{figure}

\begin{figure}[h!]
	\centering
	\includegraphics[width=\linewidth,clip]{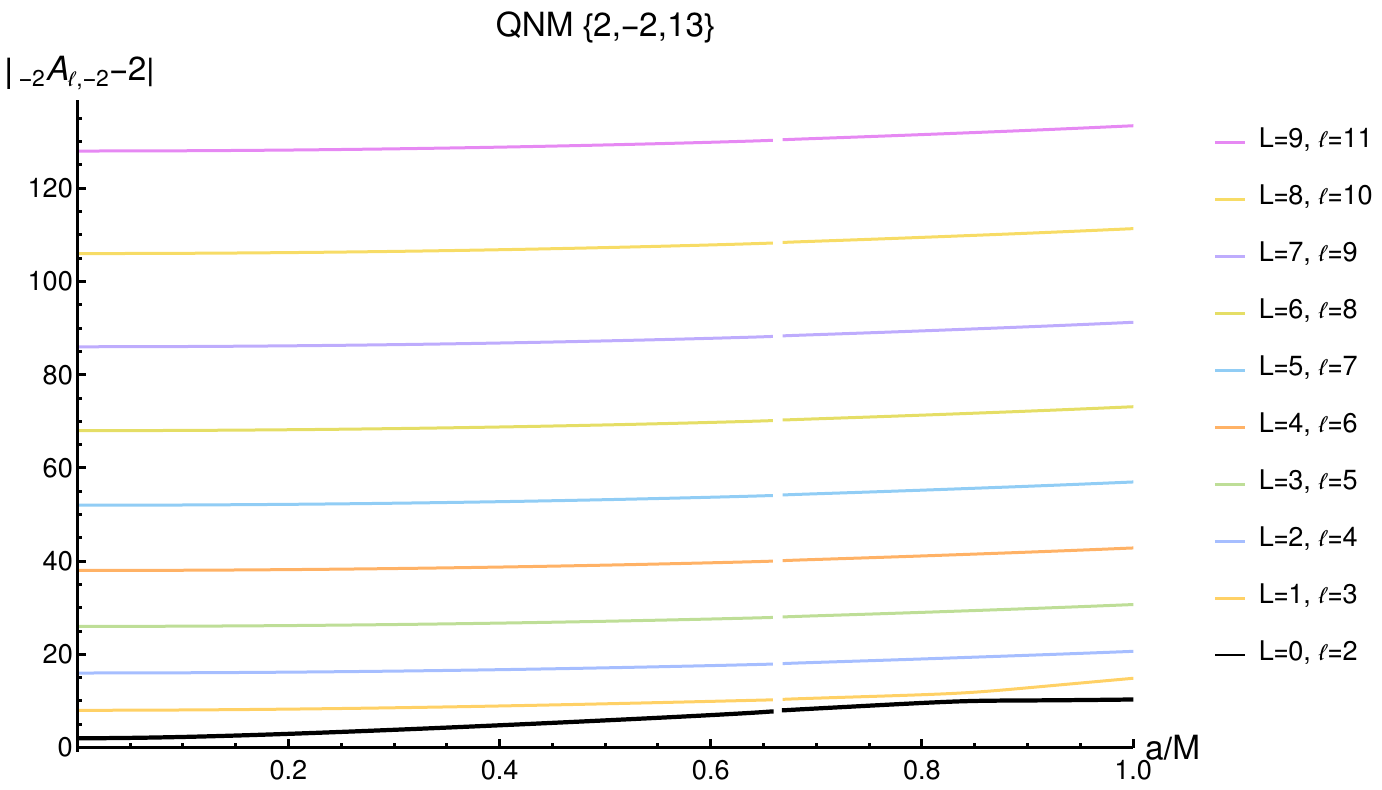}
	\caption{The first 10 eigenvalues for $s=-2$ and $m=-2$ along a sequence of values of $c$ obtained from the QNM mode sequence $\{2,-2,13\}$.  The plot displays $|\scA{-2}{\ell(-2)}{c(a)}-2|$ so that the eigenvalues appear in their sorted order.  The values are display as functions of the dimensionless angular momentum $a/M$ which is related to the oblateness parameter by $c=a\omega^+_{2(-2)13}$, where $\omega^+_{2(-2)13}$ is the complex mode frequency along the QNM sequence.  The eigenvalue sequence corresponding to the actual $\{2,-2,13\}$ mode sequences is displayed as the two thick black lines.  The left portion corresponds to the $\{2,-2,13_0\}$ multiplet, and the right portion to the $\{2,-2,13_1\}$ multiplet.  See Fig.~\ref{fig:QNM2m213omegac} for additional context.}
	\label{fig:QNM2m213Alm_a}
\end{figure}

\begin{figure}[h!]
	\centering
	\includegraphics[width=\linewidth,clip]{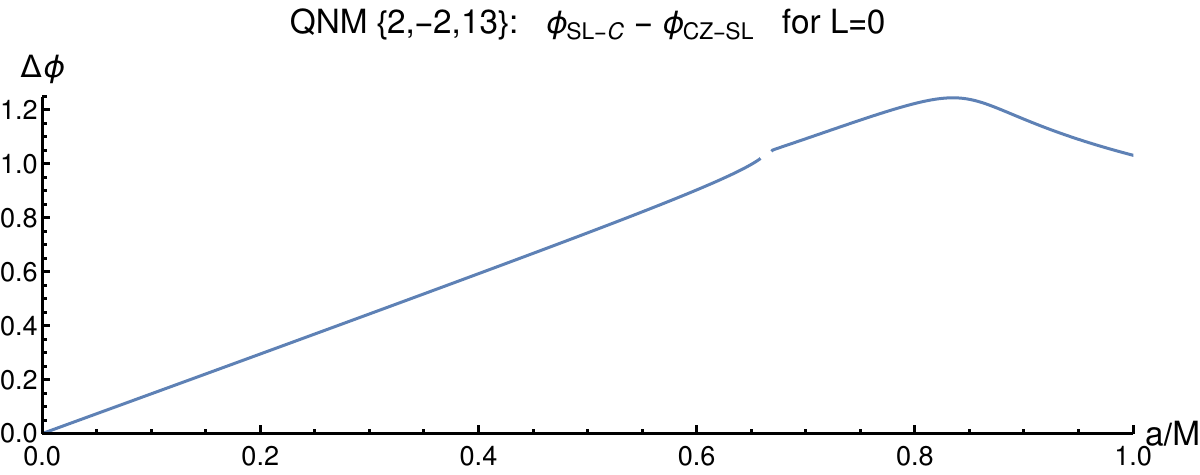}
	\includegraphics[width=\linewidth,clip]{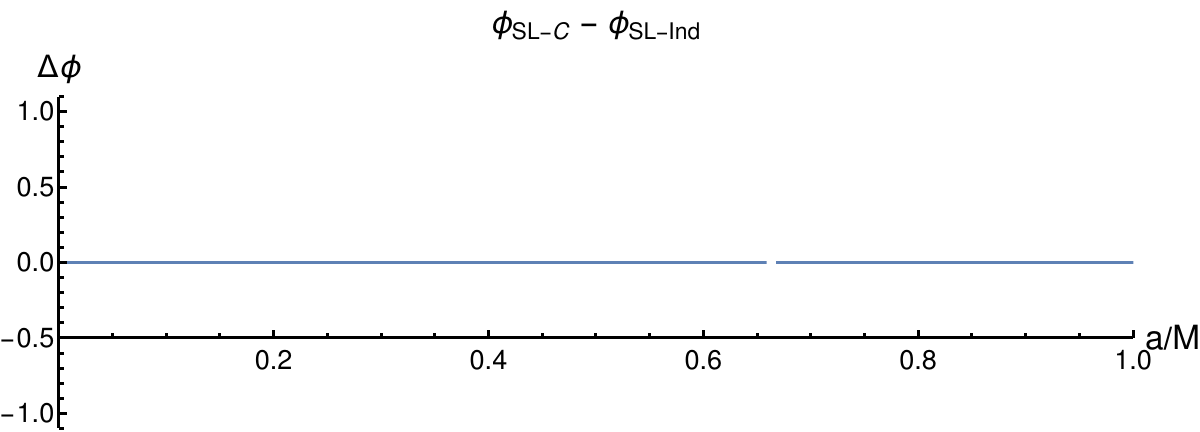}
	\caption{The phase difference between the functions $\swS{-2}{2(-2)}{x}{a\omega^+_{2(-2)13}(a)}$ which have been phase fixed using different schemes.  In the top plot, the functions have been phase fixed using $\Phase{SL-C}$ and $\Phase{CZ-SL}$.  In the bottom plot, the functions have been phase fixed using $\Phase{SL-C}$ and $\Phase{SL-Ind}$.}
	\label{fig:QNM2m213L0phasediff}
\end{figure}

\begin{figure}[h!]
	\centering
	\includegraphics[width=\linewidth,clip]{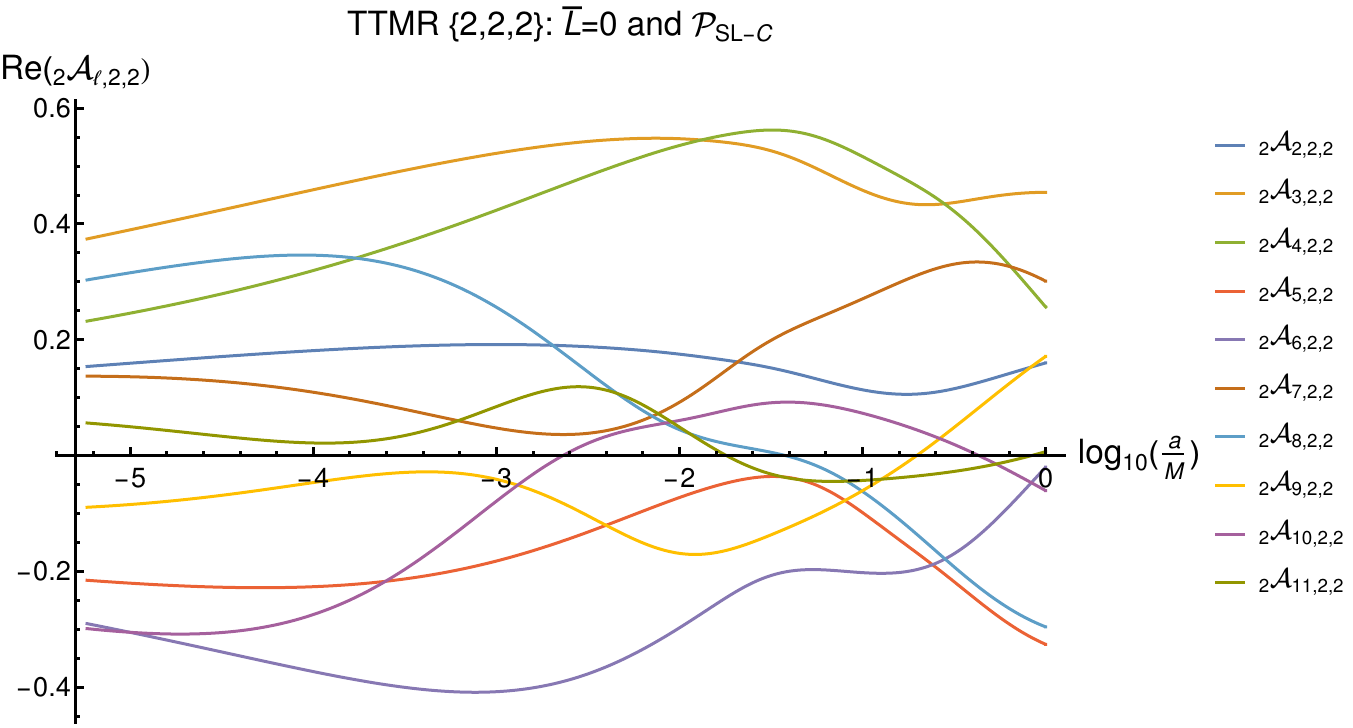}
	\caption{The real part of the first $10$ expansion coefficients $\YSH{2}{\ell22}{a\omega^+_{222}(a)}$ for $\bar{L}=0$ are plotted to demonstrate that the expansion coefficients are smooth functions of $a$ for the $\Phase{SL-C}$ phase choice.}
	\label{fig:TTMR222Lb0ExCoefRe}
\end{figure}

\begin{figure}[h!]
	\centering
	\includegraphics[width=\linewidth,clip]{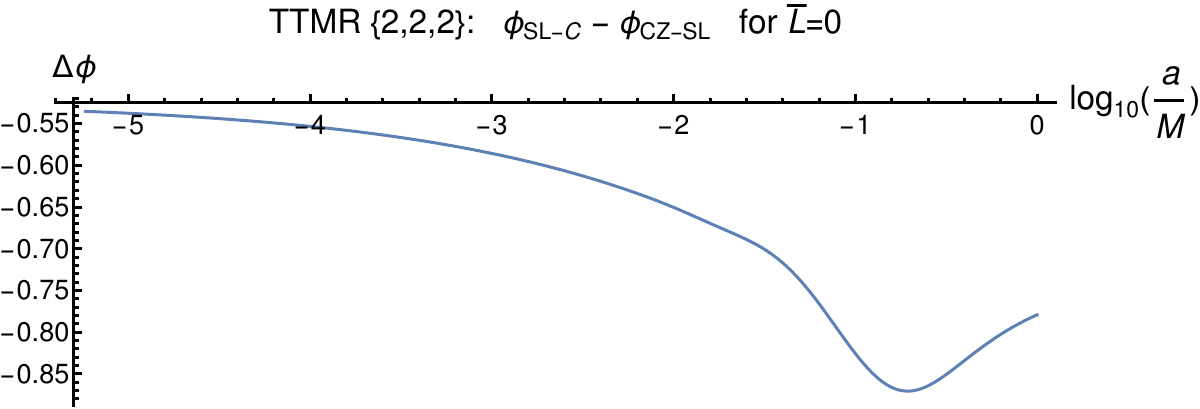}
	\includegraphics[width=\linewidth,clip]{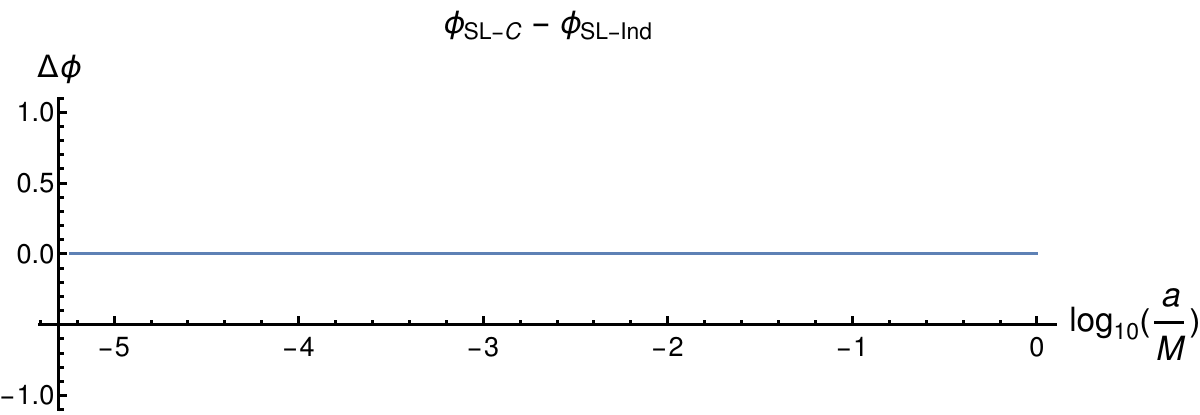}
	\caption{The phase difference between the functions $\swS{2}{22}{x}{a\omega^+_{222}(a)}$ along $\bar{L}=0$ which have been phase fixed using different schemes.  In the top plot, the functions have been phase fixed using $\Phase{SL-C}$ and $\Phase{CZ-SL}$.  In the bottom plot, the functions have been phase fixed using $\Phase{SL-C}$ and $\Phase{SL-Ind}$.}
	\label{fig:TTMR222Lb0phasediff}
\end{figure}

\begin{figure}[h!]
	\centering
	\includegraphics[width=\linewidth,clip]{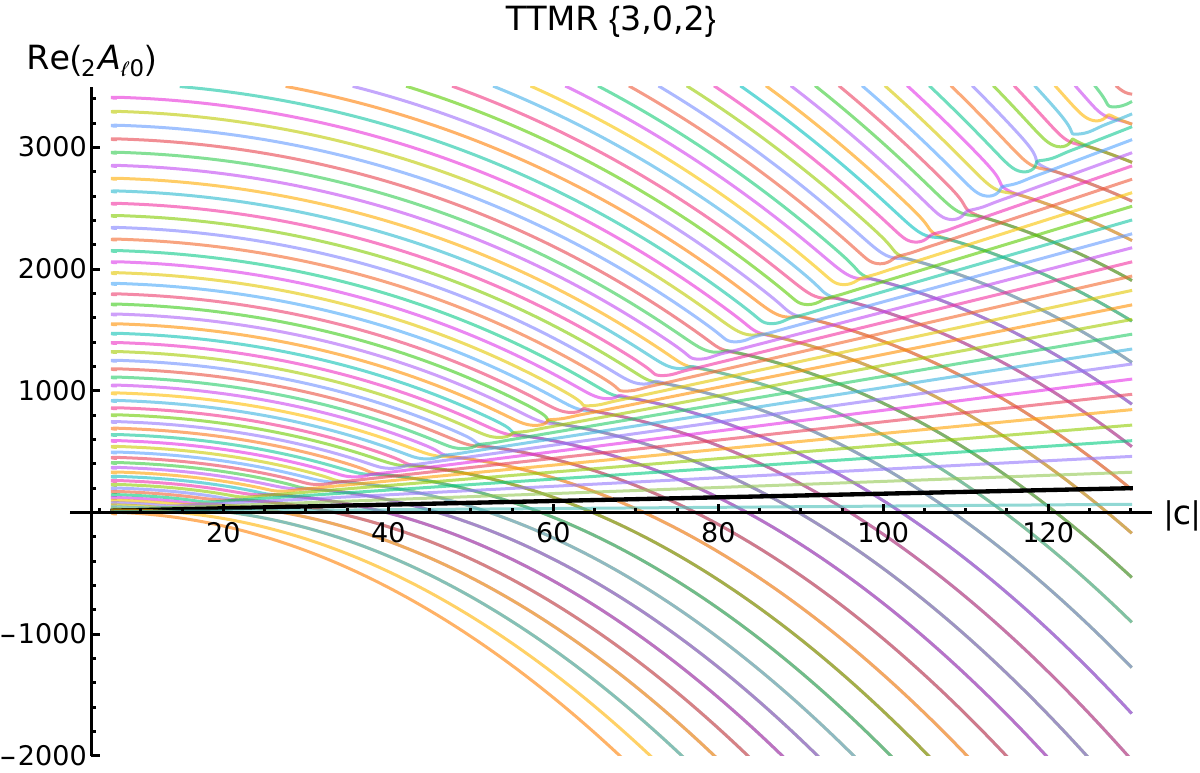}
	\caption{A large number of eigenvalues for $s=2$ and $m=0$ along a sequence of values of $c$ obtained from the TTM${}_{\rm R}$ mode sequence $\{3,0,2\}$.  The plot displays sequences of $\Re(\scA{2}{\ell0}{c})$ as functions of $|c|$ where the oblateness parameter $c=a\omega^+_{302}(a)$, and $\omega^+_{302}(a)$ is the complex mode frequency along the TTM${}_{\rm R}$ sequence.  The eigenvalue sequence corresponding to the actual $\{3,0,2\}$ mode sequences is displayed as the thick black line.}
	\label{fig:TTMR302ReAlm_cAll}
\end{figure}

\begin{figure}[h!]
	\centering
	\includegraphics[width=\linewidth,clip]{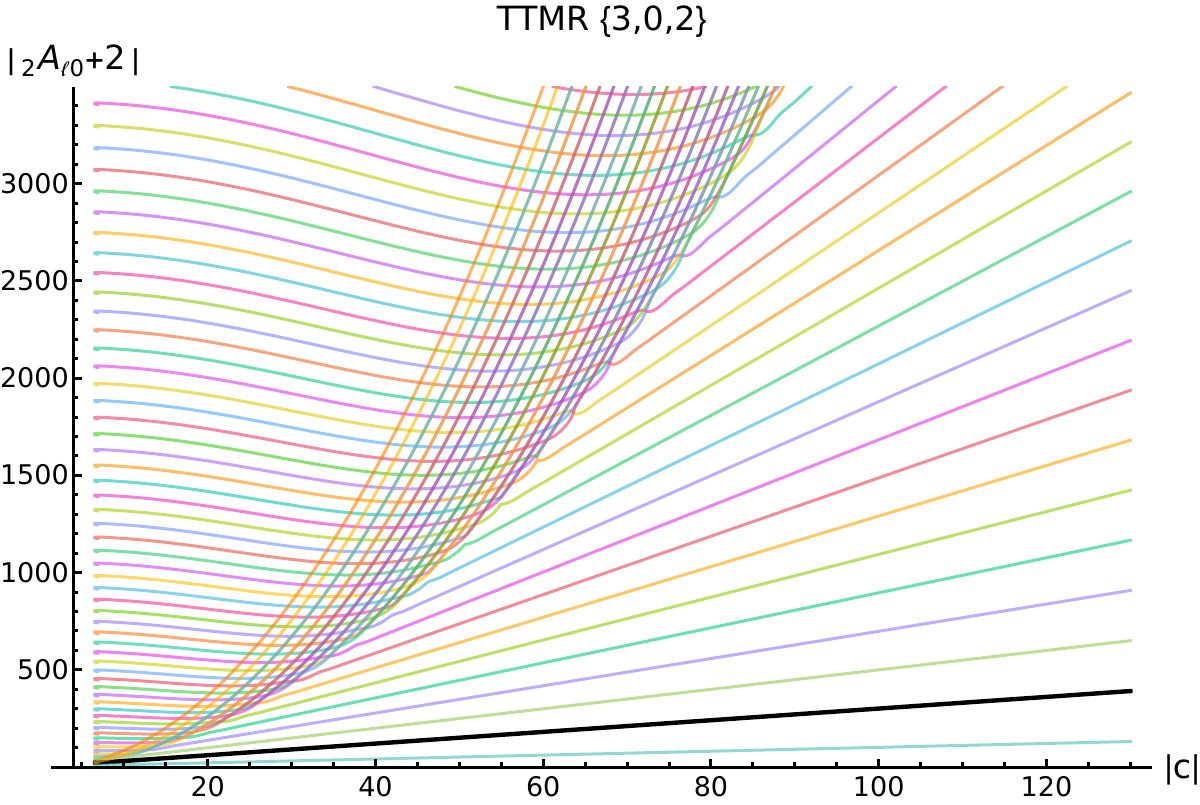}
	\caption{A large number of eigenvalues for $s=2$ and $m=0$ along a sequence of values of $c$ obtained from the TTM${}_{\rm R}$ mode sequence $\{3,0,2\}$.  The plot displays sequences of $|\scA{2}{\ell0}{c}+2|$ [see Eq.~(\ref{eqn:swSF sx A ident})] so that the eigenvalues appear in their sorted order.  The values are display as functions of $|c|$ where the oblateness parameter $c=a\omega^+_{302}(a)$, and $\omega^+_{302}(a)$ is the complex mode frequency along the TTM${}_{\rm R}$ sequence.  The eigenvalue sequence corresponding to the actual $\{3,0,2\}$ mode sequences is displayed as the thick black line.}
	\label{fig:TTMR302AbsAlm_cAll}
\end{figure}

\begin{figure}[h!]
	\centering
	\includegraphics[width=\linewidth,clip]{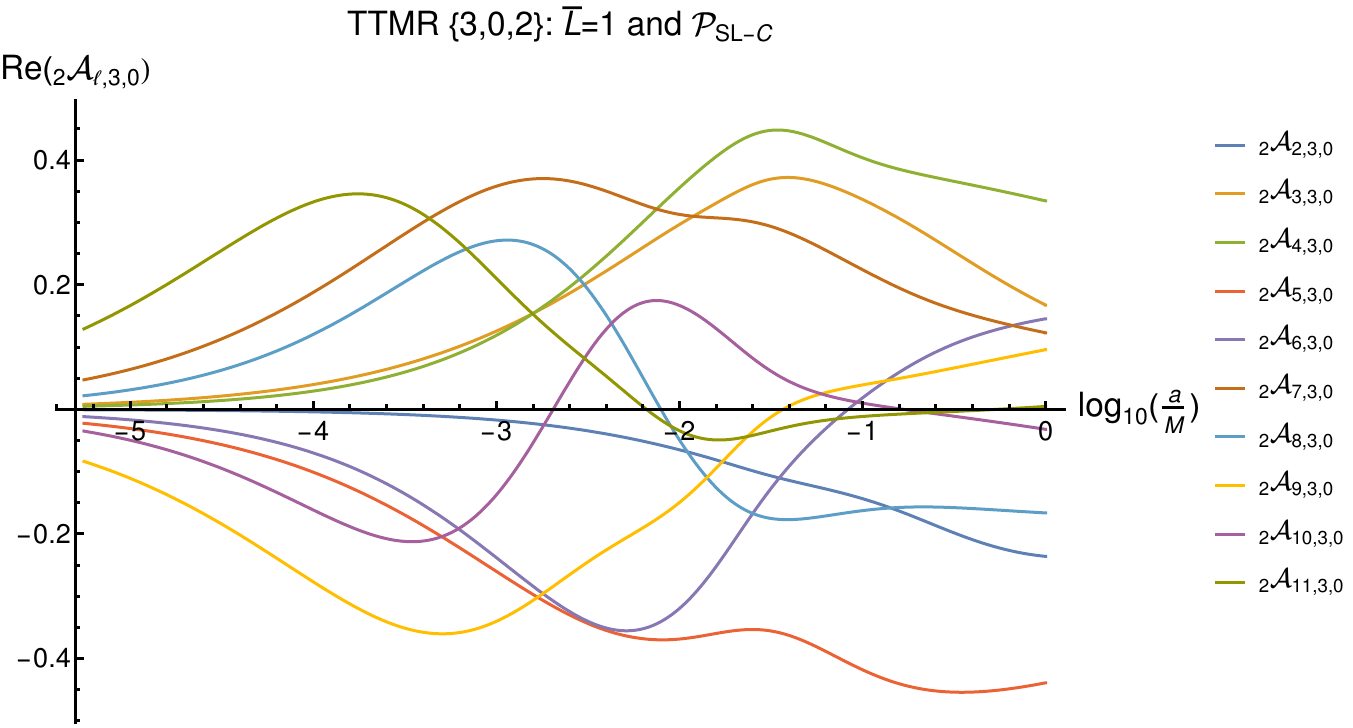}
	\caption{The real part of the first $10$ expansion coefficients $\YSH{2}{\ell30}{a\omega^+_{302}(a)}$ for $\bar{L}=1$ are plotted to demonstrate that the expansion coefficients are smooth functions of $a$ for the $\Phase{SL-C}$ phase choice.}
	\label{fig:TTMR302Lb1ExCoefRe}
\end{figure}

\begin{figure}[h!]
	\centering
	\includegraphics[width=\linewidth,clip]{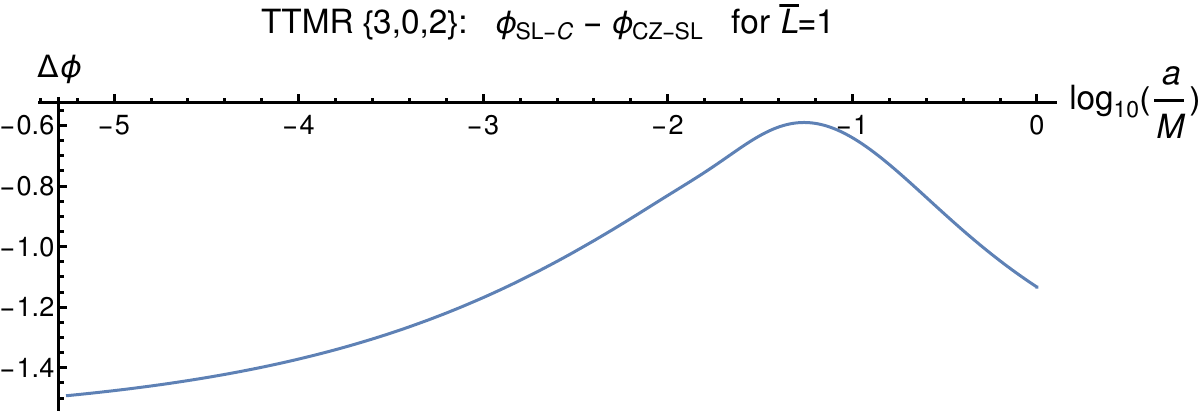}
	\includegraphics[width=\linewidth,clip]{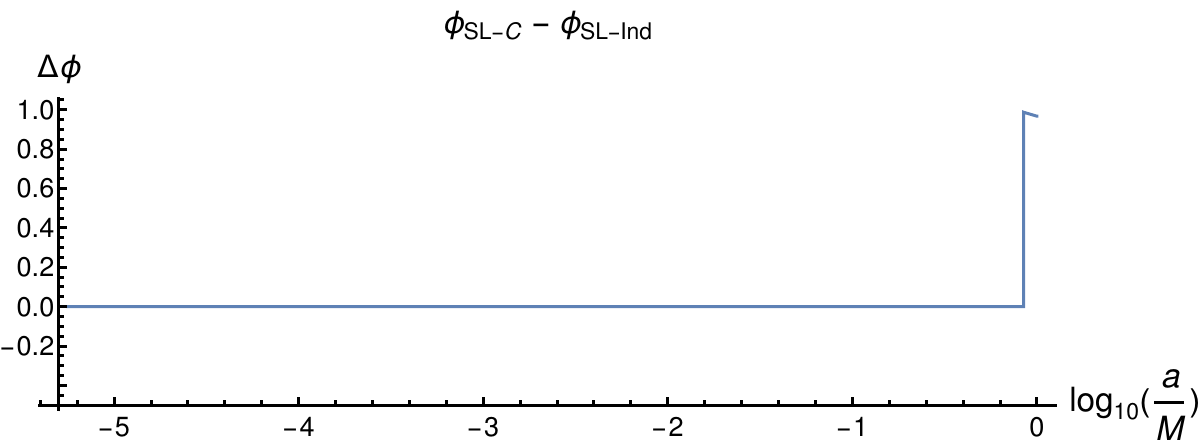}
	\caption{The phase difference between the functions $\swS{2}{30}{x}{a\omega^+_{302}(a)}$ along $\bar{L}=1$ which have been phase fixed using different schemes.  In the top plot, the functions have been phase fixed using $\Phase{SL-C}$ and $\Phase{CZ-SL}$.  In the bottom plot, the functions have been phase fixed using $\Phase{SL-C}$ and $\Phase{SL-Ind}$.}
	\label{fig:TTMR302Lb1phasediff}
\end{figure}

\begin{figure}[h!]
	\centering
	\includegraphics[width=\linewidth,clip]{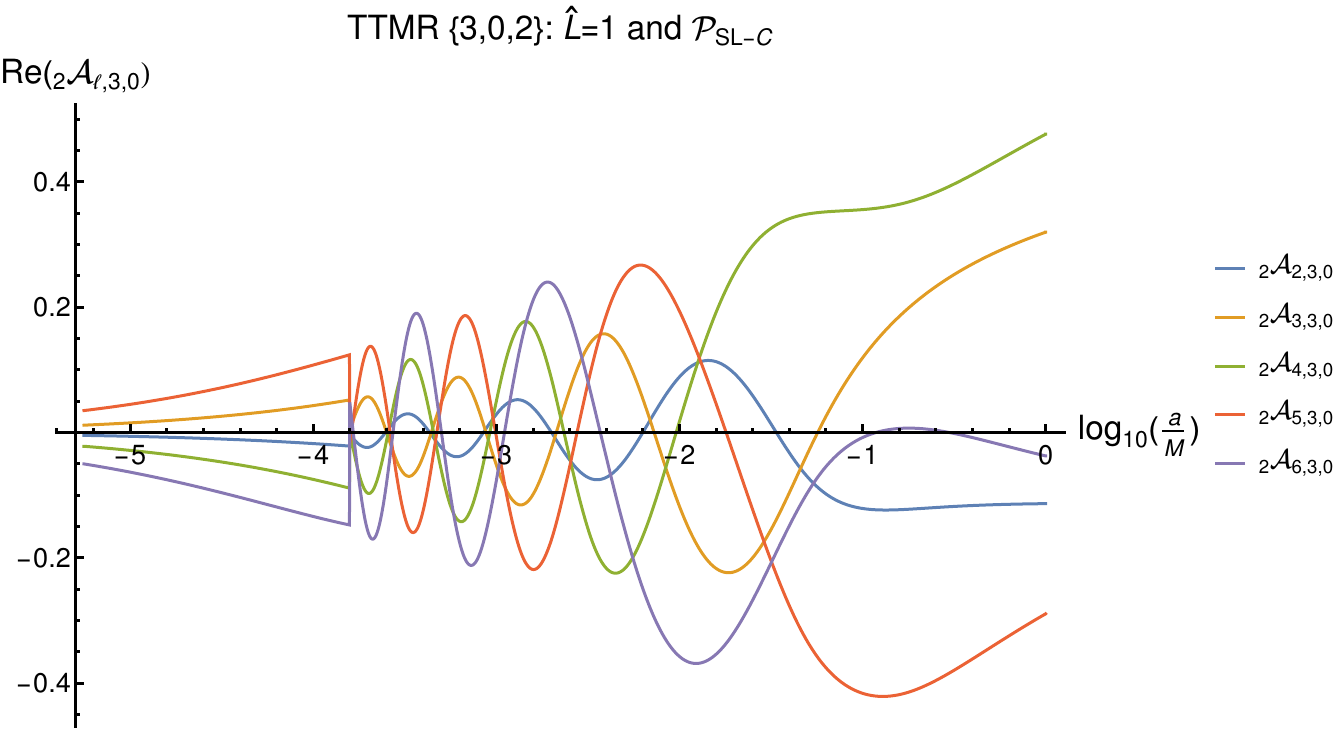}
	\caption{The real part of the first $5$ expansion coefficients $\YSH{2}{\ell30}{a\omega^+_{302}(a)}$ for $\hat{L}=1$ are plotted to demonstrate how the $\Phase{SL-C}$ scheme can lead to nonsmooth behavior along the sequence when the eigenfunction becomes very small at $x=0$.  Note the discontinuous behavior at $\log_{10}(a/M)\approx-3.8$ where the derivative of the function becomes too small to accurately set the phase.}
	\label{fig:TTMR302Lh1ExCoefRe}
\end{figure}

\begin{figure}[h!]
	\centering
	\includegraphics[width=\linewidth,clip]{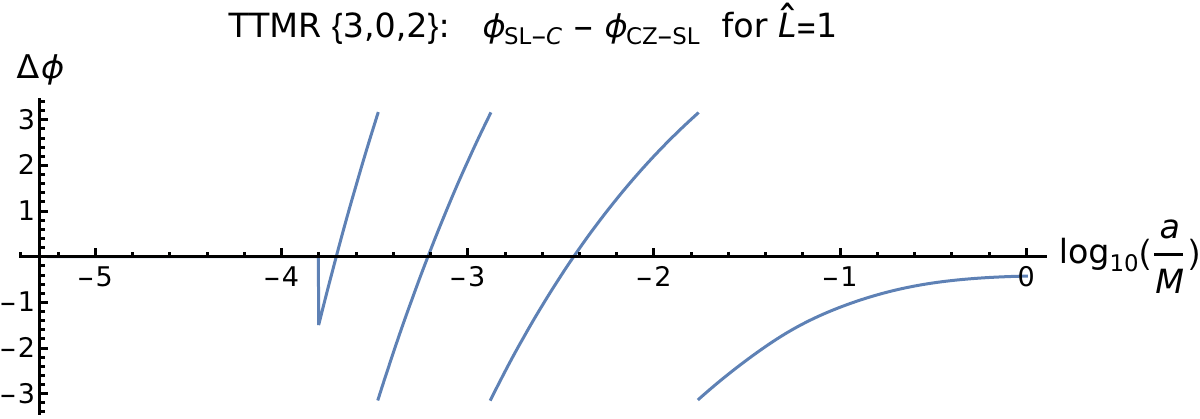}
	\includegraphics[width=\linewidth,clip]{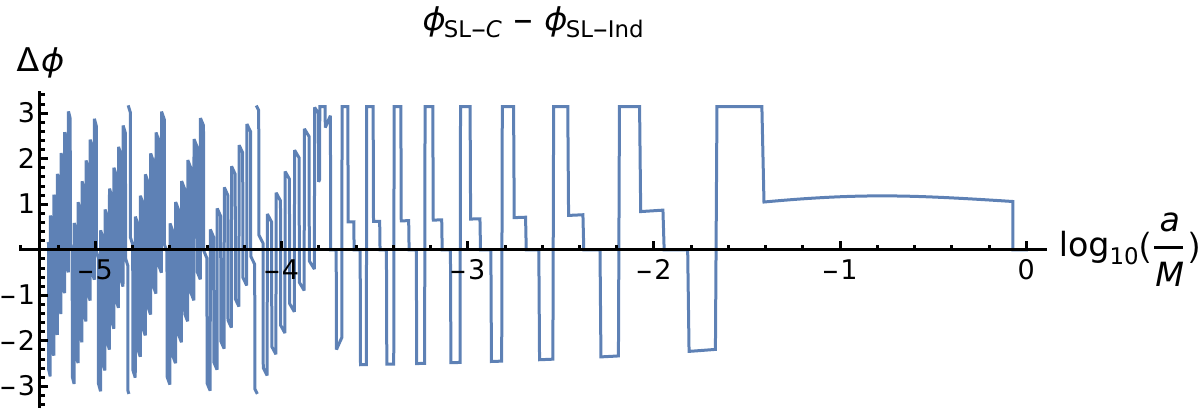}
	\caption{The phase difference between the functions $\swS{2}{30}{x}{a\omega^+_{302}(a)}$ along $\hat{L}=1$ which have been phase fixed using different schemes.  In the top plot, the functions have been phase fixed using $\Phase{SL-C}$ and $\Phase{CZ-SL}$.  In the bottom plot, the functions have been phase fixed using $\Phase{SL-C}$ and $\Phase{SL-Ind}$.}
	\label{fig:TTMR302Lh1phasediff}
\end{figure}

\section{The {\tt SWSpheroidal} {\em Mathematica} Paclet}\label{sec:SWSpheroidal}
The {\tt SWSpheroidal} {\em Mathematica} Paclet is included in the larger {\tt KerrModes} suite of publicly available {\em Mathematica} Paclets\cite{KerrModes-1.0.6}.  It includes two independent approaches for obtaining solutions of the angular Teukolsky equation Eq.~(\ref{eqn:Angular Teukolsky Equation}).  The first approach is to solve a discrete eigenvalue problem based on the spectral method developed by Cook and Zalutskiy\cite{cook-zalutskiy-2014}.  The second approach loosely follows the approach introduced by Fiziev\cite{Fiziev-2010} (see also Chen {\em et al}.\cite{Chen-etal-2025}) and uses the {\em Mathematica} routine {\tt HeunC} to directly express general solutions of the angular Teukolsky equation.  Both approaches will be described in more detail below and produce essentially identical results to very high precision.

\subsection{The spectral solver}\label{sub:spectral_solver}
Full details of the implementation of the spectral solver can be found in Ref.~\cite{cook-zalutskiy-2014}.  Summarizing, Equation~(\ref{eqn:Angular Teukolsky Equation}) is transformed into a discrete eigenvalue problem by inserting Eq.~(\ref{eqn:swSF-expansion}) after truncating the series to $N$ terms.  An $N\times N$ pentadiagonal matrix is obtained and solved numerically using the {\em Mathematica} routine {\tt Eigensystem}.  A subset of the returned eigensolutions will be highly accurate approximations to the exact solutions of Eq.~(\ref{eqn:Angular Teukolsky Equation}), with the components of the eigenvector being the expansion coefficients $\YSH{s}{\hat\ell\ell{m}}{c}$ with $\hat\ell$ corresponding to the eigenvector component index.

This approach is implemented in the routine {\tt SpinWeightedSpheroidal[m,s,c,N]}, where {\tt m} is the azimuthal index, {\tt s} is the spin weight, {\tt c} is the complex oblateness parameter, and {\tt N} is the size of the matrix.  The accuracy of solutions is limited by the precision used to express {\tt c} and by the size of the matrix.  Using {\em Mathematica}'s extended precision mathematics and a sufficiently large matrix size, highly accurate solutions can be obtained.  {\tt SpinWeightedSpheroidal} returns the eigenvalues and eigenvectors in the same standard format used by the built-in function {\tt Eigensystem}, but the solutions have been reordered based on the magnitude of the eigensolutions $|\scA{s}{\ell{m}}{c}+s|$.

Because of the finite truncation of Eq.~(\ref{eqn:swSF-expansion}) some of the eigensolutions returned by {\tt SpinWeightedSpheroidal} will not be accurate.  It is important to verify that the eigenvector for a solution is accurately representing $\swS{s}{\ell{m}}{x}{c}$ via Eq.~(\ref{eqn:swSF-expansion}) before trusting either the eigenvalue or the expansion coefficients.  If the matrix size {\tt N} has been chosen large enough for a given eigensolution to be in the convergent regime, then the magnitudes of the components of the eigenvector should be decreasing exponentially and the last few elements of the eigenvector should be negligible.

There are several additional auxiliary routines associated with the spectral solver.  Most important among these are {\tt SWSFfixphase} which computes the phase factor $\Pfactor{a}$ needed to impose a given phase-fixing scheme $\Phase{a}$ on the solutions returned by {\tt SpinWeightedSpheroidal}, and {\tt SWSFvalues} which computes values for $\swS{s}{\ell{m}}{x}{c}$ over the range $-1\le x\le1$.

\subsection{The confluent Heun solver}\label{sub:confluent_heun_solver}
The angular Teukolsky equation is an example of a confluent Heun equation.  Following the notation used in Refs.~\cite{Heun-eqn,cook-zalutskiy-2014}, the confluent Heun equation can be written as
\begin{align}
	\frac{d^2H(z)}{dz^2} + \left(4p +\frac{\gamma}{z}+\frac{\delta}{z-1} \right)\frac{dH(z)}{dz}
	& \\ + \frac{4\alpha pz-\sigma}{z(z-1)}&H(z)=0, \nonumber
\end{align}
and local solutions at the regular singular points $z=0$ and $z=1$ can be expressed in terms of the confluent Heun function $Hc^{(a)}(p,\alpha,\gamma,\delta,\sigma,z)$, while the local solution at the irregular singular point $z=\infty$ is expressed in terms of $Hc^{(r)}(p,\alpha,\gamma,\delta,\sigma,z)$.  These solutions are defined explicitly by Eqs.~(15)--(19) in Ref.~\cite{cook-zalutskiy-2014}.  Additional details on expressing the angular Teukolsky equation as a confluent Heun equations can be found in Sec.~II.B and II.D of Ref.~\cite{cook-zalutskiy-2014}.\footnote{There is an error in the text immediately preceding Eqs.~(45d) and (45e) of Ref.~\cite{cook-zalutskiy-2014}.  There are only 4 possible combinations.  The same sign choice must be made for both $s$ and $m$ in both equations.}  

Solutions of the angular Teukolsky equation which are regular at either $x=\pm1$ can be written in a number of ways in terms of the confluent Heun function $Hc^{(a)}(p,\alpha,\gamma,\delta,\sigma,z)$, and {\em Mathematica} implements this via the built-in function {\tt HeunC}.  We note that {\em Mathematica} uses a slightly different notation for the confluent Heun function.  In {\em Mathematica}, it is written as
\begin{align}
	{\tt HeunC[q,\alpha,\gamma,\delta,\epsilon,z]} = Hc^{(a)}(\epsilon/4,\alpha/\epsilon,\gamma,\delta,q,z).
\end{align}
\begin{widetext}
There are 4 fundamental ways to express the angular Teukolsky functions which are regular at $x=1$:
\begin{subequations}
\label{eqn:Sp1}
\begin{align}
\label{eqn:Sp1a}
	\swS{s}{\ell{m}}{x}{c} \propto e^{cx}(1-x)^{\frac{m+s}2}&(x+1)^{\frac{m-s}2} \times \nonumber\\
		&{\tt HeunC}\Bigl[\scA{s}{\ell{m}}{c}+c^2-2c(m+2s+1)-m(m+1)+s(s+1), \\
		&\qquad\qquad\qquad-4c(m+s+1),m+s+1,m-s+1,-4c,\frac{1-x}2\Bigr], \nonumber\\
\label{eqn:Sp1b}
	\swS{s}{\ell{m}}{x}{c} \propto e^{cx}(1-x)^{\frac{m+s}2}&(x+1)^{-\frac{m-s}2} \times \nonumber\\
		&{\tt HeunC}\Bigl[\scA{s}{\ell{m}}{c}+c^2-2c(m+2 s+1),\\
		&\qquad\qquad\qquad-4c(2s+1),m+s+1,s-m+1,-4c,\frac{1-x}2\Bigr], \nonumber\\
\label{eqn:Sp1c}
	\swS{s}{\ell{m}}{x}{c} \propto e^{cx}(1-x)^{-\frac{m+s}2}&(x+1)^{-\frac{m-s}2} \times \nonumber\\
		&{\tt HeunC}\Bigl[\scA{s}{\ell{m}}{c}+c^2+2c(m-1)-m(m-1)+s(s+1),\\
		&\qquad\qquad\qquad4c(m-s-1),-m-s+1,s-m+1,-4c,\frac{1-x}2\Bigr], \nonumber\\
\label{eqn:Sp1d}
	\swS{s}{\ell{m}}{x}{c} \propto e^{cx}(1-x)^{-\frac{m+s}2}&(x+1)^{\frac{m-s}2} \times \nonumber\\
		&{\tt HeunC}\Bigl[\scA{s}{\ell{m}}{c}+c^2+2c(m-1)+2s,\\
		&\qquad\qquad\qquad-4c,-m-s+1,m-s+1,-4c,\frac{1-x}2\Bigr]. \nonumber
\end{align}
\end{subequations}
Equations~(\ref{eqn:Sp1a}) and (\ref{eqn:Sp1b}) are regular at $x=1$ for $m+s\ge0$, while Eqns.~(\ref{eqn:Sp1c}) and (\ref{eqn:Sp1d}) are regular there for $m+s\le0$.  Using the identity
\begin{align}\label{eqn:heunident}
	{\tt HeunC}[q,\alpha,\gamma,\delta,\epsilon,z] 
		\propto e^{-\epsilon z}{\tt HeunC}[q-\epsilon \gamma,\alpha-\epsilon(\gamma+\delta),\gamma,\delta,-\epsilon,z],
\end{align}
it is possible to construct 4 additional regular solutions where the exponential term is $e^{-cx}$.  We find that Eqs.~(\ref{eqn:Sp1}) yield the most accurate results when $\Re(c)\ge0$ while the corresponding 4 equations transformed by Eq.~(\ref{eqn:heunident}) yield the most accurate results when $\Re(c)<0$.  We also find that, for any given $s$ and $m$, the most accurate results are obtained when $\swS{s}{\ell{m}}{x}{c}$ is evaluated using the version of Eqs.~(\ref{eqn:Sp1}) where the powers of the remaining prefactors are positive.  That is, when the prefactors are $(1-x)^{\frac{|m+s|}2}(x+1)^{\frac{|m-s|}2}$.

There are also 4 fundamental ways to express the angular Teukolsky functions which are regular at $x=-1$:
\begin{subequations}
\label{eqn:Sm1}
\begin{align}
\label{eqn:Sm1a}
	\swS{s}{\ell{m}}{x}{c} \propto e^{-cx}(1-x)^{\frac{m+s}2}&(x+1)^{\frac{m-s}2} \times \nonumber\\
		&{\tt HeunC}\Bigl[\scA{s}{\ell{m}}{c}+c^2-2c(m-2s+1)-m(m+1)+s(s+1), \\
		&\qquad\qquad\qquad-4c(m-s+1),m-s+1,m+s+1,-4c,\frac{x+1}2\Bigr], \nonumber\\
\label{eqn:Sm1b}
	\swS{s}{\ell{m}}{x}{c} \propto e^{-cx}(1-x)^{-\frac{m+s}2}&(x+1)^{\frac{m-s}2} \times \nonumber\\
		&{\tt HeunC}\Bigl[\scA{s}{\ell{m}}{c}+c^2-2c(m-2s+1)+2s,\\
		&\qquad\qquad\qquad4c(2s-1),m-s+1,-m-s+1,-4c,\frac{1+x}2\Bigr], \nonumber\\
\label{eqn:Sm1c}
	\swS{s}{\ell{m}}{x}{c} \propto e^{-cx}(1-x)^{-\frac{m+s}2}&(x+1)^{-\frac{m-s}2} \times \nonumber\\
		&{\tt HeunC}\Bigl[\scA{s}{\ell{m}}{c}+c^2+2c(m-1)-m(m-1)+s(s+1),\\
		&\qquad\qquad\qquad4c(m+s-1),s-m+1,-m-s+1,-4c,\frac{1+x}2\Bigr], \nonumber\\
\label{eqn:Sm1d}
	\swS{s}{\ell{m}}{x}{c} \propto e^{-cx}(1-x)^{\frac{m+s}2}&(x+1)^{-\frac{m-s}2} \times \nonumber\\
		&{\tt HeunC}\Bigl[\scA{s}{\ell{m}}{c}+c^2+2c(m-1),\\
		&\qquad\qquad\qquad-4c,s-m+1,m+s+1,-4c,\frac{1+x}2\Bigr]. \nonumber
\end{align}
\end{subequations}
Equations~(\ref{eqn:Sm1a}) and (\ref{eqn:Sm1b}) are regular at $x=-1$ for $m-s\ge0$, while Eqns.~(\ref{eqn:Sm1c}) and (\ref{eqn:Sm1d}) are regular there for $m-s\le0$.  We find that Eqs.~(\ref{eqn:Sm1}) yield the most accurate results when $\Re(c)\ge0$ while the corresponding 4 equations transformed by Eq.~(\ref{eqn:heunident}) yield the most accurate results when $\Re(c)<0$.  We again find that, for any given $s$ and $m$, the most accurate results are obtained when $\swS{s}{\ell{m}}{x}{c}$ is evaluated using the version of Eqs.~(\ref{eqn:Sm1}) where the powers of the remaining prefactors are positive.
\end{widetext}

Equations~(\ref{eqn:Sp1}) and their transformed versions yield valid solutions of the angular Teukolsky equation over the range $-1<x\le1$ for any values of $\scA{s}{\ell{m}}{c}$, but are in general indeterminate at $x=-1$.  Similarly, Eqs.~(\ref{eqn:Sm1}) and their transformed versions yield valid solutions of the angular Teukolsky equation over the range $-1\le x<1$ for any values of $\scA{s}{\ell{m}}{c}$, but are in general indeterminate at $x=+1$.  Solutions of the angular Teukolsky equation which are regular over the full range $-1\le x\le1$ are only possible for particular values of $\scA{s}{\ell{m}}{c}$.  This eigenvalue problem is solved by finding the values of $\scA{s}{\ell{m}}{c}$ for which two solutions, one regular at $x=1$ and the other at $x=-1$, become linearly dependent.  These two solutions are linearly dependent when their Wronskian vanishes.  In principle, the Wronskian can be formed from any solution from the family of Eqs.~(\ref{eqn:Sp1}) together with any solution from the family of Eqs.~(\ref{eqn:Sm1}).  In practice, the most accurate results are obtained by choosing solutions based on the values of $s$, $m$, and $c$.  For example, if $m+s\ge0$, $m-s\ge0$, and $\Re(c)\ge0$, then it is most accurate to construct the Wronskian from Eqs.~(\ref{eqn:Sp1a}) and (\ref{eqn:Sm1a}) and evaluate the Wronskian at $x=0$.

This approach is implemented in the routine {\tt SWSpheroidalWronskianRoot[s,m,c,A]}, where {\tt s} is the spin weight, {\tt m} is the azimuthal index, {\tt c} is the complex oblateness parameter, and {\tt A} is an initial guess for the value of the root $\scA{s}{\ell{m}}{c}$ which will be returned.  There are a countably infinite number of roots of the Wronskian, labeled by the polar index $\ell$, and the initial guess is necessary to select which root will be found.

Given an accurate value for $\scA{s}{\ell{m}}{c}$, the spin-weighted spheroidal function $\swS{s}{\ell{m}}{x}{c}$ can be accurately evaluated.  This is implemented in the routine {\tt SWSFHeun[norm,s,m,c,A,x]}, where {\tt s} is the spin weight, {\tt m} the azimuthal index, and {\tt c} the complex oblateness parameter corresponding to the accurate root of the Wronskian {\tt A}.  {\tt SWSFHeun} returns the value of $\swS{s}{\ell{m}}{x}{c}$ at any value of $x$ in the range $-1\le x\le1$.  Even though an accurate value for $\scA{s}{\ell{m}}{c}$ is used in evaluating 
$\swS{s}{\ell{m}}{x}{c}$, the solutions from the family of Eqs.~(\ref{eqn:Sp1}) cannot be evaluated at $x=-1$, and become less accurate close to $x=-1$.  Therefore, solutions from the family of Eqs.~(\ref{eqn:Sp1}) are only used to evaluate $\swS{s}{\ell{m}}{x}{c}$ in the range $0\le x\le1$.  Similarly, solutions from the family of Eqs.~(\ref{eqn:Sm1}) are used to evaluate $\swS{s}{\ell{m}}{x}{c}$ in the range $-1\le x<0$.  The {\em Mathematica} function {\tt HeunC} is normalized by ${\tt HeunC[q,\alpha,\gamma,\delta,\epsilon,0]}=1$, but this does not properly normalize the spin-weighted spheroidal function.  The parameter {\tt norm} is a 2-element list providing the normalization and phase factors needed to correctly evaluate the spin-weighted spheroidal function $\swS{s}{\ell{m}}{x}{c}$.

The normalization is computed by the routine {\tt SWSFHeunNorms[s,m,c,A]} which returns a 2-element list containing the normalization factors for the two different functions used to evaluate $\swS{s}{\ell{m}}{x}{c}$.  The first element is used for the region $-1\le x<0$, and the second element is for $0\le x\le1$.  The normalization factors alone do not yield a smooth function across $x=0$.  A consistent phase choice is computed by the routine {\tt SWSFHeunPhases[s,m,l,c,A]} which returns a 2-element list of complex phase factors.  {\tt SWSFHeunPhases[s,m,l,c,A]} implements the spherical limit phase choice $\Phase{SL}$ from Sec.~\ref{sec:SL-phase}, and the parameter {\tt l} is the polar index $\ell$, which can be chosen to give either the continuous $\Phase{SL-C}$ or the discontinuous $\Phase{SL-Ind}$ phase choices.  The element-wise product of the 2-element lists returned by {\tt SWSFHeunNorms} and {\tt SWSFHeunPhases} can be passed as the {\tt norm} parameter of {\tt SWSFHeun}.

Finally, the routine {\tt SWSFHeunExpansionCoefs[s,m,l,c,A]} returns a list of the expansion coefficients $\YSH{s}{\hat\ell\ell{m}}{c}$ with $\hat\ell$ corresponding to the list index.  These coefficients can be used to represent the spin-weighted spheroidal function via Eq.~(\ref{eqn:swSF-expansion}), and are computed by direct integration using the orthogonality of the spin-weighted spherical functions.  The first element in the returned list corresponds to $\hat\ell=\max(|m|,|s|)$, and additional elements are computed until $1-\sum_{\hat\ell}|\YSH{s}{\hat\ell\ell{m}}{c}|^2<\varepsilon$ and $|\YSH{s}{\hat\ell\ell{m}}{c}|<\varepsilon$ for the last two elements.  For default machine precision calculations, the default value of $\varepsilon=10^{-8}$ is used, but smaller values of $\varepsilon$ can be specified if higher precision calculations are allowed.  For example, $\varepsilon=10^{-15}$ is typically possible if a working precision of 24 decimal digits is specified.

\end{document}